\documentclass[
	final,
  british,
  a4paper,
	format=acmsmall,
	screen=true,
	natbib=true,
	authorversion=true,
	nonacm=true,
]{acmart}
\makeatletter

\newcommand*{\installoption}[2][acmart]{
	\expandafter\newif\csname if#2\endcsname
	\@ifclasswith{#1}{#2}{
		\csname#2true\endcsname
	}{}
}
\newcommand*{\obeyoption}[2]{
	\csname if#2\endcsname\else\csname #1false\endcsname\fi
}
\newcommand*{\excludeoption}[2]{
	\csname if#2\endcsname\csname #1false\endcsname\fi
}
\makeatother

\installoption{final}
\installoption{draft}
\installoption{tikzcache}

\usepackage{xparse}
\usepackage{xpatch}
\usepackage{etoolbox}
\usepackage{environ}

\makeatletter
\ifdef{\IfLabelExistsTF}{}{
\newcommand{\IfLabelExistsTF}[3]{\@ifundefined{r@#1}{#3}{#2}}
\newcommand{\IfLabelExistsT}[2]{\IfLabelExistsTF{#1}{#2}{}}
\newcommand{\IfLabelExistsF}[2]{\IfLabelExistsTF{#1}{}{#2}}}
\makeatother

\usepackage[utf8]{inputenc}
\usepackage{microtype}

\usepackage[british]{babel}
\usepackage[all]{foreign}

\usepackage{amsmath}
\usepackage{amsthm}
\usepackage{amsfonts}
\usepackage{stmaryrd}
\SetSymbolFont{stmry}{bold}{U}{stmry}{m}{n}
\usepackage{mathtools}

\allowdisplaybreaks

\usepackage{thmtools}

\usepackage{float}
\usepackage{caption}
\usepackage{subcaption}

\usepackage{tabularx,xtab,ragged2e}
\usepackage{booktabs}

\usepackage{enumitem}

\usepackage{tikz}
\usetikzlibrary{calc}
\usetikzlibrary{arrows}
\usetikzlibrary{matrix}
\usetikzlibrary{positioning}
\usetikzlibrary{intersections}
\usetikzlibrary{trees}
\usetikzlibrary{decorations.pathmorphing}
\usetikzlibrary{decorations.markings}

\tikzset{
	identity/.style={double,double equal sign distance},
	nat/.style={double,double equal sign distance},
	nat>/.style={nat,-implies},
	<nat/.style={nat,implies-},
	descr/.style={anchor=center,fill=white},
	cross/.style={preaction={draw=white, -,line width=2pt}},
	extended/.style={shorten >=-0pt, shorten <=-0pt},
	marker/.style={font=\footnotesize\normalfont},
	diagram font/.style={font=\small},
	text decoration line/.style={
		line width=.1ex,solid,
		rounded corners=0,
		round cap-round cap},
	capped line/.style={round cap-round cap},
}

\iftikzcache
	\usetikzlibrary{external}
	\tikzset{
		/pgf/images/include external/.code={\includegraphics[draft=false]{#1}},
		prevent next externalization/.style={external/export next=false},
	}
\else
	\tikzset{
		prevent next externalization/.style={},
	}
\fi

\usepackage{nameref}
\usepackage{hyperref}
\AtEndPreamble{\hypersetup{
	linktoc=all,
	final
}}
\usepackage[capitalise,noabbrev,nameinlink]{cleveref}

\makeatletter
\newcommand{\customlabel}[4][0]{%
	\protected@write\@auxout{}{\string\newlabel{#3}{{#4}{\thepage}{#4}{#3}{}}}%
	\protected@write\@auxout{}{\string\newlabel{#3@cref}{{[#2][#1][#1]#4}{\thepage}}}%
}

\newcommand{\crefv}[1]{%
	\begingroup\@cref@compressfalse\@cref@sortfalse\cref{#1}\endgroup%
}
\newcommand{\Crefv}[1]{%
	\begingroup\@cref@compressfalse\@cref@sortfalse\Cref{#1}\endgroup%
}
\newcommand{\crefabbrev}[1]{%
	\begingroup\@cref@abbrevtrue\cref{#1}\endgroup%
}
\makeatother

\AtEndPreamble{
	\theoremstyle{acmplain}

	\theoremstyle{acmdefinition}
	\newtheorem{remark}[theorem]{Remark}
}

\crefname{proof}{proof of}{proofs of}
\Crefname{proof}{Proof of}{Proofs of}

\makeatletter
\newcounter{proofatend@statemet}

\newcommand*\fixstatement[2][Proof of]{%
	\expandafter\preto\csname end#2\endcsname{%
		\global\def\proofatend@proofof{#1}%
		\stepcounter{proofatend@statemet}%
		\label{proofatend:\theproofatend@statemet}%
}}

\def\proofatend@toks{1000}
\newcounter{proofatend@stored}
\newcounter{proofatend@printed}

\let\proofatend@omit0
\let\proofatend@inplace1
\let\proofatend@postpone2

\newcommand{\omitproofs}{\global\let\proofatend@mode\proofatend@omit}
\newcommand{\postponeproofs}{\global\let\proofatend@mode\proofatend@postpone}
\newcommand{\keepproofs}{\global\let\proofatend@mode\proofatend@inplace}

\postponeproofs

\NewEnviron{proofatend}[1][Proof]{\ignorespaces%
	\if\proofatend@mode\proofatend@inplace%
		\begin{proof}[#1]\BODY\end{proof}%
	\else\if\proofatend@mode\proofatend@postpone%
		\edef\next{%
			\noexpand\begin{proof}[{\proofatend@proofof~\noexpand\cref{proofatend:\theproofatend@statemet}}]%
			\noexpand\phantomsection%
			\noexpand\label{proofatend:proof-\theproofatend@statemet}%
			\unexpanded\expandafter{\BODY}}%
		\global\toks\numexpr\proofatend@toks+\value{proofatend@stored}\relax=\expandafter{\next\end{proof}}%
		\stepcounter{proofatend@stored}%
	\fi\fi\ignorespacesafterend}

\NewDocumentCommand{\omittedproofs}{o}{%
	\ifnum\value{proofatend@printed}<\value{proofatend@stored}%
		\IfNoValueF{#1}{#1}%
		\count@=\value{proofatend@printed}%
		\loop%
			\the\toks\numexpr\proofatend@toks+\count@\relax%
			\stepcounter{proofatend@printed}%
			\ifnum\count@<\value{proofatend@stored}%
				\advance\count@\@ne%
		\repeat%
	\fi%
}
\makeatother

\AtEndPreamble{
	\fixstatement{theorem}
	\fixstatement{proposition}
	\fixstatement{corollary}
	\fixstatement{lemma}
	\fixstatement{fact}
  \fixstatement[Solution to]{exercise}
}

\makeatletter
\newcommand{\footnoteref}[1]{%
	\protected@xdef\@thefnmark{\ref{#1}}\@footnotemark%
}
\makeatother

\makeatletter
\newcommand{\Superimpose}[2]{%
	{\ooalign{$#1\@firstoftwo#2$\cr\hfil$#1\@secondoftwo#2$\hfil\cr}}}
\makeatother

\makeatletter
\newcommand{\customLabel}[2]{%
	\protected@write \@auxout {}{\string \newlabel {#1}{{#2}{\thepage}{#2}{#1}{}} }%
	\hypertarget{#1}{#2}%
}
\newcommand{\stepCounterCustomLabel}[2]{%
	\protected@write \@auxout {}{\string \newlabel {#1}{{#2}{\thepage}{#2}{#1}{}} }%
	\stepcounter{#2}%
	\the\value{#2}\relax%
}
\makeatother

\newcommand{\phantomLabel}[1]{\phantomsection\label{#1}}

\newcounter{diagrammarker}
\newcommand{\nodeCustomMarker}[3]{%
	\node[marker] at (#1) {(\customLabel{#2}{#3})};
}
\newcommand{\resetDiagMarker}{\setcounter{diagrammarker}{0}}
\newcommand{\nodeDiagMarker}[2]{%
	\refstepcounter{diagrammarker}%
	\nodeCustomMarker{#1}{#2}{\roman{diagrammarker}}
}
\newcommand{\nodeEqMarker}[2]{%
	\refstepcounter{equation}%
	\nodeCustomMarker{#1}{#2}{\theequation}
}

\makeatletter
\newcommand{\iflabelexists}[3]{\@ifundefined{r@#1}{#3}{#2}}
\makeatother

\newcommand{\defeq}{\stackrel{\vartriangle}{=}}

\DeclareMathOperator{\dom}{dom}
\DeclareMathOperator{\cod}{cod}
\DeclareMathOperator{\img}{img}
\DeclareMathOperator{\obj}{obj}

\DeclareMathOperator{\eimg}{eimg}

\DeclareMathOperator{\fix}{fix}
\DeclareMathOperator{\ifix}{\mu}
\DeclareMathOperator{\ffix}{\nu}
\DeclareMathOperator{\fseq}{fin}

\newcommand{\from}{\leftarrow}
\newcommand{\monoto}{\rightarrowtail}

\newcommand{\epito}{\twoheadrightarrow}

\newcommand{\embto}{\hookrightarrow}

\newcommand{\To}{\Rightarrow}

\newcommand{\cat}[1]{{\text{\normalfont\bfseries{#1}}}\xspace}
\newcommand{\Cat}{\cat{Cat}}
\newcommand{\VCat}[1]{\cat{\ensuremath{#1}-Cat}}
\newcommand{\CC}{\cat{C}}
\newcommand{\CD}{\cat{D}}

\newcommand{\CS}{\cat{S}}

\newcommand{\CV}{\cat{V}}
\newcommand{\argcat}[3][]{{\cat{#2}_{#1}\mspace{-.5mu}({#3})}}
\newcommand{\Set}{\cat{Set}}
\newcommand{\Pos}{\cat{Pos}}
\newcommand{\Ord}{\cat{Ord}}
\newcommand{\Cpo}{\cat{Cpo}}
\newcommand{\Cppo}{\cat{Cppo}}
\newcommand{\Cpob}{{\cat{Cpo}_{\mspace{-2mu}\bottom\mspace{-2mu}}}}
\newcommand{\Meas}{\cat{Meas}}

\newcommand{\Func}[2]{{\argcat{Fun}{#1,#2}}}
\newcommand{\VFunc}[3]{{\argcat{\ensuremath{#1}-Fun}{#2,#3}}}
\newcommand{\Endo}[1]{\argcat{End}{#1}}
\newcommand{\VEndo}[2]{\argcat{\ensuremath{#1}-End}{#2}}
\newcommand{\Mnd}[1]{\argcat{Mnd}{#1}}
\newcommand{\MndEndo}[1]{\argcat{MndEnd}{#1}}
\newcommand{\Alg}[1]{\argcat{Alg}{#1}}
\newcommand{\Coalg}[1]{\argcat{Coalg}{#1}}
\newcommand{\Kl}[1]{\argcat{Kl}{#1}}

\newcommand{\Sh}[2][]{\argcat[#1]{Sh}{#2}}
\newcommand{\PSh}[2][]{\argcat[#1]{PSh}{#2}}

\newcommand{\Id}{{Id}}%
\newcommand{\id}{{id}}%

\newcommand{\res}[2]{{\iota_{#1,#2}}}

\newcommand{\str}{{str}}%
\newcommand{\cstr}{{cstr}}%
\newcommand{\dstr}{{dstr}}%

\newcommand{\later}{{\blacktriangleright}}
\newcommand{\latercat}[1]{{_\later#1}}
\newcommand{\nxt}{{next}}%

\newcommand{\bottom}{\perp}

\newcommand{\op}{^{o\mspace{-1mu}p}}

\newcommand{\carr}[1]{{|#1|}}

\newcommand{\underlying}[1]{{\lfloor#1\rfloor}}

\newcommand{\opens}[1][]{\mathcal{O}_{#1}}
\newcommand{\bopens}[1][]{\mathcal{B}_{#1}}
\newcommand{\alexT}[1]{{\mathcal{A}(#1)}}

\ifdraft
	\newcommand{\extP}[1]{{\underline{#1}}}
	\newcommand{\liftKl}[1]{\overline{#1}}
	\newcommand{\extPliftKl}[1]{\liftKl{\extP{#1}}}
	\newcommand{\liftKlextP}[1]{\liftKl{\extP{#1}}}
	\newcommand{\symF}[1]{\overline{#1}}
\else
	\makeatletter
	\def\extP#1{{\mathpalette\extP@{#1}}}
	\def\extP@#1#2{{%
		\tikzset{prevent next externalization}%
		\tikz[baseline=(n.base)]{
			\node[inner sep=1pt,outer sep=0pt] (n) {\(\m@th#1#2\)};
			\draw[text decoration line, capped line] ($(n.south west)+(.1ex,.0ex)$) -- ($(n.south west)+(.2ex,-.1ex)$) -- ($(n.south east)+(-.2ex,-.1ex)$) -- ($(n.south east)+(-.1ex,.0ex)$);
	}}}
	
	\def\liftKl#1{{\mathpalette\liftKl@{#1}}}
	\def\liftKl@#1#2{{%
	\tikzset{prevent next externalization}%
	\tikz[baseline=(n.base)]{
			\node[inner sep=1pt,outer sep=0pt] (n) {\(\m@th#1#2\)};
			\draw[text decoration line, capped line] ($(n.north west)+(.1ex,.0ex)$) -- ($(n.north west)+(.2ex,.1ex)$) -- ($(n.north east)+(-.2ex,.1ex)$) -- ($(n.north east)+(-.1ex,.0ex)$);
	}}}
	
	\def\liftKlextP#1{{\mathpalette\eplk@{#1}}}
	\def\extPliftKl#1{{\mathpalette\eplk@{#1}}}
	\def\eplk@#1#2{{%
		\tikzset{prevent next externalization}%
		\tikz[baseline=(n.base)]{
			\node[inner sep=1pt,outer sep=0pt] (n) {\(\m@th#1#2\)};
			\draw[text decoration line, capped line] ($(n.south west)+(.1ex,.0ex)$) -- ($(n.south west)+(.2ex,-.1ex)$) -- ($(n.south east)+(-.2ex,-.1ex)$) -- ($(n.south east)+(-.1ex,.0ex)$);
			\draw[text decoration line, capped line] ($(n.north west)+(.1ex,.0ex)$) -- ($(n.north west)+(.2ex,.1ex)$) -- ($(n.north east)+(-.2ex,.1ex)$) -- ($(n.north east)+(-.1ex,.0ex)$);
	}}}
	
	\def\symF#1{{\mathpalette\symF@{#1}}}
	\def\symF@#1#2{{%
		\tikzset{prevent next externalization}%
		\tikz[baseline=(n.base)]{
			\node[inner sep=1pt,outer sep=0pt] (n) {\(\m@th#1#2\)};
			\draw[text decoration line, capped line] 
				($(n.north west)+(.1ex,.-.1ex)$) 
				-- 
				($(n.north west)+(.2ex,.1ex)$) 
				--
				($(n.north)+(-0.3ex,.1ex)$) arc (180:0:0.3ex)
				--
				($(n.north east)+(-.1ex,.1ex)$) 
				-- 
				($(n.north east)+(-.0ex,-.1ex)$);
	}}}
	\makeatother
\fi

\title[Infinite Traces by Finality]{Infinite Traces by Finality: a Sheaf-Theoretic Approach}

\author{Marco Peressotti}
\orcid{0000-0002-0243-0480}
\affiliation{
	\department{Department of Mathematics and Computer Science}
  \institution{University of Southern Denmark}
  \streetaddress{Campusvej 55}
  \city{Odense}
  \postcode{5230}
  \country{Denmark}
}
\email{peressotti@imada.sdu.dk}

\begin{abstract}
	Kleisli categories have long been recognised as a setting for modelling the linear behaviour of various types of systems. However, the final coalgebra in such settings does not, in general, correspond to a fixed notion of linear semantics. While there are well-understood conditions under which final coalgebras capture finite trace semantics, a general account of infinite trace semantics via finality has remained elusive.

	In this work, we present a sheaf-theoretic framework for infinite trace semantics in Kleisli categories that systematically constructs final coalgebras capturing infinite traces. 
	Our approach combines Kleisli categories, sheaves over ordinals, and guarded (co)recursion, enabling infinite behaviours to emerge from coherent families of finite approximations via amalgamation. 
	We introduce the notion of guarded behavioural functor and show that, under mild conditions, their final coalgebras directly characterise infinite traces.
\end{abstract}

\begin{document}

\maketitle

\section{Introduction}
Since the seminal paper \cite{pt:entcs1999}, Kleisli categories have been recognised as the context where to model the linear semantics of several types of transition systems as shown by a plethora of works such as \cite{jacobs:cmcs2004,hjs:lmcs2007,cirstea:entcs2010,kk:lmcs2013,hu:calco2015,hsu:concur2016,bp:lmcs2019}. The key idea behind this approach falls under the motto ``change the category not the definition'' and can be traced back to \citeauthor{moggi:lics1989}'s modelling of side effects in Kleisli categories of monads \cite{moggi:lics1989,ms:jfc2001,mp:entcs1999,bhm:appsem2000}.
Roughly speaking, systems are modelled as $TF$-coalgebras where $T$ is a monad describing the ``branching type'' (\eg partiality, non-determinism, probabilistic) and $F$ is an endofunctor describing the ``linear type'' (\eg labelled transitions). Coalgebras of this type form a wide subcategory of coalgebras for certain endofunctors obtained as suitable extensions of $F$ to the Kleisli category of $T$ and called \emph{Kleisli liftings}. In this setting, objects modelling systems are the same of $TF$-coalgebras whereas coalgebra homomorphisms abstract from branching (the computational effect associated to the monad $T$) and hence capture the linear behaviour of systems under scrutiny. In general, final semantics for coalgebras of Kleisli liftings may not coincide with any established notion of trace semantics: there are instances where this results in finite, possibly infinite, infinite only traces, or none at all (see \eg \cite{kk:lmcs2013}).
In \cite{hjs:lmcs2007} \citeauthor{hjs:lmcs2007} presented general and sufficient conditions that ensure finite trace semantics is captured by the final semantics of coalgebras of certain Kleisli liftings. Although there are works recovering (possibly) infinite trace semantics via canonical maps to weakly final coalgebras (see \eg \cite{jacobs:cmcs2004,hjs:lmcs2007,cirstea:entcs2010,hu:calco2015}), a general account on par with those of finite trace semantics is still missing.
In this work we propose a general approach to infinite trace semantics via final semantics.

Our proposal combines three main ingredients: Kleisli liftings, sheaves on ordinals (equipped with the Alexandrov topology), and guarded (co)recursion. Although each of them is widely studied, their combination is the key novelty that allows us to systematically capture infinite traces by finality. Clearly, the r\^ole of Kleisli lifting is to abstract branching. The r\^ole of sheaves and guarded (co)recursion becomes clear when one notes that an infinite object (\eg a stream) is equivalently described by an infinite family of coherent approximations (\eg the countable family of its prefixes).
This is exactly how a global section is characterised from local sections via amalgamation.
If we replace infinite traces and finite traces for streams and words, respectively, then infinite traces are \emph{global observations}, finite traces are partial or \emph{local observations}, and \emph{amalgamation} is the mechanism for obtaining the former from coherent families of the latter. 
From this perspective, observations are naturally organised in sheaves over the Alexandrov topology of an ordinal number---$\omega$, in the case of the example of streams and words considered above.

We introduce the notion of guarded behavioural functor and \emph{guarded coalgebras} as a way to capture local and global observations at once.
Guarded behavioural endofunctors are systematically derived from any behavioural endofunctor while preserving the associated final semantics: there is an inclusion functor that exhibits the category of coalgebras as a (coreflective) subcategory of that of guarded coalgebras. We prove that the final semantics of guarded coalgebras in Kleisli categories always capture infinite trace semantics of the systems under scrutiny.

\paragraph{Synopsis}
In \cref{sec:kleisli-liftings-and-linear-semantics} we shortly describe the modelling of linear and trace semantics via Kleisli liftings and propose a notion of morphisms for relating such models.
In \cref{sec:extp-functor} we consider category-valued sheaves over ordinals equipped with the Alexandrov topology and study extensions of behavioural endofunctors and Kleisli liftings to this setting while preserving the original semantics.
In \cref{sec:guarded-kleisli-corecursion} we consider locally contractive endofunctors over Kleisli categories of monads obtained by pointwise extension as the technical foundation for guarded coalgebras in Kleisli categories.
In \cref{sec:infinite-trace-semantics} we combine all these techniques into the notion of guarded coalgebras and guarded Kleisli liftings and prove that this combination is suitable for capturing infinite trace semantics.
Concluding remarks are in \cref{sec:trace-semantics-remarks}.
Preliminaries on sheaf categories and algebraically compact functors are in \cref{sec:sheaves-sites,sec:algebraically-compact-functors}. 
Omitted proofs are in \cref{sec:omittedproofs}.
This report draws on the results presented by the author in Chapter 3 of his dissertation \cite{peressotti:phdthesis}.

\section{Linear and trace semantics via Kleisli liftings}
\label{sec:kleisli-liftings-and-linear-semantics}

In this section we recall the approach to modelling linear-time semantics (or linear semantics, for short) introduced in \cite{pt:entcs1999} and its application to trace semantics \cite{jacobs:cmcs2004,hjs:lmcs2007}. The section is organised as follows: 
in \cref{sec:kleisli-liftings} we describe endofunctor extensions to Kleisli categories known as Kleisli liftings; 
in \cref{sec:kleisli-coinduction} 
we study coalgebras for Kleisli liftings and their associated notions of final semantics; %
in \cref{sec:kleisli-lifting-distributive-law-morphisms} we propose a notion of morphism between models of linear semantics with the property that they preserve linear bisimulations---hence trace equivalences.

\subsection{Kleisli liftings}
\label{sec:kleisli-liftings}

Let $(T,\mu,\eta)$ be a monad over a category $\CC$ and write $(K \dashv L)\colon \Kl{T} \to \CC$ for the canonical adjunction presenting $\CC$ as a subcategory of $\Kl{T}$. We are interested in extending an endofunctor $F$ from $\CC$ to $\Kl{T}$ in a way that preserves its action on (the image of) $\CC$. This intuition is formalised by the following definition:

\begin{definition}[\cite{mulry:mfps1993}]
	\label{def:kleisli-liftings}
	Let $(T,\mu,\eta)$ be a monad and $F$ an endofunctor, both over some category $\CC$. A \emph{Kleisli lifting} of $F$ to $\Kl{T}$ is any endofunctor $\liftKl{F}$ over $\Kl{T}$ such that the diagram below commutes.
	\begin{equation}	
		\label{eq:kleisli-lifting}
		\begin{tikzpicture}[
				auto, xscale=1.8, yscale=1.2, diagram font,
				baseline=(current bounding box.center)
			]	
			\node (n0) at (0,1) {\(\Kl{T}\)};	
			\node (n1) at (1,1) {\(\Kl{T}\)};	
			\node (n2) at (0,0) {\(\CC\)};	
			\node (n3) at (1,0) {\(\CC\)};	
	
			\draw[right hook->] (n2) to node {\(K\)} (n0);
			\draw[right hook->] (n3) to node[swap] {\(K\)} (n1);
			\draw[->] (n0) to node {\(\liftKl{F}\)} (n1);
			\draw[->] (n2) to node[swap] {\(F\)} (n3);
			
		\end{tikzpicture}
	\end{equation}
\end{definition}

As noted in \cite{mulry:mfps1993}, Kleisli liftings of an endofunctor $F$ along $K\colon \CC \to \Kl{T}$ are uniquely characterised by suitable natural transformations that distribute the monad $T$ over the endofunctor $F$. Formally:

\begin{definition}
	\label{def:kleisli-lifting-dist-law}
	Let $(T,\mu,\eta)$ be a monad and $F$ an endofunctor, both over some category $\CC$. A law distributing $(T,\mu,\eta)$ over $F$ is a natural transformation $\lambda\colon FT \to TF$ compatible with the monad structure of $T$ as stated by the commuting diagrams below.
	\begin{equation*}
	\begin{tikzpicture}[
			 auto, scale=1.6, diagram font,
			 baseline=(current bounding box.center)
		 ]	
		 
		 \foreach \x in {0,1,...,4}{
			 \pgfmathsetmacro\a{162-\x * 72}
			 \coordinate (j\x) at (\a:1.1);
		 }
		 \node (m0) at (j0) {\(F \circ T \circ T\)};
		 \node (m1) at (j1) {\(T \circ F \circ T\)};
		 \node (m2) at (j2) {\(T \circ T \circ F\)};
		 \node (m3) at (j3) {\(T \circ F\)};
		 \node (m4) at (j4) {\(F \circ T\)};
		 
		 \draw[->] (m0) to node {\(\lambda \circ \id_{T}\)} (m1);
		 \draw[->] (m1) to node {\(\id_{T} \circ \lambda\)} (m2);
		 \draw[->] (m2) to node {\(\mu \circ \id_{F}\)} (m3);
		 \draw[<-] (m3) to node {\(\lambda\)} (m4);
		 \draw[<-] (m4) to node {\(\id_{F} \circ \mu\)} (m0);
		 
		 \nodeEqMarker{0,0}{eq:law-dist-mnd-end-mu}
	\end{tikzpicture}
	\qquad
	\begin{tikzpicture}[
			auto, scale=1.6, diagram font,
			baseline=(current bounding box.center)
		]	
		
		\foreach \x in {0,1,...,4}{
			\pgfmathsetmacro\a{-90-\x * 72}
			\coordinate (i\x) at (\a:1);
		}
		
		\node (n0) at (i0) {\(F\)};
		\node (n1) at (i1) {\(F \circ \Id_{\CC}\)};
		\node (n2) at (i2) {\(F \circ T\)};
		\node (n3) at (i3) {\(T \circ F\)};
		\node (n4) at (i4) {\(\Id_{\CC} \circ F\)};
		
		\draw[identity] (n0) -- (n1);
		\draw[->] (n1) to node {\(\id_{F} \circ \eta\)} (n2);
		\draw[->] (n2) to node {\(\lambda\)} (n3);
		\draw[<-] (n3) to node {\(\eta \circ \id_{F}\)} (n4);
		\draw[identity] (n4) -- (n0);
		
		\nodeEqMarker{0,0}{eq:law-dist-mnd-end-eta}
	\end{tikzpicture}
	\end{equation*}
\end{definition}

A precise notation would require to write distributive laws as triples such as $((T,\mu,\eta),F,\lambda)$ in order to keep the monad structure explicit; as common practice, we will often write $\lambda\colon FT \to TF$ or just $\lambda^{T,F}$ akin to how we write $T$ for a monad and denote its multiplication and unit by $\mu^{T}$ and $\eta^{T}$, respectively.

The following fact is stated in \cite[Theorem~2.2]{mulry:mfps1993}; see also \cite{lpw:entcs2000,lpw:tcs2004} for further details and generalisations.
\begin{proposition}
	\label{thm:kleisli-liftings-dist-laws}
	For $(T,\mu,\eta)$ and $F$ a monad and an endofunctor both over a category $\CC$, Kleisli liftings of $F$ to $\Kl{T}$ are in bijective correspondence with laws distributing $(T,\mu,\eta)$ over $F$.
\end{proposition}

\begin{proofatend}
	Let $\lambda$ be a distributive law of $(T,\mu,\eta)$ over $F$.
	For $X$ an object and $f\colon X \to Y$ a morphism of $\Kl{T}$, the assignments
	\begin{equation*}
		X \mapsto FX \qquad f \mapsto \lambda_{Y} \circ Ff
	\end{equation*}
	define an endofunctor $\liftKl{F}$ over $\Kl{T}$. This functor is a Kleisli lifting since:
	\begin{equation*}
		KFf = \eta_{FY} \circ Ff = 
		\lambda_Y \circ F\eta_{Y} \circ Ff = \liftKl{F}Kf
	\end{equation*}
	for any $f\colon X \to Y$ in $\CC$.
		
	For the converse assume $\liftKl{F}$ Kleisli lifting of $F$ and let $\varepsilon$ denote the counit of $(K \dashv L)$. Since $\liftKl{F}K = KF$, define $\lambda'\colon F L \to L \liftKl{F}$ as the transpose of $\liftKl{F} \varepsilon \colon \liftKl{F}KL \to \liftKl{F}$. Then, $\lambda'$ is $(\id_{L\liftKl{F}} \circ \varepsilon) \bullet (\eta \circ \id_{FL})$ where $\circ$ and $\bullet$ denote horizontal and vertical composition of natural transformations in $\Cat$.
	Since $LK = T$ and $\liftKl{F}K = KF$, $\lambda'K$ is a natural transformation of type $FLK \to LKF$ as needed.
	Compatibility with $(T,\mu,\eta)$ and $F$ follows by diagram chasing.
\end{proofatend}

\paragraph{Canonical liftings via tensorial strength}

Existence of distributive laws of $(T,\mu,\eta)$ over $F$ is not unusual. In fact, there are several classes of monads and functors of interest, w.r.t.~linear and trace semantics, for which Kleisli liftings can be constructed in a canonical way. In particular we mention strong (commutative) monads and polynomial functors \cite{mulry:mfps1993}.

Assume $(\CC,\otimes,I)$ to be a monoidal category and let $a$, $l$, and $r$ denote its associator, left unitor, and right unitor. A monad $(T,\mu,\eta)$ on $\CC$ is called \emph{strong} if it is equipped with a family of morphisms:
\phantomLabel{def:monad-strength}
\begin{equation*}
	\{\str_{X,Y} \colon X \otimes TY \to T(X \otimes Y)\}_{X,Y \in \CC}
	\text{,}
\end{equation*}
called \emph{(tensorial) strength}, which is natural in both components 
and is coherent with the structure of monads and monoidal categories, \ie:
\begin{gather*}
	\mu_{X \otimes Y} \circ T(\str_{X,Y})\circ\str_{X,TY} = 
	\str_{X,Y} \circ (\id_X \otimes \mu_Y)
	\qquad
	\eta_{X\otimes Y} = \str_{X,Y} \circ (\id_{X} \otimes \eta_Y)
	\\
	l_{TX} = T(l_X) \circ \str_{I,X} 
	\qquad
	T(a_{X,Y,Z}) \circ \str_{X \otimes Y, Z} = 
	\str_{X,Y\otimes Z} \circ id_X \otimes \str_{Y,Z} \circ a_{X,Y,TZ}
	\text{.}
\end{gather*}
Dually, a \emph{costrength} for a monad is family:
\phantomLabel{def:monad-costrength}
\begin{equation*}
	\{\cstr_{X,Y} \colon TX \otimes Y \to T(X\otimes Y)\}_{X,Y \in \CC}
	\text{,}
\end{equation*}
which is natural in both $X$ and $Y$ and 
coherent with respect to the structure of $T$ and $\CC$.
Every strong monad on a symmetric monoidal category
has a costrength given on each component as
$
	\cstr_{X,Y}	= T\phi_{Y,X} \circ \str_{Y,X} \circ \phi_{TX,Y}
$
where $\phi = \{\phi_{X,Y}\colon X \otimes Y \cong Y \otimes X\}_{X,Y \in \CC}$ is the braiding natural isomorphism for the symmetric monoidal category $(\CC,\otimes,I)$.
A strong monad on a symmetric monoidal category is called 
\emph{commutative} whenever:
\begin{equation*}
	\mu_{X \otimes Y} \circ T(\str_{X,Y}) \circ \cstr_{X,TY} = 
	\mu_{X \otimes Y} \circ T(\cstr_{X,Y}) \circ \str_{TX,Y}
	\text{.}
\end{equation*}
Every strong commutative monad is a symmetric monoidal monad
(and \emph{vice versa}). In fact, its \emph{double strength}: 
\phantomLabel{def:monad-double-strength}
\begin{equation*}
	\{\dstr_{X,Y} \colon TX \otimes TY \to T(X \otimes Y)\}_{X,Y \in \CC}
	\text{,}
\end{equation*}
can be defined in terms of its (co)strength as follows:
\begin{equation*}
\dstr_{X,Y} = \mu_{X \otimes Y} \circ T(\str_{X,Y}) \circ \cstr_{X,TY} = 
	\mu_{X \otimes Y} \circ T(\cstr_{X,Y}) \circ \str_{TX,Y}\text{.}
\end{equation*}
Conversely, $\str_{X,Y} = \dstr_{X,Y} \circ (\eta_X \otimes id_{TY})$ and
$\cstr_{X,Y} = \dstr_{X,Y} \circ (id_{TX} \otimes \eta_{Y})$.

Kleisli categories of (symmetric) monoidal monads have a canonical (symmetric) monoidal structure, induced by the monoidal structure of the monad and such that the canonical adjunction is a monoidal adjunction with respect to this structure.
For $(T,\mu,\eta,\dstr)$ a strong commutative monad over $(\CC,\otimes,I)$, define $(-\liftKl{\otimes}-)\colon \Kl{T} \times \Kl{T} \to \Kl{T}$ as $(X \otimes X')$, on each pair of objects $X$ and $X'$, and as $f \mathbin{\liftKl{\otimes}} f' = \dstr_{Y,Y'}  \circ (f\times f')$, on each pair of morphisms $f\colon X \to Y$ and $f'\colon X' \to Y'$.
This functor is a lifting of $(-\otimes-)$ along $K\colon \CC \to \Kl{T}$ and forms, together with the unit $I$ of $\otimes$, a monoidal structure on the Kleisli category of $T$. 
We refer the reader to \cite{kock:adm1972,kock:adm1970} for further details on strong and monoidal monads.

\begin{example}
	The powerset functor $\mathcal P$ assigns to any set the set $\mathcal{P}X$ of all its subsets and to any function $f\colon X \to Y$ the function $\mathcal{P}(f)(X') = \{f(x) \mid x \in X'\}$; it admits a monad structure $(\mathcal{P},\mu,\eta)$ whose multiplication and unit are given on each component $X$ as $\mu_X(Y) = \bigcup Y$ and $\eta_X(x) = \{x\}$. This monad is equipped with strength, costrength and double strength given, on each component, as:
	\begin{equation*}
		\str_{X,Y}\left(x,Y'\right) = \{x\} \times Y' \qquad
		\cstr_{X,Y}\left(X',y\right) = X' \times \{y\} \qquad
		\dstr_{X,Y}\left(X',Y'\right) = X' \times Y'
	\end{equation*}
	and hence is strong and commutative (see \eg \cite{kock:adm1972,kock:adm1970,hjs:lmcs2007}).
\end{example}

\begin{example}
	The probability distribution functor $\mathcal{D}$ assigns to any set $X$ the set $\mathcal{D}X = \{{\phi\colon X\to [0,1]}\mid \sum_{x\in X} \phi(x) = 1\}$ of discrete measures and to any function $f\colon X\to Y$ the function $\mathcal{D}f(\phi)(y) = \sum_{f(x) = y}\phi(x)$; it admits a monad structure $(\mathcal{D},\mu,\eta)$ whose multiplication and unit are given on each component $X$ as $\mu_X(\phi)(x) = \sum_{\psi}\psi(x)\cdot\phi(\psi)$ and $\eta_{X}(x) = \delta_x$ where $\delta_x\colon X \to [0,1]$ is the Dirac's delta function. This monad is equipped strength, costrength and double strength given, on each component, as: (see \eg \cite{hjs:lmcs2007}):
	\begin{gather*}
	\str_{X,Y}(x,\psi)(x',y') = \delta_x(x')\cdot\psi(y')\qquad
	\cstr_{X,Y}(\phi,y)(x',y') = \phi(x')\cdot\delta_y(y')\\
	\dstr_{X,Y}(\phi,\psi)(x',y') = \phi(x')\cdot\psi(y')
	\end{gather*}
	and hence is strong and commutative (see \eg \cite{bmp:jlamp2015,hjs:lmcs2007}).
\end{example}

An endofunctor over a category with products and coproducts is called \emph{polynomial} whenever it is formed by constants, products, and coproducts.
Assume $\CC$ has coproducts of cardinality $\kappa$, a polynomial endofunctor over $\CC$ is any endofunctor $F$ generated by the grammar:
\begin{equation*}\textstyle
	F \Coloneqq \Id_\CC \mid A \mid \coprod_{i \in I} F_i \mid F_0 \times F_1
\end{equation*}
where $A$ ranges over $\obj(\CC)$ and $I$ has cardinality at most $\kappa$.
Kleisli liftings for polynomial functors can be constructed by structural recursion: all cases are trivial except for products which require the additional assumption that $T$ is a symmetric monoidal monad with respect to the structure $(\CC,\times,1)$. 
\begin{itemize}
	\item If $F = \Id_\CC$ or $F = A$ then, define $\liftKl{F}$ as $\Id_{\Kl{T}}$ and $A$, respectively. 
	\item If $F = \coprod_{i \in I} F_i$ then, define its Kleisli lifting as the coproduct $\coprod_{i \in I} \liftKl{F_i}$ where each $\liftKl{F_i}$ is the lifting of $F_i$ obtained via this recursive procedure---this yields a Kleisli lifting by construction of each $\liftKl{F_i}$ and by $K\colon\CC \to \Kl{T}$ preserving coproducts. 
	\item If $F = F_0 \times F_1$, define $\liftKl{F}$ as $\liftKl{F_0} \mathbin{\liftKl{\times}} \liftKl{F_1}$ where $\liftKl{\times}$ is the tensor product induced by $\times$ and the monoidal structure of $T$, $\liftKl{F_0}$ and $\liftKl{F_1}$ are obtained via this recursive procedure.
\end{itemize}

\label{sec:kleisli-liftings-examples}

\begin{example}
	LTSs with labels in a given set $A$ (see e.g~\cite{sangiorgi:bisbook}) can be viewed as coalgebras for the endofunctor $\mathcal{P}(A\times \Id)\colon \Set\to \Set$ \cite{rutten:tcs2000}. Since $A \times \Id$ is polynomial and $\mathcal{P}$ is strong and commutative, it is possible to apply the above procedure and construct $\liftKl{A\times \Id}$ canonical Kleisli lifting of $A \times \Id$; this endofunctor over $\Kl{\mathcal{P}}$ acts as $A \times \Id$ on objects and as $\dstr(\eta_A \times \Id)$ on morphisms. 
	For any object $X$ we have:
	\begin{equation*}
		(\liftKl{A\times \Id}) X = A \times X
	\end{equation*}
	and for any morphism $f\colon X\to Y$ in $\Kl{\mathcal{P}}$ we have:
	\begin{equation*}
		\pushQED{\qed} 
		(\liftKl{A\times \Id})(f)(a,x) = \{(a,y)\mid y\in f(x)\}
		\text{.}
		\qedhere
	\end{equation*}
\end{example}

\begin{example}
	Fully-probabilistic systems \cite{vsst:lics1990} are modelled as 
	coalgebras for the endofunctor $\mathcal{D}(A \times \Id)$ on \Set
	\cite{sokolova:tcs2011}.
	Since $A \times \Id$ is polynomial and $\mathcal{D}$ is strong and commutative, it is possible to construct $\liftKl{A\times \Id}$ canonical Kleisli lifting of $A \times \Id$; this endofunctor over $\Kl{\mathcal{D}}$ acts as $A \times \Id$ on objects and as $\dstr(\eta_A \times \Id)$ on morphisms. 
	Since the probability distribution monad $\mathcal{D}$ is strong and commutative, the endofunctor $(A \times \Id)$ has a canonical Kleisli lifting
	$\liftKl{A\times \Id}$ to $\Kl{\mathcal{D}}$ acting as $A \times \Id$ on objects	and as $\dstr(\eta_A \times \Id)$ on morphisms. 
	In particular, for any object $X$ we have:
	\begin{equation*}
		(\liftKl{A\times \Id}) X = A \times X 
	\end{equation*}
	and for any morphism $f\colon X\to Y$ in $\Kl{\mathcal{D}}$ we have:
	\begin{equation*}
		\pushQED{\qed}
		(\liftKl{A\times \Id})(f)(a,x)(b,y) = \delta_{a}(b) \cdot f(x)(y)
		\text{.}
		\qedhere
	\end{equation*}
\end{example}

\subsection{Kleisli coinduction, linear and trace semantics}
\label{sec:kleisli-coinduction}

For $\liftKl{F}$ a Kleisli lifting of $F$ along $K\colon \CC \to \Kl{T}$, the category $\Coalg{TF}$ is a wide subcategory of $\Coalg{\liftKl{F}}$:
the inclusion functor $K$ lifts along the forgetful functors for $\Coalg{TF}$ and $\Coalg{\liftKl{F}}$, as shown in the diagram below, to a functor that acts as the identity on coalgebras and as $K$ on morphisms:
\begin{equation}
	\label{eq:lift-k-coalg}	
	\begin{tikzpicture}[
			auto, xscale=2, yscale=1.2, diagram font,
			baseline=(current bounding box.center)
		]	
		\node[] (n0) at (0,0) {\(\CC\)};	
		\node[] (n1) at (1,0) {\(\Kl{T}\)};	
		\node[] (n2) at (0,1) {\(\Coalg{TF}\)};	
		\node[] (n3) at (1,1) {\(\Coalg{\liftKl{F}}\)};	
		
		\draw[->] (n0) to node[swap] {\(K\)} (n1);
		\draw[<-] (n0) -- (n2);
		\draw[<-] (n1) -- (n3);
		\draw[->] (n2) -- (n3);
	\end{tikzpicture}
\end{equation}
Although $TF$-coalgebras are precisely $\liftKl{F}$-coalgebras, their morphisms capture a different kind of relations between the systems under scrutiny: $TF$-coalgebra homomorphisms are \emph{functional} bisimulations \cite[Theorem~2.5]{rutten:tcs2000} whereas $\liftKl{F}$-coalgebra homomorphisms are functional \emph{linear} bisimulations \cite[Proposition~2.8]{pt:entcs1999}. 
Here the term linear is intended in a broad sense generalising from non-determinism (\cite{pt:entcs1999} considers LTSs only) to effects modelled by an arbitrary monad $T$.

This difference becomes clear when the definition of $\liftKl{F}$-coalgebra homomorphisms is expressed as a diagram in $\CC$, the category underlying $\Kl{T}$. To this end, let $f\colon (X,h) \to (Y,k)$ be a coalgebra homomorphism with underlying morphisms $f\colon X \to Y$ and consider its associated diagram in $\Kl{T}$:
\begin{equation*}	
	\begin{tikzpicture}[
			auto, xscale=1.8, yscale=1.2, diagram font,
			baseline=(current bounding box.center)
		]	
		\node (n0) at (0,1) {\(X\)};	
		\node (n1) at (0,0) {\(\liftKl{F}X\)};	
		\node (n2) at (1,1) {\(Y\)};	
		\node (n3) at (1,0) {\(\liftKl{F}Y\)};	
	
		\draw[->] (n0) to node[swap] {\(h\)} (n1);
		\draw[->] (n2) to node {\(k\)} (n3);
		
		\draw[->] (n0) to node {\(f\)} (n2);
		\draw[->] (n1) to node[swap] {\(\liftKl{F}f\)} (n3);
	\end{tikzpicture}
\end{equation*}
This diagram corresponds to the following diagram in $\CC$:
\begin{equation*}	
	\begin{tikzpicture}[
			auto, xscale=2, yscale=1.2, diagram font,
			baseline=(current bounding box.center)
		]	
		\node (n0) at (0,2) {\(X\)};	
		\node (n1) at (0,0) {\(TFX\)};	
		\node (n2) at (3,2) {\(TY\)};	
		\node (n3) at (3,0) {\(TFY\)};	

		\node (m0) at (3,1) {\(TTFY\)};
		\node (m1) at (1,0) {\(TFTY\)};
		\node (m2) at (2,0) {\(TTFY\)};

		\draw[->] (n0) to node[swap] {\(h\)} (n1);
		\draw[->] (n2) to node {\(Tk\)} (m0);
		\draw[->] (m0) to node {\(\mu_{FY}\)} (n3);
		
		\draw[->] (n0) to node {\(f\)} (n2);
		\draw[->] (n1) to node[swap] {\(TFf\)} (m1);
		\draw[->] (m1) to node[swap] {\(T\lambda_Y\)} (m2);
		\draw[->] (m2) to node[swap] {\(\mu_{FY}\)} (n3);
				
	\end{tikzpicture}
\end{equation*}
For instance, consider non-deterministic transition systems by taking $T$ and $F$ to be $\mathcal{P}$ and $A \times \Id$, respectively. The diagram above commutes if, and only if, for any label $a \in A$ and states $x, x' \in X$ it holds that:
\begin{equation*}
	(a,x') \in h(x) \iff \forall y' \in f(x') \exists y \in f(x) ((a,y') \in k(y))
	\text{.}
\end{equation*}
Intuitively, $\lambda\colon TF \to FT$ distributes the ``branching'' part of the behaviour (\ie the computational effects modelled by $T$) over the observable ``linear'' part of it (characterised by the endofunctor $F$) while $\mu$ collects and combines effects thus forgetting when and how branching occurred. 

From this perspective, final $TF$-coalgebras, and their associated coinduction principle, capture branching semantics whereas final $\liftKl{F}$-coalgebras, and their coinduction principle, capture linear semantics. 
\begin{definition}
	Let $\liftKl{F}$ be a lifting of $F$ to $\Kl{T}$.	For $h\colon X \to TFX$, its linear semantics is the unique morphisms $lbeh_h$:
	\begin{equation*}	
		\begin{tikzpicture}[
				auto, xscale=2.2, yscale=1.2, diagram font,
				baseline=(current bounding box.center)
			]	
			\node (n0) at (0,1) {\(X\)};	
			\node (n1) at (0,0) {\(\liftKl{F}X\)};	
			\node (n2) at (1,1) {\(Z\)};	
			\node (n3) at (1,0) {\(\liftKl{F}Z\)};	
	
			\draw[->] (n0) to node[swap] {\(h\)} (n1);
			\draw[->] (n2) to node {\(\ffix\liftKl{F}\)} (n3);
			
			\draw[dashed,->] (n0) to node {\(lbeh_h\)} (n2);
			\draw[->] (n1) to node[swap] {\(\liftKl{F}lbeh_h\)} (n3);
		\end{tikzpicture}
	\end{equation*}
\end{definition}

In general, the linear semantics described by final $\liftKl{F}$-coalgebra homomorphisms may not capture a known notion of trace semantics for systems modelled as $TF$-coalgebras. For instance, if $T$ is powerset monad $\mathcal{P}$ and $F$ is the labelling functor $A \times \Id$ then the final $\liftKl{A \times \Id}$-coalgebra coincides with the initial one and hence has the emptyset as its carrier \cite{hjs:lmcs2007,pt:entcs1999}. Even when final $\liftKl{F}$-coalgebras capture some notion of trace semantics, this is not unique across the range of choices for $T$ and $F$.
In fact, there are several examples in the literature where finality characterises finite, possibly infinite, infinite only traces or none at all.
For instance, in \cite{kk:lmcs2013}, \citeauthor{kk:lmcs2013} investigate trace semantics for continuous probabilistic transition systems. 
To this end they consider different combinations of monads and endofunctors over $\cat{Meas}$, the category of continuous functions between measurable spaces. In particular, they take $T$ to be either the probability measure monad $\mathcal{G}$ (a.k.a.~Giry monad \cite{prakash:ic2002,prakash:markovprocs,doberkat:stoclogic}) or the sub-probability measure monad $\mathcal{G}_{\leq}$, and $F$ as either $A \times \Id + 1$ or $A \times \Id$ \ie labelling endofunctors with or without explicit termination. For each combination they compute the final $\liftKl{F}$-coalgebra and determine whether it captures some established notion of trace semantics. The results are summarised by the table below.
\begin{table}[!h]
\begin{tabular}{*{4}{c}}
	\toprule
	$T$ & $F$ & $\ffix \liftKl{F}$ & trace semantics\\
	\midrule
	$\mathcal{G}_{\leq}$ & $A \times \Id$ & $K\ifix F$ & none\\
	$\mathcal{G}_{\leq}$ & $A \times \Id + 1$ & $K\ifix F$ & finite\\
	$\mathcal{G}$ & $A \times \Id$ & $K\ffix F$ & infinite\\
	$\mathcal{G}$ & $A \times \Id + 1$ & $K\ffix F$ & possibly infinite\\
	\bottomrule
\end{tabular}
\end{table}

In the wake of the examples above, finite and (possibly) infinite trace semantics are abstractly defined by lifting initial $F$-algebras and final $F$-coalgebras. %

\begin{definition}
	\label{def:trace-as-linear}
	The final $\liftKl{F}$-coalgebra, whenever it exists, is said to capture:
	\begin{itemize}
		\item 
			finite trace semantics if $\ffix\liftKl{F} \cong K(\ifix F)^{-1}$,
		\item 
			(possibly) infinite trace semantics if $\ffix\liftKl{F} \cong K\ffix F$.
	\end{itemize}
\end{definition}

In \cite{hjs:lmcs2007} \citeauthor{hjs:lmcs2007} present general and sufficient conditions for capturing finite traces via Kleisli coinduction based on suitable order-enrichment. Below we propose a modest generalisation of this seminal result by rephrasing it in terms of algebraic compactness.
\begin{proposition}
	\label{thm:kleisli-initial-invariants}
	Let $\liftKl{F}$ be a Kleisli lifting of an endofunctor $F$ to $\Kl{T}$.
	Assume that $F$ is algebraically complete and that $\liftKl{F}$ is algebraically compact. The initial $F$-algebra lifts (along the inclusion $K$) to the final $\liftKl{F}$-coalgebra:
	\begin{equation*}
		 \ffix\liftKl{F} \cong K(\ifix F)^{-1}\text{.}
	\end{equation*}
\end{proposition}
\begin{proofatend}
	Recall from \cite[Theorem.~2.14]{hj:ic1998} that a law distributing $(T,\mu,\eta)$ over $F$ induces a lifting (along the obvious forgetful functors) of the canonical adjunction $(K\dashv L)\colon \Kl{T} \to \CC$ to an adjunction between the categories of algebras for $F$ and $\liftKl{F}$, receptively, as shown in the following diagram:
	\begin{equation*}	
		\begin{tikzpicture}[
				auto, xscale=2, yscale=1.2, diagram font,
				baseline=(current bounding box.center)
			]	
			\node[] (n0) at (0,0) {\(\CC\)};	
			\node[] (n1) at (1,0) {\(\Kl{T}\)};	
			\node[] (n2) at (0,1) {\(\Alg{F}\)};	
			\node[] (n3) at (1,1) {\(\Alg{\liftKl{F}}\)};	
			
			\draw[->,bend left] (n0) to node[pos=.6] {\(K\)} (n1);
		 	\draw[->,bend left] (n1) to node[pos=.4] {\(L\)} (n0);
		 	\node[rotate=90] at ($(n0)!.5!(n1)$) {\(\vdash\)};
		 	\draw[<-] (n0) -- (n2);
		 	\draw[<-] (n1) -- (n3);
		 	\draw[->,bend left] (n2) to (n3);
		 	\draw[->,bend left] (n3) to (n2);
		 	\node[rotate=90] at ($(n2)!.5!(n3)$) {\(\vdash\)};
			
		\end{tikzpicture}
	\end{equation*}
	In particular, the lifting of $K$ maps an $F$-algebra $g\colon FX \to X$
	to the $\liftKl{F}$-algebra $(\eta_{FX} \circ g)\colon \liftKl{F}X \to X$ and an $F$-algebra homomorphism $f\colon (X,g) \to (Y,h)$ to $(\eta_{Y} \circ f)\colon (X,\eta_{FX}\circ g) \to (Y,\eta_{FY} \circ h)$.
	Because of the above adjoint situation, $K\ifix F = \eta_{\carr{\ifix F}} \circ \ifix F$ is an initial $\liftKl{F}$-algebra.
	By algebraic compactness any initial $\liftKl{F}$-algebra is canonically isomorphic to a final $\liftKl{F}$-coalgebra hence $K(\ifix F)^{-1} = (K\ifix F)^{-1}$ is (up to isomorphism) the required final $\liftKl{F}$-coalgebra.
\end{proofatend}

The result presented in \cite{hjs:lmcs2007} is readily recovered as an instance of the above: assume an initial $F$-algebra can be computed via an initial sequence indexed by $\omega$, that both $\Kl{T}$ and $\liftKl{F}$ are enriched over the category of continuous maps between $\omega$-CPOs with bottom elements
\phantomLabel{def:cppocat}%
$\Cppo$ (\ie the full image of the inclusion $\Cpob \hookrightarrow \Cpo$), and that composition in $\Kl{T}$ is left-strict. In fact, under these assumptions, $\liftKl{F}$ is algebraically compact and hence \cref{thm:kleisli-initial-invariants} applies. These hypothesis are met by polynomial functors and monads modelling several computational effects of interests and encompassing non-deterministic transition systems, weighted transition systems, discrete and continuous probabilistic transition systems among others (\cf \cite{hjs:lmcs2007,bmp:jlamp2015} and \cref{sec:kleisli-liftings-examples}).

A general account of infinite trace semantics on par with the finite case is currently missing. Several works investigated this issue, see \eg \cite{jacobs:cmcs2004,hjs:lmcs2007,hu:calco2015,hsu:concur2016,cirstea:entcs2010}, but in all of them liftings of final $F$-coalgebras are not final $\liftKl{F}$-coalgebras. They are \emph{weakly final} ones.
Although these characterisations are weakly universal in $\Coalg{\liftKl{F}}$, they can be uniquely defined by means of some other properties: \cite{hjs:lmcs2007,hu:calco2015,hsu:concur2016} assume an order-enriched setting and identify infinite trace semantics as the maximal $\liftKl{F}$-coalgebras morphism to $K(\ffix F)$. Likewise, \cite{cirstea:entcs2010} defines infinite trace semantics as the maximal among the mediating maps to $K(\ffix F)$ arising from a suitable weak limit in $\Kl{T}$.
Finally, we remark that all these works assume $F$ admits a final coalgebra computable via the final sequence construction and in $\omega$ steps. It follows that infinite traces are implicitly defined as $\omega$-indexed sequences in contrast to more general definitions such as 
transfinite traces \cite[Chapter~11]{dr:traces}.
Nonetheless, these constructions capture (countably) infinite traces for labelled non-deterministic systems, discrete and continuous labelled probabilistic systems.

\subsection{Distributive law morphisms}
\label{sec:kleisli-lifting-distributive-law-morphisms}

Distributive laws can be organised into categories by means of several notions of distributive law morphisms depending on the kind of structures being distributed as discussed in \cite{lpw:entcs2000,watanabe:entcs2002,pw:tcs2002,lpw:tcs2004}. In \cite{lpw:entcs2000}, these notions are introduced as part of different 2-categorical contexts where to analyse distributive laws arising in operational and denotational semantics. This effort, and especially works such as \cite{watanabe:entcs2002,kn:calco2015}, resulted in the proposal of (suitable formulations of) distributive law morphisms as the abstract understanding of translations between SOS specifications. As a consequence, this result extends the theory of abstract GSOS \cite{klin:tcs2011,tp:lics1997} with morphisms able to relate models while preserving structures of interest.

The study of Kleisli coinduction faces a situation similar to that of abstract GSOS prior to the aforementioned works: this theory lacks morphisms between models. To this end, we propose the use of following notion of distributive law morphisms because they induce functors between categories of coalgebras and transformations between Kleisli liftings that are coherent with respect to bisimulation and the relevant structures of Kleisli categories such as the canonical inclusion of their underlying category.

\begin{definition}
	\label{def:dist-law-morphism}
	Let $\lambda\colon FT \to TF$ and $\lambda'\colon F'T' \to T'F'$ be distributive laws of monads over endofunctors over $\CC$. A distributive law morphism from $\lambda$ to $\lambda'$ is a pair $(\theta,\upsilon)$ composed by a monad morphism $\theta\colon (T,\mu^{T},\eta^{T}) \to (T',\mu^{T'},\eta^{T'}) \in \Mnd{\cat{\CC}}$ and an endofunctor morphism $\upsilon\colon F \to F' \in \Endo{\cat{\CC}}$ subject to the following coherence condition:
	\begin{equation}
		\label{eq:law-dist-mnd-end-morph}	
		\begin{tikzpicture}[
				auto, xscale=1.8, yscale=1.2, diagram font,
				baseline=(current bounding box.center)
			]	
			
			\node (n0) at (0,1) {\(F\circ T\)};	
			\node (n1) at (1,1) {\(T\circ F\)};	
			\node (n2) at (0,0) {\(F'\circ T'\)};	
			\node (n3) at (1,0) {\(T'\circ F'\)};	
	
			\draw[->] (n0) to node {\(\lambda\)} (n1);
			\draw[->] (n0) to node[swap] {\(\upsilon \circ \theta\)} (n2);
			\draw[->] (n1) to node {\(\theta \circ \upsilon\)} (n3);
			\draw[->] (n2) to node[swap] {\(\lambda'\)} (n3);
			
		\end{tikzpicture}
	\end{equation}
\end{definition}
\phantomLabel{def:mndendocat}%
Distributive laws of monads over endofunctors on $\CC$ together with their morphisms for the category $\MndEndo{\CC}$. Forgetting either component of distributive law morphisms gives rise to two functors from $\MndEndo{\CC}$ to $\Mnd{\CC}$ and $\Endo{\CC}$, respectively. Both these forgetful functors have sections since any monad distributes over the identity functor and the identity monad distributes over any endofunctor.

An interesting property of distributive law morphisms (as defined above) is that they induce functors between categories of coalgebras modelling branching and linear semantics that are coherent with respect to the canonical inclusion into Kleisli categories of their underlying category and to linear semantics (\cf \cref{thm:dist-law-morphisms-coherent-kl}). Let $(\theta,\upsilon)$ be a distributive law morphism from $\lambda\colon FT \to TF$ to $\lambda'\colon F'T' \to T'F'$. Consider the assignments mapping each $TF$-coalgebra $(X,g)$ and homomorphism $f\colon (X,g) \to (Y,h)$ as follows:
\begin{equation}
	\label{eq:dist-law-morph-indices-coalg-functor-1}
	(X, g) \mapsto (X,(\theta\circ\upsilon)_X \circ g)
	\qquad
	f \mapsto f
\end{equation}
It follows from naturality of $\theta$ and $\upsilon$ that these assignments define the functor 
\[\Coalg{\theta\circ\upsilon}\colon \Coalg{TF} \to \Coalg{T'F'}\text{.}\]
For a $\liftKl{F}$-coalgebra $g\colon X \to \liftKl{F}X$ and a $\liftKl{F}$-coalgebra homomorphism $f\colon (X,g) \to (Y,h)$, consider the assignments
\begin{equation}
	\label{eq:dist-law-morph-indices-coalg-functor-2}
	(X,g) \mapsto (X,(\theta\circ\upsilon)_X \circ g)
	\qquad
	f \mapsto \theta_Y \circ f
	\text{.}
\end{equation}
These assignments map $\liftKl{F}$-coalgebras to $\liftKl{F'}$-coalgebras and homomorphisms accordingly as the following commuting diagram asserts:
\resetDiagMarker
\begin{equation*}	
	\begin{tikzpicture}[
			auto, xscale=2.3, yscale=1.6, diagram font,
			baseline=(current bounding box.center)
		]	
		
		\foreach \c in {0,1,...,5}{
			\foreach \r in {0,1,...,4}{
				\coordinate (i\r\c) at (\c,\r);
			};
		};
		
		\node (n00) at (i00) {\(T'F'X\)};
		\node (n01) at (i01) {\(T'F'TY\)};
		\node (n02) at (i02) {\(T'F'T'Y\)};
		\node (n03) at (i03) {\(T'T'F'Y\)};
		\node (n04) at (i05) {\(T'F'Y\)};
		\node (n05) at (i10) {\(T'FX\)};
		\node (n06) at (i11) {\(T'FTY\)};
		\node (n07) at (i12) {\(T'TFY\)};
		\node (n08) at (i13) {\(T'T'FY\)};
		\node (n09) at (i14) {\(T'FY\)};
		\node (n10) at (i20) {\(TFX\)};
		\node (n11) at (i21) {\(TFTY\)};
		\node (n12) at (i22) {\(TTFY\)};
		\node (n13) at (i23) {\(TFY\)};
		\node (n14) at (i25) {\(T'T'FY\)};
		\node (n15) at (i15) {\(T'T'F'Y\)};
		\node (n16) at (i33) {\(TTFY\)};
		\node (n17) at (i35) {\(T'TFY\)};
		\node (n18) at (i40) {\(X\)};
		\node (n19) at (i43) {\(TY\)};
		\node (n20) at (i45) {\(T'Y\)};
		
		\draw[->] (n00) to node[swap] {\(T'F'f\)}(n01);
		\draw[->] (n01) to node[swap] {\(T'F'\theta_Y\)}(n02);
		\draw[->] (n02) to node[swap] {\(T'\lambda'_Y\)}(n03);
		\draw[->] (n03) to node[swap] {\(\mu'_{F'Y}\)}(n04);
		
		\draw[->] (n05) to node {\(T'Ff\)}(n06);
		\draw[->] (n06) to node {\(T'\lambda_Y\)}(n07);
		\draw[->] (n07) to node {\(T'\theta_{FY}\)}(n08);
		\draw[->] (n08) to node {\(\mu'_{F'Y}\)}(n09);
		
		\draw[->] (n05) to node {\(T'\upsilon_X\)}(n00);
		\draw[->] (n06) to node {\(T'\upsilon_{TY}\)}(n01);		
		\draw[->] (n08) to node {\(T'T'\upsilon_{Y}\)}(n03);		
		\draw[->] (n09) to node[swap,pos=.3] {\(T'\upsilon_{Y}\)}(n04);
		
		\draw[->] (n10) to node {\(TFf\)} (n11);
		\draw[->] (n11) to node {\(T\lambda_Y\)} (n12);
		\draw[->] (n12) to node {\(\mu_{FY}\)}(n13);
		\draw[->] (n14) to node[swap,pos=.6] {\(T'\upsilon_{Y}\)}(n15);
		
		\draw[->] (n10) to node {\(\theta_{FX}\)} (n05);
		\draw[->] (n11) to node {\(\theta_{FTY}\)} (n06);		
		\draw[->] (n12) to node {\(\theta_{TFY}\)} (n07);		
		\draw[->] (n13) to node {\(\theta_{FY}\)} (n09);			
		\draw[->] (n14) to node[swap] {\(\mu'_{FY}\)} (n09);	
		\draw[->] (n15) to node[swap,pos=.3] {\(\mu'_{F'Y}\)} (n04);
		
		\draw[->] (n16) to node {\(\theta_{TFY}\)} (n17);
		\draw[->] (n16) to node {\(\mu_{FY}\)} (n13);
		\draw[->] (n17) to node[swap] {\(T'\theta_{FY}\)} (n14);
		
		\draw[->] (n18) to node {\(f\)} (n19);
		\draw[->] (n19) to node {\(\theta_Y\)} (n20);

		\draw[->] (n18) to node {\(h\)} (n10);
		\draw[->] (n19) to node {\(Tk\)} (n16);
		\draw[->] (n20) to node[swap] {\(T'k\)} (n17);
		
		\nodeDiagMarker{$(i40)!.5!(i23)$}{diag:dist-law-morph-coalg-functor-20};
		\nodeDiagMarker{$(i11)!.5!(i03)$}{diag:dist-law-morph-coalg-functor-21};
		\nodeDiagMarker{$(i22)!.5!(i14)$}{diag:dist-law-morph-coalg-functor-22};
		\nodeDiagMarker{$(i33)!.5!(i25)$}{diag:dist-law-morph-coalg-functor-23};
					
	\end{tikzpicture}
\end{equation*}
In order to check that the diagram above indeed commutes note that \eqref{diag:dist-law-morph-coalg-functor-20} is the expansion in $\CC$ of the diagram asserting that $f$ is a $\liftKl{F}$-coalgebra homomorphism, that \eqref{diag:dist-law-morph-coalg-functor-21} is a component of \eqref{eq:law-dist-mnd-end-morph} under $T$, that \eqref{diag:dist-law-morph-coalg-functor-22} and \eqref{diag:dist-law-morph-coalg-functor-23} follow from the fact that $\theta$ is a monad morphism, and that all remaining sub-diagrams are naturality squares.
As a consequence of the fact that the above diagram commutes and the assumption of $(\theta,\upsilon)$ that is a distributive law morphism, the assignments \eqref{eq:dist-law-morph-indices-coalg-functor-2} define the functor \[\Coalg{\theta,\upsilon}\colon \Coalg{\liftKl{F}} \to \Coalg{\liftKl{F}}\text{.}\]
Moreover, this functor is a lifting of $\Kl{\theta}\colon \Kl{T} \to \Kl{T'}$ along the forgetful functors depicted in the following diagram:
\begin{equation*}	
	\begin{tikzpicture}[
			auto,
			xscale=4,yscale=1.2,
			diagram font,
			baseline=(current bounding box.center)
		]	
		\node[] (n0) at (0,0) {\(\Kl{T}\)};	
		\node[] (n1) at (1,0) {\(\Kl{T'}\)};	
		\node[] (n2) at (0,1) {\(\Coalg{\liftKl{F}}\)};	
		\node[] (n3) at (1,1) {\(\Coalg{\liftKl{F'}}\)};	
	
		\draw[->] (n0) to node[swap] {\(\Kl{\theta}\)} (n1);
		\draw[->] (n2) to node {\(\Coalg{\theta,\upsilon}\)} (n3);			
		
	 	\draw[<-] (n0) -- (n2);
	 	\draw[<-] (n1) -- (n3);
	 	
	\end{tikzpicture}
\end{equation*}
From a more abstract perspective, this situation can be seen to follow from lifting $\upsilon\colon F \to F'$ to a natural transformation exchanging $\Kl{\theta}$ with Kleisli liftings as depicted in the diagram below. In fact, any distributive law morphism $(\theta,\upsilon)$ determines a natural transformation $\liftKl{\upsilon}$ such that:
\begin{equation*}	
	\begin{tikzpicture}[
			auto, scale=1.7, rotate=-45, diagram font,
			baseline=(current bounding box.center)
		]	
		\node (n0) at (0,1) {\(\Kl{T}\)};	
		\node (n1) at (1,1) {\(\Kl{T'}\)};	
		\node (n2) at (0,0) {\(\Kl{T}\)};	
		\node (n3) at (1,0) {\(\Kl{T'}\)};	

		\draw[->] (n0) to node[]     {\(\Kl{\theta}\)} (n1);
		\draw[->] (n2) to node[]     {\(\liftKl{F}\)} (n0);
		\draw[->] (n2) to node[swap] {\(\Kl{\theta}\)} (n3);
		\draw[->] (n3) to node[swap] {\(\liftKl{F'}\)} (n1);
		
		\draw[nat>] ($(n0)!.3!(n3)$) to node[swap] {\(\liftKl{\upsilon}\)} ($(n0)!.7!(n3)$);
	\end{tikzpicture}
\end{equation*}
The natural transformation $\liftKl{\upsilon}\colon \Kl{\theta} \circ \liftKl{F} \to \liftKl{F'} \circ \Kl{\theta}$ is given on each object $X$ as $K' \circ \upsilon_X$. To see that this is indeed a natural transformation it suffices to note that, for any $f\colon X \to TY$, the corresponding naturality square reduces to the  following diagram in the underlying category $\CC$:
\begin{equation*}	
	\begin{tikzpicture}[
			auto, xscale=1.8, yscale=1.2, diagram font,
			baseline=(current bounding box.center)
		]	
		
		\node (n0) at (0,3) {\(FX\)};	
		\node (n1) at (0,2) {\(FTY\)};	
		\node (n2) at (0,1) {\(TFY\)};	
		\node (n3) at (0,0) {\(T'FY\)};	
		\node (n4) at (1,3) {\(F'X\)};	
		\node (n5) at (1,2) {\(F'TY\)};	
		\node (n6) at (1,1) {\(F'T'Y\)};	
		\node (n7) at (1,0) {\(T'F'Y\)};	

		\draw[->] (n0) to node[swap] {\(Ff\)} (n1);
		\draw[->] (n1) to node[swap] {\(\lambda_Y\)} (n2);
		\draw[->] (n2) to node[swap] {\(\theta_{FY}\)} (n3);

		\draw[->] (n4) to node {\(F'f\)} (n5);
		\draw[->] (n5) to node {\(F'\theta_{Y}\)} (n6);
		\draw[->] (n6) to node {\(\lambda'_Y\)} (n7);

		\draw[->] (n0) to node {\(\upsilon_X\)} (n4);
		\draw[->] (n1) to node {\(\upsilon_{TY}\)} (n5);
		\draw[->] (n3) to node[swap] {\(T'\upsilon_{Y}\)} (n7);
					
		\draw[rounded corners,->] (n0) -- ($(n0)+(-.65,0)$) to node[swap] {\((\Kl{\theta}\circ\liftKl{F})f\)} ($(n3)+(-.65,0)$) -- (n3);
		\draw[rounded corners,->] (n4) -- ($(n4)+(.65,0)$) to node {\((\liftKl{F'}\circ\Kl{\theta})f\)} ($(n7)+(.65,0)$) -- (n7);
	\end{tikzpicture}
\end{equation*}
The diagram above commutes since $(\theta,\upsilon)$ is a distributive law morphism. Note that the diagram above lies in the lower part of the diagram unfolding \eqref{eq:dist-law-morph-indices-coalg-functor-2}.

Actually, the functor $\Coalg{\theta \circ \upsilon}$ is the restriction of $\Coalg{\theta,\upsilon}$ to the wide subcategory determined by $TF$-coalgebra homomorphisms. 

\begin{proposition}
	\label{thm:dist-law-morphisms-coherent-kl}
	For $(\theta,\upsilon)$ a distributive law morphism from $\lambda\colon FT \to TF$ to $\lambda'\colon F'T' \to T'F'$, the diagram below commutes:
	\begin{equation*}	
		\begin{tikzpicture}[
				auto,
				diagram font,
				baseline=(current bounding box.center)
			]	
			\node[] (n0) at (0,0,3) {\(\CC\)};	
			\node[] (n1) at (0,0,0) {\(\Kl{T}\)};	
			\node[] (n2) at (0,2,3) {\(\Coalg{TF}\)};	
			\node[] (n3) at (0,2,0) {\(\Coalg{\liftKl{F}}\)};	
			
			\node[] (m0) at (4.5,0,3) {\(\CC\)};	
			\node[] (m1) at (4.5,0,0) {\(\Kl{T'}\)};	
			\node[] (m2) at (4.5,2,3) {\(\Coalg{T'F'}\)};	
			\node[] (m3) at (4.5,2,0) {\(\Coalg{\liftKl{F'}}\)};	
		
			\draw[nat] (n0) to (m0);
			\draw[->] (n1) to node {\(\Kl{\theta}\)} (m1);
			\draw[->] (n2) to node {\(\Coalg{\theta\circ\upsilon}\)} (m2);
			\draw[->] (n3) to node {\(\Coalg{\theta,\upsilon}\)} (m3);			
			
			\draw[->] (n0) to node[pos=.6] {\(K\)} (n1);
		 	\draw[<-] (n0) to (n2);
		 	\draw[<-] (n1) to (n3);
			\draw[->] (n2) to (n3);
			
			\draw[->] (m0) to node[swap] {\(K'\)} (m1);
		 	\draw[<-] (m0) to (m2);
		 	\draw[<-] (m1) to (m3);
			\draw[->] (m2) to (m3);
			
			\draw[cross,->] (n2) to (m2);
		 	\draw[<-,cross] (m0) to (m2);
		\end{tikzpicture}
	\end{equation*}
\end{proposition}

\begin{proofatend}
	Left and right faces assert that the inclusions $K$ and $K'$ lift along the forgetful functors for the coalgebra categories involved and this holds true for any Kleisli lifting as per \eqref{eq:lift-k-coalg}. The front face of the diagram commutes by definition of $\Coalg{\theta \circ \upsilon}$ since, as clear from \eqref{eq:dist-law-morph-indices-coalg-functor-1}, this functor acts as the identity on coalgebra homomorphisms. The back face of the diagram commutes by definition of $\Coalg{\theta,\upsilon}$ since, as clear from \eqref{eq:dist-law-morph-indices-coalg-functor-2}, this functor composes coalgebra homomorphisms to the opportune components of $\theta$ \ie it acts as $\Kl{\theta}$. The bottom of the diagram commutes since $\theta\colon T \to T'$ is a monad morphism and by definition of the functor $\Kl{\theta}$. To see that the top of the diagram commutes as well note that	
	$\Coalg{TF}$ and $\Coalg{T'F'}$ are wide subcategories of $\Coalg{\liftKl{F}}$ and $\Coalg{\liftKl{F'}}$, respectively and that on coalgebras $\Coalg{\theta \circ \upsilon}$ and $\Coalg{\theta,\upsilon}$ act in the very same way, whereas on coalgebra homomorphisms the first acts as the identity and the second as $\Kl{\theta}\colon \Kl{T} \to \Kl{T'}$. Therefore, the top diagram commutes since $\Kl{\theta} \circ K = K'$ ($\theta \bullet \eta = \eta'$). Finally, the whole diagram commutes since all of its faces are so.
\end{proofatend}

The situation described by \cref{thm:dist-law-morphisms-coherent-kl} %
confirms \cref{def:dist-law-morphism} as a suitable notion of morphisms since these coherently induce functors between categories of coalgebras capturing the linear and branching semantics of systems.

\section{Pointwise extensions to sheaf categories}
\label{sec:extp-functor}

In this section we consider $\CC$-valued sheaves over ordinal numbers equipped with the Alexandrov topology and study the pointwise extension of endofunctors, monads, and their distributive laws to this setting. Hereafter, let $\alpha$ be limit ordinal and assume that the constant sheaf adjunction $(\Delta \dashv \Gamma) \colon \Sh[\CC]{\alpha} \to \CC$ is defined (\cf \cref{sec:sheaves-sites}).

\subsection{The pointwise extension functor}

For $F$ an endofunctor over $\CC$ consider the endofunctor $\Func{\Id}{F}$ over $\PSh[\CC]{\alpha}$ defined on any presheaf $X$, morphism $f$, and stages $\beta \leq \beta'$ as follows:
\begin{align*}
	\Func{\Id}{F}X_{\beta} &= FX_{\beta} 
	&
	\Func{\Id}{F}X_{\res{\beta}{\beta'}} &= FX_{\res{\beta}{\beta'}}
	&
	\Func{\Id}{F}f_{\beta} &= Ff_{\beta}
\end{align*}
Because this functor acts on values as $F$ we call $\Func{\Id}{F}$ the \emph{pointwise extension (to presheaves) of $F$}.

This endofunctor need not to preserve sheaves since $F$ may not preserve the necessary limits (which are pointwise in $\PSh[\CC]{\alpha}$). Therefore, to obtain an extension of $F$ to the category of sheaves we need to apply the associated sheaf functor $\mathbf{a}$ that, together with its right adjoint $\mathbf{i}$, yields the endofunctor:
\begin{equation*}
	\mathbf{a} \circ \Func{\Id}{F} \circ \mathbf{i}\colon \Sh[\CC]{\alpha} \to \Sh[\CC]{\alpha}\text{.}
\end{equation*}
We call this functor the \emph{pointwise extension (to sheaves) of $F$} and denote it as $\extP{F}$. This functor takes any sheaf $X$ and any morphism $f\colon X \to Y$ to:
\begin{align*}
	\extP{F}X_{\beta} &= FX_{\beta} 
	&
	\extP{F}X_{\res{\beta}{\beta+1}} &= FX_{\res{\beta}{\beta+1}}
	&
	\extP{F}f_{\beta} &= Ff_{\beta}
	\\
	\extP{F}X_{\gamma} &= \lim_{\beta < \gamma} FX_{\beta}
	&
	\extP{F}X_{\res{\beta}{\gamma}} &= \pi_\beta
	&
	\extP{F}f_{\gamma} &= \rho_\gamma
\end{align*}
where $\beta$ is a successor ordinal, $\gamma$ a limit one, $\pi_\beta\colon \lim_{\beta' < \gamma} FX_{\beta'} \to FX_{\beta}$ is the component at $\beta$ of the limiting cone, and $\rho_\gamma\colon \lim_{\beta < \gamma} FX_{\beta} \to \lim_{\beta < \gamma} FY_{\beta}$ is the mediating map for the cone $\{f_{\beta}\circ\pi_{\beta}\}_{\beta < \gamma}$.

The endofunctor $\extP{F}$ is \emph{an} extension of $F$ (actually of $\Delta \circ F$) along the constant sheaf functor $\Delta$ since it makes the diagram below commute.
\begin{equation}	
	\label{eq:extp-extension-along-delta}
	\begin{tikzpicture}[
			auto, xscale=1.8, yscale=1.2, diagram font,
			baseline=(current bounding box.center)
		]	
		\node (n0) at (0,1) {\(\CC\)};	
		\node (n1) at (1,1) {\(\Sh[\CC]{\alpha}\)};	
		\node (n3) at (0,0) {\(\CC\)};	
		\node (n4) at (1,0) {\(\Sh[\CC]{\alpha}\)};	

		\draw[->] (n0) to node[] {\(\Delta\)} (n1);
		\draw[->] (n0) to node[swap] {\(F\)} (n3);
		\draw[->] (n1) to node[] {\(\extP{F}\)} (n4);
		\draw[->] (n3) to node[swap] {\(\Delta\)} (n4);
		
	\end{tikzpicture}
\end{equation}
Intuitively, this diagram abstractly describes the idea that $\extP{F}$ acts as $F$ on the image of the subcategory $\CC$ of $\Sh[\CC]{\alpha}$, especially on the final object $1$. As a consequence, the final sequence for the endofunctor $F$ extends along $\Delta$ to the final sequence for $\extP{F}$. 
Formally, the constant sheaf functor $\Delta\colon \CC \to \Sh[\CC]{\alpha}$ lifts along the forgetful functors for $\Coalg{F}$ and $\Coalg{\extP{F}}$ as in the diagram below.
\begin{equation*}
	\begin{tikzpicture}[
			auto, xscale=2.2, yscale=1.3, diagram font,
			baseline=(current bounding box.center)
		]	
		\node (n0) at (0,1) {\(\Coalg{F}\)};	
		\node (n1) at (1,1) {\(\Coalg{\extP{F}}\)};	
		\node (n3) at (0,0) {\(\CC\)};	
		\node (n4) at (1,0) {\(\Sh[\CC]{\alpha}\)};	

		\draw[->] (n0) to (n1);
		\draw[->] (n0) to (n3);
		\draw[->] (n1) to (n4);
		\draw[->] (n3) to node {\(\Delta\)} (n4);
		
	\end{tikzpicture}
\end{equation*}
The lifted functor is an inclusion functor and takes final $F$-coalgebras to final $\extP{F}$-coalgebras.

\begin{proposition}
	\label{thm:extp-final-invariants}
	For $F$ an endofunctor on $\CC$, the constant sheaf functor $\Delta$ takes final $F$-coalgebras to final $\extP{F}$-coalgebras.
\end{proposition}

\begin{proofatend}
	First note that $\Delta$ lifts along the forgetful functors for $\Coalg{F}$ to $\Coalg{\extP{F}}$ to an inclusion functor from $\Coalg{F}$ to $\Coalg{\extP{F}}$. For $(X,h)$ a $F$-coalgebra, $\Delta h$ has type $\Delta h \to \Delta F X$ and $\Delta F = \extP{F} \Delta$ from which we conclude that $\Delta h$ is a $\extP{F}$-coalgebra. Likewise, $\Delta$ maps $F$-coalgebra homomorphisms to $\extP{F}$-coalgebra homomorphisms.
	Assume $\ffix F$ exists, we show its image final. For $(Y,k)$ a $\extP{F}$-coalgebra and $\beta$ a successor ordinal in $\alpha$, the component $k_\beta$ is a $F$-coalgebra and hence there is a unique $F$-coalgebra homomorphism $!_{k_\beta}\colon k_\beta \to \ffix F$. For any successor ordinal $\beta'$ such that $\beta \leq \beta'$, the restriction morphism $Y_{\res{\beta}{\beta'}}\colon Y_{\beta'} \to Y_{\beta}$ carries a $F$-coalgebra homomorphism from $k_{\beta'}$ to $k_{\beta}$ such that ${!}_{k_{\beta'}} = {!}_{k_{\beta}} \circ Y_{\res{\beta}{\beta'}}$. Reworded, for $\beta$ and $\beta'$ successor ordinals, $!_{k_{\beta}}$ and $!_{k_{\beta'}}$ satisfy naturality. Since successor ordinals form a base for $\alexT{\alpha}$, the family of $F$-coalgebra homomorphisms $\{!_{k_\beta}\}$ uniquely extends to a $\extP{F}$-coalgebra homomorphism from $k$ to $\Delta \ffix F$. This homomorphism is necessarily unique by assumption on $\ffix F$ and hence exhibits $\Delta \ffix F$ as final in $\Coalg{\extP{F}}$.
\end{proofatend}

\begin{remark}
	Assume that $\CC$ is bicomplete with respect to $\alpha$-sequences hence that the constant sheaf functor $\Delta$ is both left and right adjoint; examples of this situation are the categories $\Set$ and $\Meas$. Since under these assumption $\CC$ is a (co)reflective subcategory of $\Sh[\CC]{\alpha}$, \cref{thm:extp-final-invariants} follows from a result known as ``\citeauthor{freyd:ct1991}'s Reflective Subcategory Lemma'' \cite{freyd:ct1991}: final coalgebras for endofunctors which restrict to reflective subcategories lie in such categories and coincide with final coalgebras for their restrictions.
\end{remark}

Symmetrically to the situation described in \eqref{eq:extp-extension-along-delta}, the endofunctor $F$ is the right extension of $\extP{F}$ (actually of $\Gamma \circ \extP{F}$) along the global section functor $\Gamma$ since there is a (unique) 2-cell $\varrho$ such that:
\begin{equation}	
	\label{eq:extp-lift-along-gamma}	
	\begin{tikzpicture}[
			auto, xscale=1.8, yscale=1.2, diagram font,
			baseline=(current bounding box.center)
		]	
		\node (n1) at (1,1) {\(\Sh[\CC]{\alpha}\)};	
		\node (n2) at (2,1) {\(\CC\)};	
		\node (n4) at (1,0) {\(\Sh[\CC]{\alpha}\)};	
		\node (n5) at (2,0) {\(\CC\)};	

		\draw[nat>] ($(n2)!.3!(n4)$) to node[swap] {\(\varrho\)} ($(n2)!.7!(n4)$);

		\draw[->] (n1) to node[] {\(\Gamma\)} (n2);
		\draw[->] (n1) to node[swap] {\(\extP{F}\)} (n4);
		\draw[->] (n2) to node[] {\(F\)} (n5);
		\draw[->] (n4) to node[swap] {\(\Gamma\)} (n5);
		
	\end{tikzpicture}
\end{equation}
In particular, $(F,\varrho) \cong Ran_{\Gamma}(\Gamma \extP{F})$.
The natural transformation $\varrho$ is given on each sheaf $X$ as the mediating map $\varrho_X\colon F \lim X \to \lim FX$ which is clearly unique.
This characterisation might appear backward since we obtained $F$ from $\extP{F}$--one might prefer to call $\extP{F}$ a lifting of $F$ along $\Gamma$. Nonetheless, the natural transformation $\varrho$ and its sections are be of relevance for lifting the adjunction $(\Delta \dashv \Gamma)$ to categories of coalgebras as stated by \cref{thm:extp-delta-gamma-coalg} below. 

\begin{proposition}
	\label{thm:extp-delta-gamma-coalg}
	If $\varrho\colon F\Gamma \to \Gamma\extP{F}$ from  \eqref{eq:extp-lift-along-gamma} is a retraction then the constant sheaf adjunction $(\Delta \dashv \Gamma)\colon \Sh[\CC]{\alpha} \to \CC$ lifts along the forgetful functors for $\Coalg{F}$ and $\Coalg{\extP{F}}$ as shown in the diagram below.
	\begin{equation*}	
		\begin{tikzpicture}[
				auto, xscale=2.1, yscale=1.2, diagram font,
				baseline=(current bounding box.center)
			]	
			\node[] (n0) at (0,0) {\(\CC\)};	
			\node[] (n1) at (1,0) {\(\Sh[\CC]{\alpha}\)};	
			\node[] (n2) at (0,1) {\(\Coalg{F}\)};	
			\node[] (n3) at (1,1) {\(\Coalg{\extP{F}}\)};	
			
			\draw[->,bend left] (n0) to node[pos=.6] (l0) {\(\Delta\)} (n1);
		 	\draw[->,bend left] (n1) to node[pos=.4] (l1) {\(\Gamma\)} (n0);
		 	\node[rotate=90] at ($(l0)!.5!(l1)$) {\(\vdash\)};
		 	\draw[<-] (n0) -- (n2);
		 	\draw[<-] (n1) -- (n3);
		 	\draw[->,bend left] (n2) to (n3);
		 	\draw[->,bend left] (n3) to (n2);
		 	\node[rotate=90] at ($(n2)!.5!(n3)$) {\(\vdash\)};
			
		\end{tikzpicture}
	\end{equation*}
\end{proposition}
\begin{proofatend}
	Let $\eta$ and $\varepsilon$ denote the unit and counit of $(\Delta \dashv \Gamma)$, respectively. It follows from \cite[Theorem~2.5]{kkw:cmcs2014} that natural transformations $\vartheta\colon \Delta F \to \extP{F}\Delta$ and $\varsigma\colon \Gamma\extP{F} \to F\Gamma$ define a lifting of $(\Delta \dashv \Gamma)$ along the forgetful functors for $\Coalg{F}$ and $\Coalg{\liftKl{F}}$ whenever the diagrams below commute:
	\resetDiagMarker
	\begin{equation*}	
		\begin{tikzpicture}[
				auto, xscale=2.8, yscale=1.6, diagram font,
				baseline=(current bounding box.center)
			]	
			\node[] (n0) at (0,0) {\(\Gamma\circ \Delta\circ F\)};	
			\node[] (n1) at (1,0) {\(\Gamma\circ \extP{F}\circ \Delta\)};	
			\node[] (n2) at (0,1) {\(F\)};	
			\node[] (n3) at (1,1) {\(F\circ \Gamma \circ \Delta\)};	
			
			\draw[->] (n0) to node[swap] {\(\id_\Gamma \circ \vartheta\)} (n1);
			\draw[->] (n1) to node[swap] {\(\varsigma \circ \id_\Delta\)} (n3);
			\draw[->] (n2) to node[swap] {\(\eta \circ \id_F\)} (n0);
			\draw[->] (n2) to node {\(\id_F \circ \eta\)} (n3);
			
			\nodeDiagMarker{$(n0)!.5!(n3)$}%
				{diag:extp-delta-gamma-coalg-1};
		\end{tikzpicture}
		\qquad
		\begin{tikzpicture}[
				auto, xscale=2.8, yscale=1.6, diagram font,
				baseline=(current bounding box.center)
			]	
			\node[] (n0) at (0,0) {\(\extP{F}\circ \Delta \circ \Gamma\)};	
			\node[] (n1) at (1,0) {\(\Delta\circ F\circ \Gamma\)};	
			\node[] (n2) at (0,1) {\(\extP{F}\)};	
			\node[] (n3) at (1,1) {\(\Delta\circ \Gamma\circ \extP{F}\)};	
			
			\draw[<-] (n0) to node[swap] {\(\vartheta \circ \id_\Gamma\)} (n1);
			\draw[<-] (n1) to node[swap] {\(\id_\Delta \circ \varsigma\)} (n3);
			\draw[<-] (n2) to node[swap] {\(\id_{\extP{F}} \circ \varepsilon\)} (n0);
			\draw[<-] (n2) to node {\(\varepsilon \circ \id_{\extP{F}}\)} (n3);
			
			\nodeDiagMarker{$(n0)!.5!(n3)$}%
				{diag:extp-delta-gamma-coalg-2};
		\end{tikzpicture}
	\end{equation*}
	In this setting, the desired lifting for $\Delta$ is given, on each $F$-coalgebra $(X,g)$ and homomorphism $f$, by the assignments: 
	\begin{equation*}
		(X,g) \mapsto (\Delta X, \vartheta_X \circ \Delta g) 	\qquad f \mapsto \Delta f
		\text{.}
	\end{equation*}
	Its right adjoint, that is the desired lifting for $\Gamma$, is given on each $\extP{F}$-coalgebra $(Y,h)$ and homomorphism $f$, by the assignments: 
	\begin{equation*}
		(Y,h) \mapsto (\Gamma Y,\varsigma_Y \circ \Gamma h)  \qquad f \mapsto \Gamma f
		\text{.}
	\end{equation*}	
	We apply \cite[Theorem~2.5]{kkw:cmcs2014} by choosing $\vartheta$ and $\varsigma$ as the identity and as any section for $\varrho$, respectively. This choice is justified by the fact that $\extP{F}$ is the pointwise extension of $F$ hence an extension along $\Delta$ as per \eqref{eq:extp-extension-along-delta} and $\varrho$ is a retraction by hypothesis. The natural transformation $\varrho \circ \id_\Delta\circ \Gamma \extP{F} \Delta\to F \Gamma \Delta$ is an identity since for any object $X$ of $\CC$ we have that $\Gamma \extP{F} \Delta X = \Gamma \Delta F X$ by \eqref{eq:extp-extension-along-delta} and that $\Gamma \Delta F X = \lim \Delta FX  = FX$. It follows that $\varsigma \circ \id_\Delta$ is an identity as well and that diagram \eqref{diag:extp-delta-gamma-coalg-1} is made from identities.
	Note that for $X$ a sheaf, the component of $\varepsilon_{X}$ at any stage $\beta$ is the restriction arrow $X_{\res{\beta}{\alpha}}$, and that the sheaves $\Delta \Gamma \extP{F}X$ and $\Delta F \Gamma X$ take values $\lim_{\beta' < \alpha} FX_{\beta'}$ and $F \lim_{\beta' < \alpha} X_{\beta'}$ at any stage, respectively. Thus, for $\beta$ a stage, the component at stage $\beta$ of \eqref{diag:extp-delta-gamma-coalg-2} is the following square in $\CC$:
	\begin{equation*}
		\begin{tikzpicture}[
				auto, xscale=2.8, yscale=1.2, diagram font,
				baseline=(current bounding box.center)
			]	
			\node[] (n0) at (0,0) {\(FX_{\beta}\)};	
			\node[] (n1) at (1,0) {\(FX_{\beta}\)};	
			\node[] (n2) at (0,1) {\(\lim_{\beta' < \alpha} FX_{\beta'}\)};
			\node[] (n3) at (1,1) {\(F\lim_{\beta' < \alpha} X_{\beta'}\)};
			
			\draw[nat] (n0) to (n1);
			\draw[->] (n2) to node {\(\varsigma_{X}\)} (n3);
			\draw[->] (n2) to node[swap] {\(\extP{F}X_{\res{\beta}{\alpha}}\)} (n0);
			\draw[->] (n3) to node {\(FX_{\res{\beta}{\alpha}}\)} (n1);
		\end{tikzpicture}
	\end{equation*}	
	For $X$ a sheaf, the components at $X$ of $\varepsilon \circ \id_F$ and $(\id_F \circ \varepsilon) \bullet (\vartheta \circ \id_\Gamma)$ shown in \eqref{diag:extp-delta-gamma-coalg-2} describe two cones for the $\alpha$-sequence $(\extP{F}X_{\res{\beta}{\beta'}}\colon \extP{F}X_{\beta'} \to \extP{F}X_{\beta})_{\beta \leq \beta' < \alpha}$ (since $\extP{F}X_{\res{\beta+1}{\beta'+1}} = FX_{\res{\beta}{\beta}}$) and such cones can be safely restricted to the successors base for $\alexT{\alpha}$ (\cf \cref{sec:sheaves}). We conclude by noting that the first of these cones is limiting, the associated mediating map is exactly the component at $X$ of $\varrho$, and $\varsigma_X\varrho_X = \id_X$ by hypothesis.	
\end{proofatend}

Note that for $\varrho$ to be a retraction entails for $F$ to weakly preserve limits of $\alpha$-sequences since any section of $\varrho$ represents a coherent choice of weak mediating morphisms. In particular, $\varrho$ is a natural isomorphism if and only if the endofunctor $F$ (strongly) preserves limits of $\alpha$-sequences. This is a mild assumption in the context of this work: \cref{thm:extp-delta-gamma-coalg} will be applied to choices of $F$ and $\alpha$ such that the final sequence of $F$ is stable at $\alpha$ (\cf \cref{sec:guarded-coalgebras}).

Taking an endofunctor to its pointwise extension is a functorial operation. There is a functor: 
\begin{equation*}
	(\extP{-})\colon \Endo{\CC} \to \Endo{\Sh[\CC]{\alpha}}
\end{equation*}
defined as follows:
\begin{equation*}
	(\extP{-}) \defeq \mathbf{a} \circ \Func{\Id}{-} \circ \mathbf{i}
	\text{.}
\end{equation*}
We call the \emph{pointwise extension functor} (to sheaves).
In the remaining of this section we show that this functor lifts to distributive laws and that it enriches over the category of sheaves.

\subsection{Pointwise extension of distributive laws}

The pointwise extension functor preserves the identity functor over $\CC$ up to isomorphism: the unit of the reflection $(\mathbf{a}\dashv \mathbf{i})$ defines the isomorphism
\begin{equation*}
	\label{eq:extp-monoidal-psi}
	\psi\colon \Id_{\Sh[\CC]{\alpha}} \cong \extP{\Id_{\CC}}
	\qquad
	\Id_{\Sh[\CC]{\alpha}} \cong  \mathbf{a} \circ \mathbf{i} = \extP{\Id_{\CC}}
	\textbf{.}
\end{equation*}
Likewise, $(\extP{-})$ preserves endofunctor composition up to isomorphism:
for $F$ and $G$ in $\Endo{\CC}$ there is an isomorphism
\begin{equation*}
	\label{eq:extp-monoidal-phi}
	\phi_{F,G}\colon \extP{F} \circ \extP{G} \cong \extP{F\circ G}
\end{equation*}
natural in $F$ and $G$. For $\beta$ and $\gamma$ a successor and a limit ordinal, the component $\beta$ of $\phi_{F,G}$ is defined as:
\begin{equation*}
	(\extP{F}\circ\extP{G})X_\beta = F (\extP{G}X)_\beta = (F\circ G)X_\beta = (\extP{F \circ G})X_\beta
\end{equation*}
and the component $\gamma$ as:
\begin{equation*}
	(\extP{F}\circ\extP{G})X_\gamma = \lim_{\beta < \gamma}F \extP{G}X_\beta \stackrel{\ddagger}{\cong} \lim_{\beta < \gamma}(F\circ G)X_\beta = (\extP{F \circ G})X_\gamma
\end{equation*}
where $(\ddagger)$ easily follows by restriction to the successors base of $\alexT{\alpha}$.
As any category of endofunctors, $\Endo{\CC}$ and $\Endo{\Sh[\CC]{\alpha}}$ are (strict) monoidal categories whose tensor product and unit are endofunctor composition and the identity functor, respectively.
The isomorphisms $\psi$ and $\phi$ from above render the functor $(\extP{-})$ a monoidal functor.	

\begin{theorem}
	\label{thm:extp-monoidal-functor}
	The following data defines a strong monoidal functor from  $(\Endo{\CC},\circ,\Id_\CC)$ to $(\Endo{\Sh[\CC]{\alpha}},\circ,\Id_{\Sh[\CC]{\alpha}})$:
	\begin{itemize}
		\item the pointwise extension functor $(\extP{-})\colon \Endo{\CC} \to \Endo{\Sh[\CC]{\alpha}}$,
		\item the natural isomorphism $\phi_{F,G}\colon \extP{F}\circ\extP{G} \cong \extP{F\circ G}$, and
		\item the isomorphism	$\psi\colon \Id_{\Sh[\CC]{\alpha}} \cong \extP{\Id_{\CC}}$.
	\end{itemize}
\end{theorem}

\begin{proofatend}
	The monoidal structure of an endofunctor category under composition is strict since its associator, left unitor, and right unitor are all identities---in fact $(F\circ G) \circ H = F \circ (G \circ H)$, $F \circ \Id = F$, and $\Id \circ F = F$ for any $F$, $G$, and $H$. 
	Therefore, the coherence diagrams stating the compatibility of $(\extP{-})$, $\phi$, and $\psi$ with respect to the associator, left and right unitor (\cf \cite[Section~X.2]{maclane:cats}) instantiate as follows:
	\begin{equation*}
		\begin{tikzpicture}[
				auto, xscale=1.8, yscale=1.6, diagram font,
				baseline=(current bounding box.center)
			]	
			\node (n0) at (0,2) {\(\extP{F} \circ (\extP{G} \circ \extP{H})\)};
			\node (n1) at (2,2) {\((\extP{F} \circ \extP{G}) \circ \extP{H}\)};
			\node (n2) at (-1,1) {\(\extP{F} \circ (\extP{G \circ H})\)};
			\node (n3) at (3,1) {\((\extP{F \circ G}) \circ \extP{H}\)};
			\node (n4) at (0,0) {\(\extP{F \circ (G \circ H)}\)};
			\node (n5) at (2,0) {\(\extP{(F \circ G) \circ H}\)};
			\draw[identity] (n0) -- (n1);
			\draw[identity] (n4) -- (n5);
			\draw[->] (n0) to node[swap] {\(\id_{\extP{F}} \circ \phi_{G,H}\)} (n2);
			\draw[->] (n1) to node {\(\phi_{F,G} \circ \id_{\extP{H}}\)} (n3);
			\draw[->] (n2) to node[swap] {\(\phi_{F, G\circ H}\)} (n4);
			\draw[->] (n3) to node {\(\phi_{F \circ G, H}\)} (n5);
		\end{tikzpicture}
	\end{equation*}\begin{equation*}
		\begin{tikzpicture}[
				auto, xscale=2.5, yscale=1.6, diagram font,
				baseline=(current bounding box.center)
			]	
			\node (n0) at (0,1) {\(\extP{F} \circ \Id_{\Sh[\CC]{\alpha}} \)};
			\node (n1) at (1,1) {\(\extP{F}\)};
			\node (n2) at (0,0) {\(\extP{F} \circ \extP{\Id_\CC}\)};
			\node (n3) at (1,0) {\(\extP{G \circ \Id_\CC}\)};
			\draw[identity] (n0) -- (n1);
			\draw[identity] (n3) -- (n1);
			\draw[->] (n0) to node[swap] {\(\id_{\extP{F}} \circ \psi\)} (n2);
			\draw[->] (n2) to node[swap] {\(\phi_{G,\Id_\CC}\)} (n3);
		\end{tikzpicture}
		\qquad
		\begin{tikzpicture}[
				auto, xscale=2.5, yscale=1.6, diagram font,
				baseline=(current bounding box.center)
			]	
			\node (n0) at (0,1) {\(\Id_{\Sh[\CC]{\alpha}} \circ \extP{F}\)};
			\node (n1) at (1,1) {\(\extP{F}\)};
			\node (n2) at (0,0) {\(\extP{\Id_\CC} \circ \extP{F}\)};
			\node (n3) at (1,0) {\(\extP{\Id_\CC \circ G}\)};
			\draw[identity] (n0) -- (n1);
			\draw[identity] (n3) -- (n1);
			\draw[->] (n0) to node[swap] {\(\psi \circ \id_{\extP{F}}\)} (n2);
			\draw[->] (n2) to node[swap] {\(\phi_{\Id_\CC,G}\)} (n3);
		\end{tikzpicture}
	\end{equation*}
	Let $\beta$ a successor ordinal. At stage $\beta$ any component of the isomorphisms $\phi$ and $\psi$ is an identity and thus each corresponding component of the diagrams above has only identities as arrows. Stages associated to limit ordinals follow by universality and the coherent choice of limits inherent in fixing $\mathbf{a}$.
\end{proofatend}

A defining property of monoidal functors is that they send monoids to monoids. In the case of $(\extP{-})$ this means that if $T$ carries a monad structure, then its extension $\extP{T}$ carries a monad structure derived from the extensions of $\mu^{T}$ and $\eta^{T}$.
Note that $\extP{\mu^{T}}$ and $\extP{\eta^{T}}$ have types $\extP{T\circ T} \to \extP{T}$ and $\extP{\Id_\CC} \to \extP{T}$ instead of $\extP{T}\circ\extP{T} \to \extP{T}$ and $\Id_{\Sh[\CC]{\alpha}} \to \extP{T}$ (expected from any multiplication and unit for $\extP{T}$). The necessary gluing is provided by the isomorphisms $\phi$ and $\psi$ which allows us to derive a multiplication and unit for $\extP{T}$ from those of $T$ as follows:
\begin{equation}
	\label{eq:extp-monad}
	\mu^{\extP{T}} \defeq \extP{\mu^T} \bullet \phi_{T,T}
	\qquad
	\eta^{\extP{T}} \defeq \extP{\eta^T} \bullet \phi_{T,T}
	\text{.}
\end{equation}
It follows from simple diagram chasing that $\mu^{\extP{T}}$ and $\eta^{\extP{T}}$ satisfy the usual diagrams of associativity and unit (\cf \cref{thm:extp-monad-functor} below) and hence define a monad structure on $\extP{T}$. We call the monad $(\extP{T},\mu^{\extP{T}},\eta{\extP{T}})$ the pointwise extension of $(T,\mu^{T},\eta{T})$.
This structure is uniquely defined and hence by assigning to each monad its extension we obtain a functor $\Mnd{\extP{-}}$ between monad categories that is the restriction to monads of the pointwise extension functor $\extP{-}$ for endofunctors.

\begin{corollary}
	\label{thm:extp-monad-functor}
	The pointwise extension functor restricts to a functor between the categories of monads over $\CC$ and over $\Sh[\CC]{\alpha}$ as illustrated by the diagram below.
	\begin{equation*}	
		\begin{tikzpicture}[
				auto, xscale=3.5, yscale=1.2, diagram font,
				baseline=(current bounding box.center)
			]	
			\node (n0) at (0,1) {\(\Endo{\CC}\)};	
			\node (n1) at (1,1) {\(\Endo{\Sh[\CC]{\alpha}}\)};	
			\node (n2) at (0,0) {\(\Mnd{\CC}\)};	
			\node (n3) at (1,0) {\(\Mnd{\Sh[\CC]{\alpha}}\)};	
	
			\draw[->] (n0) to node[] {\((\extP{-})\)} (n1);
			\draw[->] (n2) to node[swap] {\(\Mnd{\extP{-}}\)} (n3);
			\draw[right hook->] (n2) -- (n0);
			\draw[right hook->] (n3) -- (n1);
			
		\end{tikzpicture}
	\end{equation*}	
\end{corollary}

\begin{proofatend}
	Let $(T,\mu^{T},\eta^{T})$ be a monad over $\CC$ and consider the following diagram asserting that $\mu^{\extP{T}} = \extP{\mu^T} \bullet \phi_{T,T}$ is associative:
 	\resetDiagMarker
	\begin{equation*}	
		\begin{tikzpicture}[
				auto, xscale=2.1, yscale=2.1, diagram font,
				rotate=-45,
				baseline=(current bounding box.center)
			]	
			\node (n0) at (0,2) {\(\extP{T}\circ \extP{T}\circ \extP{T}\)};	
			\node (n1) at (0,1) {\(\extP{T}\circ \extP{T\circ T}\)};	
			\node (n2) at (0,0) {\(\extP{T}\circ \extP{T}\)};	
			\node (n3) at (1,2) {\(\extP{T\circ T} \circ \extP{T}\)};	
			\node (n4) at (1,1) {\(\extP{T \circ T \circ T}\)};	
			\node (n5) at (1,0) {\(\extP{T \circ T}\)};	
			\node (n6) at (2,2) {\(\extP{T} \circ \extP{T}\)};	
			\node (n7) at (2,1) {\(\extP{T\circ T}\)};	
			\node (n8) at (2,0) {\(\extP{T}\)};	
			
			\draw[->] (n0) to node[swap] {\(\phi_{T,T\circ T}\)} (n1);
			\draw[->] (n0) to node {\(\phi_{T\circ T,T}\)} (n3);
			\draw[->] (n1) to node[swap] {\(\id_{\extP{T}} \circ \extP{\mu^T}\)} (n2);
			\draw[->] (n1) to node[swap] {\(\phi_{T,T}\)} (n4);
			\draw[->] (n2) to node[swap] {\(\phi_{T,T}\)} (n5);
			\draw[->] (n3) to node {\(\phi_{T,T}\)} (n4);
			\draw[->] (n3) to node {\(\extP{\mu^T} \circ \id_{\extP{T}}\)} (n6);
			\draw[->] (n4) to node[swap] {\(\extP{\id_T \circ \mu^T}\)} (n5);
			\draw[->] (n4) to node {\(\extP{\mu^T \circ \id_T}\)} (n7);
			\draw[->] (n5) to node[swap] {\(\extP{\mu^T}\)} (n8);
			\draw[->] (n6) to node {\(\phi_{T,T}\)} (n7);
			\draw[->] (n7) to node {\(\extP{\mu^T}\)} (n8);
			
			\draw[rounded corners,->] (n0) -- ($(n0)+(0,.7)$) to node {\(\id_{\extP{T}} \circ \mu^{\extP{T}}\)} ($(n6)+(0,.7)$) -- (n6);
			\draw[rounded corners,->] (n0) -- ($(n0)+(-.7,0)$) to node[swap] {\(\mu^{\extP{T}}\circ \id_{\extP{T}}\)} ($(n2)+(-.7,0)$) -- (n2);
			\draw[rounded corners,->] (n2) -- ($(n2)+(0,-.7)$) to node[swap] {\(\mu^{\extP{T}}\)} ($(n8)+(0,-.7)$) -- (n8);
			\draw[rounded corners,->] (n6) -- ($(n6)+(.7,0)$) to node {\(\mu^{\extP{T}}\)} ($(n8)+(.7,0)$) -- (n8);
			
			\nodeDiagMarker{$(n4)!.5!(n8)$}%
				{diag:extp-monad-mu-mu};
			\nodeDiagMarker{$(n1)!.5!(n5)$}%
				{diag:extp-monad-mu-phi-0};
			\nodeDiagMarker{$(n0)!.5!(n4)$}%
				{diag:extp-monad-mu-phi-1};
			\nodeDiagMarker{$(n3)!.5!(n7)$}%
				{diag:extp-monad-mu-phi-2};
		\end{tikzpicture}
	\end{equation*}
	The diagram above commutes since:
	\eqref{diag:extp-monad-mu-mu} commutes by hypothesis on $(T,\mu,\eta)$, squares (\ref{diag:extp-monad-mu-phi-0}-\ref{diag:extp-monad-mu-phi-2}) follow from naturality of $\phi$ and $\mu^{T}$, and all remaining diagrams commute by definition of $\mu^{\extP{T}}$.
	Consider the following decomposition of the diagram asserting that $\eta^{\extP{T}}$ is a right unit for $\mu^{\extP{T}}$:
 	\resetDiagMarker
	\begin{equation*}	
		\begin{tikzpicture}[
				auto, xscale=2.2, yscale=1.8, diagram font,
				baseline=(current bounding box.center)
			]	
			\node (n1) at (0,0) {\(\extP{T}\circ \Id_{\Sh[\CC]{\alpha}}\)};
			\node (n2) at (0,1) {\(\extP{T} \circ \extP{\Id_\CC}\)};	
			\node (n3) at (0,2) {\(\extP{T} \circ \extP{T}\)};	
			\node (n4) at (1,1) {\(\extP{T \circ \Id_\CC}\)};	
			\node (n5) at (1,2) {\(\extP{T \circ T}\)};	
			\node (n6) at (2,2) {\(\extP{T}\)};
			\node (n7) at (2,0) {\(\extP{T}\)};
			
			\draw[->] (n1) to node {\(\id_{\extP{T}}\circ\psi\)} (n2);
			\draw[->] (n2) to node {\(\id_{\extP{T}}\circ\extP{\eta^{T}}\)} (n3);
			\draw[->] (n2) to node[swap] {\(\phi_{T,\Id_\CC}\)} (n4);
			\draw[->] (n3) to node {\(\phi_{T,T}\)} (n5);
			\draw[->] (n4) to node[swap] {\(\extP{\id_T \circ \eta^{T}}\)} (n5);
			\draw[->] (n5) to node {\(\extP{\mu^T}\)} (n6);

			\draw[nat] (n7) to (n4);
			\draw[nat] (n7) to (n6);
			\draw[nat] (n7) to (n1);
			
			\draw[rounded corners,->] (n1) -- ($(n1)+(-.75,0)$) to node {\(\id_{\extP{T}}\circ\eta^{\extP{T}}\)} ($(n3)+(-.75,0)$) -- (n3);
			
			\draw[rounded corners,->] (n3) -- ($(n3)+(0,.5)$) to node {\(\mu^{\extP{T}}\)} ($(n6)+(0,.5)$) -- (n6);
			
			\nodeDiagMarker{$(n5)!.5!(n7)$}%
					{diag:extp-monad-eta-r-eta};
			\nodeDiagMarker{$(n2)!.5!(n7)$}%
					{diag:extp-monad-eta-r-psi};
			\nodeDiagMarker{$(n3)!.5!(n4)$}%
					{diag:extp-monad-eta-r-phi};			
			
		\end{tikzpicture}
	\end{equation*}
	This diagram commutes for: \eqref{diag:extp-monad-eta-r-eta} asserts that $\eta^{T}$ is a right unit of $\mu^{T}$, \eqref{diag:extp-monad-eta-r-psi} is states the compatibility of $\psi$ with the right unitor and $\phi$, and \eqref{diag:extp-monad-eta-r-phi} is a naturality square for $\phi$.
	The diagram describing $\eta^{\extP{T}}$ as a left unit for $\mu^{\extP{T}}$ can be shown to commute by a similar decomposition:
	\begin{equation*}	
		\begin{tikzpicture}[
				auto, xscale=2.2, yscale=1.8, diagram font,
				baseline=(current bounding box.center)
			]					
			\node (n1) at (2,0) {\(\Id_{\Sh[\CC]{\alpha} \circ \extP{T}}\)};
			\node (n2) at (2,1) {\(\extP{\Id_\CC} \circ \extP{T}\)};	
			\node (n3) at (2,2) {\(\extP{T} \circ \extP{T}\)};	
			\node (n4) at (1,1) {\(\extP{\Id_\CC \circ T}\)};	
			\node (n5) at (1,2) {\(\extP{T \circ T}\)};	
			\node (n6) at (0,2) {\(\extP{T}\)};
			\node (n7) at (0,0) {\(\extP{T}\)};
			
			\draw[->] (n1) to node[swap] {\(\psi \circ \id_{\extP{T}}\)} (n2);
			\draw[->] (n2) to node[swap] {\(\extP{\eta^{T}} \circ \id_{\extP{T}}\)} (n3);
			\draw[->] (n2) to node {\(\phi_{\Id_\CC,T}\)} (n4);
			\draw[->] (n3) to node[swap] {\(\phi_{T,T}\)} (n5);
			\draw[->] (n4) to node {\(\extP{ \eta^{T} \circ \id_T}\)} (n5);
			\draw[->] (n5) to node[swap] {\(\extP{\mu^T}\)} (n6);
			
			\draw[nat] (n7) to (n4);
			\draw[nat] (n7) to (n6);
			\draw[nat] (n7) to (n1);
			
			\draw[rounded corners,->] (n1) -- ($(n1)+(.75,0)$) to node[swap] {\(\eta^{\extP{T}} \circ \id_{\extP{T}}\)} ($(n3)+(.75,0)$) -- (n3);
			
			\draw[rounded corners,->] (n3) -- ($(n3)+(0,.5)$) to node[swap] {\(\mu^{\extP{T}}\)} ($(n6)+(0,.5)$) -- (n6);
		\end{tikzpicture}
	\end{equation*}
	Thus, $\extP{T}$ equipped with $\mu^{\extP{T}}$ and $\eta^{\extP{T}}$ as defined in \eqref{eq:extp-monad} is a monad over $\Sh[\CC]{\alpha}$.
	
	For $\theta\colon T \to T'$ a monad morphism, the natural transformation $\extP{\theta}\colon \extP{T} \to \extP{T'}$ is a morphism between the extensions for $T$ and $T'$ as the following commuting diagrams assert:
	\begin{equation*}	
		\begin{tikzpicture}[
				auto, xscale=2.2, yscale=1.8, diagram font,
				baseline=(current bounding box.center)
			]	
			\node (n0) at (0,2) {\(\extP{T} \circ \extP{T}\)};
			\node (n1) at (1,2) {\(\extP{T'} \circ \extP{T'}\)};
			\node (n2) at (0,1) {\(\extP{T \circ T}\)};
			\node (n3) at (1,1) {\(\extP{T' \circ T'}\)};
			\node (n4) at (0,0) {\(\extP{T}\)};
			\node (n5) at (1,0) {\(\extP{T'}\)};
			
			\draw[->] (n0) to node {\(\extP{\theta} \circ \extP{\theta}\)} (n1);
			\draw[->] (n2) to node {\(\extP{\theta \circ \theta}\)} (n3);
			\draw[->] (n4) to node[swap] {\(\extP{\theta}\)} (n5);
			
			\draw[->] (n0) to node[swap] {\(\phi_{T,T}\)} (n2);
			\draw[->] (n2) to node[swap] {\(\mu^T\)} (n4);
			
			\draw[->] (n1) to node {\(\phi_{T',T'}\)} (n3);
			\draw[->] (n3) to node {\(\mu^{T'}\)} (n5);	
			
			\draw[rounded corners,->] (n0) -- ($(n0)+(-.65,0)$) to node[swap] {\(\mu^{\extP{T}}\)} ($(n4)+(-.65,0)$) -- (n4);		
			
			\draw[rounded corners,->] (n1) -- ($(n1)+(.65,0)$) to node {\(\mu^{\extP{T}}\)} ($(n5)+(.65,0)$) -- (n5);
			
		\end{tikzpicture}
		\qquad
		\begin{tikzpicture}[
				auto, xscale=2.2, yscale=1.8, diagram font,
				baseline=(current bounding box.center)
			]	
			
			\node (n0) at (0,1) {\(\extP{\Id_\CC}\)};
			\node (n1) at (-.6,0) {\(\extP{T}\)};
			\node (n2) at (.6,0) {\(\extP{T}\)};
			\node (n3) at (0,2) {\(\Id_{\Sh[\CC]{\alpha}}\)};
			
			\draw[->] (n0) to node[pos=.2,swap] {\(\extP{\eta^{T}}\)} (n1);
			\draw[->] (n0) to node[pos=.2] {\(\extP{\eta^{T'}}\)} (n2);
			\draw[->] (n2) to node {\(\extP{\theta}\)} (n1);
			\draw[->] (n3) to node {\(\psi\)} (n0);
			\draw[rounded corners,->] (n3) -| node[swap,pos=.7] {\(\eta^{\extP{T}}\)} (n1);
			\draw[rounded corners,->] (n3) -| node[pos=.7] {\(\eta^{\extP{T'}}\)} (n2);
		\end{tikzpicture}
	\end{equation*}
	We conclude by noting that the functorial assignments above act on monad morphisms and on the functorial part of monads as the pointwise extension functor $(\extP{-})\colon \Endo{\CC} \to \Endo{\Sh[\CC]{\alpha}}$.
\end{proofatend}

For notational convenience, we will often write just $(\extP{-})$ instead of $\Mnd{\extP{-}}$.

\begin{remark}[Generalised writer monad]
\looseness=-1
Let $(T,\mu,\eta)$ be a monad over $\CC$. The endofunctor $\Func{\Id_{\CD}}{T}$ over $\Func{\CD}{\CC}$ carries a monad structure whose multiplication and unit are derived from $\mu$ and $\eta$ as follows:
\begin{equation*}
	(\mu \circ -)\colon (T\circ T \circ -) \to (T\circ -)
	\qquad
	(\eta \circ -)\colon (\Id \circ -) \to (T\circ -)
\end{equation*}
\looseness=-1
The usual coherence diagrams can be directly checked by simple diagram chasing. This construction is an instance of the \emph{writer monad transformer} \cite{jones:afp1995} where the monoid of writes is $(T,\mu,\eta)$ and the transformed monad is the identity on $\Func{\CD}{\CC}$. The pointwise extension of $T$ is the ``sheafified'' version of the above.
\end{remark}

Let $(T,\mu^T,\eta^T)$ be a monad, $F$ an endofunctor over $\CC$ and $\lambda^{T,F}$ be a distributive law for them. The pointwise extension of the natural transformation $\lambda^{T,F}$ is not a distributive law of the extensions of $(T,\mu^T,\eta^T)$ over $F$, yet. In fact, the type of $\extP{\lambda^{T,F}}$ is $\extP{F\circ T} \to \extP{T\circ F}$ instead of $\extP{F}\circ\extP{T} \to \extP{T}\circ\extP{F}$ required by distributive laws of $\extP{T}$ over $\extP{F}$.
Similarly to \eqref{eq:extp-monad}, the isomorphisms $\phi$ and $\psi$ that render $(\extP{-})$ a monoidal functor, provide us with the required gluing: for $\lambda^{T,F}$ a distributive law, define its pointwise extension as the natural transformation $\lambda^{\extP{T},\extP{F}}\colon \extP{F}\circ\extP{T} \To \extP{T}\circ\extP{F}$ defined as follows:
\begin{equation*}
	\lambda^{\extP{T},\extP{F}}	\defeq \phi^{-1}_{T,F}\bullet\extP{\lambda^{T,F}}\bullet\phi_{F,T}
	\text{.}
\end{equation*}
Recall that by \cref{thm:extp-dist-law-functor} the application of $(\extP{-})$ to natural transformations underlying monad morphisms yields monad morphisms for the extended monads. Therefore, the component-wise application of $(\extP{-})$ to a distributive law morphism $(\theta,\upsilon)$ yields the pair $(\extP{\theta},\extP{\upsilon})$ whose components are those of a distributive law morphisms. It follows from simple diagram chasing that the pair $(\extP{\theta},\extP{\upsilon})$ makes the necessary diagram commute and hence is a distributive law morphism. By assigning to each distributive law its extension and to each morphism the pair formed by the extensions of its components we obtain a functor between categories of distributive laws
$\MndEndo{\extP{-}}\colon \MndEndo{\CC} \to \MndEndo{\Sh[\CC]{\alpha}}$ that projects on the categories of monads and endofunctors as $\Mnd{\extP{-}}$ and $(\extP{-})$.

\begin{corollary}
	\label{thm:extp-dist-law-functor}
	The pointwise extension of distributive laws is functorial and commutes with the projections to the categories of monads and endofunctors as shown by the diagram below.
	\begin{equation*}	
		\begin{tikzpicture}[
				auto, xscale=4.8, yscale=1.3, diagram font,
				baseline=(current bounding box.center)
			]	
			\node (n0) at (0,2) {\(\Endo{\CC}\)};	
			\node (n1) at (1,2) {\(\Endo{\Sh[\CC]{\alpha}}\)};	
			\node (n2) at (0,0) {\(\Mnd{\CC}\)};	
			\node (n3) at (1,0) {\(\Mnd{\Sh[\CC]{\alpha}}\)};	
			\node (n4) at (0,1) {\(\MndEndo{\CC}\)};	
			\node (n5) at (1,1) {\(\MndEndo{\Sh[\CC]{\alpha}}\)};

			\draw[->] (n0) to node[] {\((\extP{-})\)} (n1);
			\draw[->] (n2) to node[swap] {\(\Mnd{\extP{-}}\)} (n3);
			\draw[->] (n4) to node[swap] {\(\MndEndo{\extP{-}}\)} (n5);
			\draw[->] (n4) -- (n0);
			\draw[->] (n4) -- (n2);
			\draw[->] (n5) -- (n1);
			\draw[->] (n5) -- (n3);
			
		\end{tikzpicture}
	\end{equation*}
	
\end{corollary}

\begin{proofatend}	
 	\resetDiagMarker
	First we prove that, for $\lambda\colon FT \to TF$ a distributive law, the natural transformation $\lambda^{\extP{T},\extP{F}}\colon \extP{F}\extP{T} \to \extP{T}\extP{F}$ is compatible with the pointwise extensions of $(T,\mu,\eta)$ and $F$.
	Consider the following decomposition of the compatibility diagram for $\lambda^{\extP{T},\extP{F}}\colon \extP{F}\extP{T} \to \extP{T}\extP{F}$ and $\mu^{\extP{T}}\colon \extP{T}\extP{T} \to \extP{T}$:
	\begin{equation*}
		\label{diag:extp-distlaw-mu-dec}
		\begin{tikzpicture}[
				auto, scale=1.6, diagram font,
				baseline=(current bounding box.center)
			]	
			
			\foreach \x in {0,1,...,4}{
				\pgfmathsetmacro\a{162-\x * 72}
				\pgfmathsetmacro\ak{162-18-\x * 72}				
				\pgfmathsetmacro\al{162+18-\x * 72}
	  			\coordinate (i\x) at (\a:1.1);
				\coordinate (j\x) at (\a:3.7);
				\coordinate (k\x) at (\ak:2.3);
				\coordinate (l\x) at (\al:2.3);
			}
			
			\node (n0) at (i0) {\(\extP{F \circ T \circ T}\)};
			\node (n1) at (i1) {\(\extP{T \circ F \circ T}\)};
			\node (n2) at (i2) {\(\extP{T \circ T \circ F}\)};
			\node (n3) at (i3) {\(\extP{T \circ F}\)};
			\node (n4) at (i4) {\(\extP{F \circ T}\)};
			
			\draw[->] (n0) to node {\(\extP{\lambda \circ \id_{T}}\)} (n1);
			\draw[->] (n1) to node {\(\extP{\id_{T} \circ \lambda}\)} (n2);
			\draw[->] (n2) to node {\(\extP{\mu^{T} \circ \id_{F}}\)} (n3);
			\draw[<-] (n3) to node {\(\extP{\lambda}\)} (n4);
			\draw[<-] (n4) to node {\(\extP{\id_{F} \circ \mu^{T}}\)} (n0);
			
			\node (m0) at (j0) {\(\extP{F} \circ \extP{T} \circ \extP{T}\)};
			\node (m1) at (j1) {\(\extP{T} \circ \extP{F} \circ \extP{T}\)};
			\node (m2) at (j2) {\(\extP{T} \circ \extP{T} \circ \extP{F}\)};
			\node (m3) at (j3) {\(\extP{T} \circ \extP{F}\)};
			\node (m4) at (j4) {\(\extP{F} \circ \extP{T}\)};
			
			\draw[->] (m0) to node {\(\lambda^{\extP{T},\extP{F}} \circ \id_{\extP{T}}\)} (m1);
			\draw[->] (m1) to node {\(\id_{\extP{T}} \circ \lambda^{\extP{T},\extP{F}}\)} (m2);
			\draw[->] (m2) to node {\(\mu^{\extP{T}} \circ \id_{\extP{F}}\)} (m3);
			\draw[<-] (m3) to node {\(\lambda^{\extP{T},\extP{F}}\)} (m4);
			\draw[<-] (m4) to node {\(\id_{\extP{F}} \circ \mu^{\extP{T}}\)} (m0);
			
			\draw[<-] (m3) to node {\(\phi^{-1}_{F,T}\)} (n3);
			\draw[->] (m4) to node[swap] {\(\phi_{T,F}\)} (n4);
			
			\node (o0) at (k0) {\(\extP{F \circ T} \circ \extP{T}\)};
			\node (o1) at (k1) {\(\extP{T} \circ \extP{F \circ T}\)};
			\node (o2) at (k2) {\(\extP{T \circ T} \circ \extP{F}\)};
			
			\node (p0) at (l0) {\(\extP{F} \circ \extP{T \circ T}\)};
			\node (p1) at (l1) {\(\extP{T \circ F} \circ \extP{T}\)};
			\node (p2) at (l2) {\(\extP{T} \circ \extP{T \circ F}\)};

			\draw[<-] (o0) to node {\(\cong\)} (m0);
			\draw[->] (m0) to node {\(\cong\)} (p0);
			\draw[<-] (n0) to node {\(\cong\)} (o0);
			\draw[->] (p0) to node {\(\cong\)} (n0);
			
			\draw[->] (o1) to node {\(\cong\)} (m1);
			\draw[<-] (m1) to node {\(\cong\)} (p1);
			\draw[<-] (n1) to node {\(\cong\)} (o1);
			\draw[->] (p1) to node {\(\cong\)} (n1);
			
			\draw[<-] (o2) to node {\(\cong\)} (m2);
			\draw[<-] (m2) to node {\(\cong\)} (p2);
			\draw[->] (n2) to node {\(\cong\)} (o2);
			\draw[->] (p2) to node {\(\cong\)} (n2);
			
			\draw[->] (o2) to node[swap] {\(\extP{\mu^{T}} \circ \id_{\extP{F}}\)} (m3);
			\draw[->] (p0) to node {\(\id_{\extP{F}} \circ \extP{\mu^{T}}\)} (m4);
			
			\draw[->] (o0) to node[] {\(\extP{\lambda} \circ \extP{\id_{T}}\)} (p1);
			\draw[->] (o1) to node[] {\(\id_{\extP{T}} \circ \extP{\lambda}\)} (p2);

			\nodeDiagMarker{0,0}				{diag:extp-distlaw-mu-def-lambda};
			\nodeDiagMarker{-90:2}{diag:extp-distlaw-mu-def-ext-lambda};
			\nodeDiagMarker{$(m0)!.6!(m1)!.1!(0,0)$}{diag:extp-distlaw-mu-def-ext-lambda-r};
			\nodeDiagMarker{ $(m1)!.4!(m2)!.1!(0,0)$}{diag:extp-distlaw-mu-def-ext-lambda-l};
			\nodeDiagMarker{$(m4)!.5!(m0)!.1!(0,0)$}{diag:extp-distlaw-mu-def-ext-mu-l};
			\nodeDiagMarker{$(m3)!.5!(m2)!.1!(0,0)$}{diag:extp-distlaw-mu-def-ext-mu-r};
			\nodeDiagMarker{$(m4)!.6!(n0)$}				{diag:extp-distlaw-mu-phi-mu-l};
			\nodeDiagMarker{$(m3)!.6!(n2)$}{diag:extp-distlaw-mu-phi-mu-r};
			\nodeDiagMarker{$(o0)!.5!(n1)$}{diag:extp-distlaw-mu-phi-lambda-r};
			\nodeDiagMarker{$(o1)!.5!(n2)$}{diag:extp-distlaw-mu-phi-lambda-l};
			\foreach \x in {0,1,2}{
				\nodeDiagMarker{$(m\x)!.5!(n\x)$}{diag:extp-distlaw-mu-phi-\x};
			}
			
		\end{tikzpicture}
	\end{equation*}
	The diagram \eqref{diag:extp-distlaw-mu-def-lambda} is \eqref{eq:law-dist-mnd-end-mu} and commutes since $\lambda\colon TF \to FT$ is assumed compatible with the structure of the monad $T$.
	Diagrams  (\ref{diag:extp-distlaw-mu-def-ext-lambda}--\ref{diag:extp-distlaw-mu-def-ext-lambda-l}) and (\ref{diag:extp-distlaw-mu-def-ext-mu-l}, \ref{diag:extp-distlaw-mu-def-ext-mu-r}) commute by distributivity of horizontal and vertical composition of natural transformations and by definition of $\lambda^{\extP{T},\extP{F}}\colon \extP{F}\extP{T} \to \extP{T}\extP{F}$ and $\mu^{\extP{T}}\colon \extP{T}\extP{T} \to \extP{T}$, respectively. 
	Diagrams \eqref{diag:extp-distlaw-mu-phi-mu-l} and \eqref{diag:extp-distlaw-mu-phi-lambda-r} are naturality squares for $\phi$.
	Diagrams \eqref{diag:extp-distlaw-mu-phi-0} and \eqref{diag:extp-distlaw-mu-phi-2}
	follow by coherence of $\phi$ with the monoidal associator.
	Thus the whole diagram commutes and in particular the outer pentagon \ie the compatibility diagram for $\lambda^{\extP{T},\extP{F}}$ and $\mu^{\extP{T}}$:	
 	\begin{equation*}
		\label{diag:extp-distlaw-mu}
		\begin{tikzpicture}[
			 auto, scale=1.7, diagram font,
			 baseline=(current bounding box.center)
		 ]	
		 
		 \foreach \x in {0,1,...,4}{
			 \pgfmathsetmacro\a{162-\x * 72}
			 \coordinate (j\x) at (\a:1.1);
		 }
		 \node (m0) at (j0) {\(\extP{F} \circ \extP{T} \circ \extP{T}\)};
		 \node (m1) at (j1) {\(\extP{T} \circ \extP{F} \circ \extP{T}\)};
		 \node (m2) at (j2) {\(\extP{T} \circ \extP{T} \circ \extP{F}\)};
		 \node (m3) at (j3) {\(\extP{T} \circ \extP{F}\)};
		 \node (m4) at (j4) {\(\extP{F} \circ \extP{T}\)};
		 
		 \draw[->] (m0) to node {\(\lambda^{\extP{T},\extP{F}} \circ \id_{\extP{T}}\)} (m1);
		 \draw[->] (m1) to node {\(\id_{\extP{T}} \circ \lambda^{\extP{T},\extP{F}}\)} (m2);
		 \draw[->] (m2) to node {\(\mu^{\extP{T}} \circ \id_{\extP{F}}\)} (m3);
		 \draw[<-] (m3) to node {\(\lambda^{\extP{T},\extP{F}}\)} (m4);
		 \draw[<-] (m4) to node {\(\id_{\extP{F}} \circ \mu^{\extP{T}}\)} (m0);
		 
		\end{tikzpicture}
	\end{equation*}
	
	\resetDiagMarker
	Consider the following decomposition of the compatibility diagram for $\lambda^{\extP{T},\extP{F}}\colon \extP{F}\extP{T} \to \extP{T}\extP{F}$ and $\eta^{\extP{T}}\colon \Id_{\Sh[\CC]{\alpha}} \to \extP{T}$:
	\begin{equation*}	
		\label{diag:extp-distlaw-eta}
		\begin{tikzpicture}[
				auto, scale=1.5, diagram font,
				baseline=(current bounding box.center)
			]	
			
			\foreach \x in {0,1,...,4}{
				\pgfmathsetmacro\a{-90-\x * 72}
  			\coordinate (i\x) at (\a:1);
				\coordinate (j\x) at (\a:2.5);
			}
			
			\node[alias=m0] (n0) at (i0) {\(\extP{F}\)};
			\node (n1) at (i1) {\(\extP{F \circ \id_{\CC}}\)};
			\node (n2) at (i2) {\(\extP{F \circ T}\)};
			\node (n3) at (i3) {\(\extP{T \circ F}\)};
			\node (n4) at (i4) {\(\extP{\Id_{\CC} \circ F}\)};
			
			\draw[identity] (n0) -- (n1);
			\draw[->] (n1) to node {\(\extP{\id_{F} \circ \eta^{T}}\)} (n2);
			\draw[->] (n2) to node {\(\extP{\lambda}\)} (n3);
			\draw[<-] (n3) to node {\(\extP{\eta^{T} \circ \id_{F}}\)} (n4);
			\draw[identity] (n4) -- (n0);

			\node (m1) at (j1) {\(\extP{F} \circ \extP{\Id_{\CC}}\)};
			\node (m2) at (j2) {\(\extP{F} \circ \extP{T}\)};
			\node (m3) at (j3) {\(\extP{T} \circ \extP{F}\)};
			\node (m4) at (j4) {\(\extP{\Id_{\CC}} \circ \extP{F}\)};
						
			\node (o1) at (240:2.5) {\(\extP{F} \circ \Id_{\Sh[\CC]{\alpha}}\)};
			\node (o4) at (300:2.5) {\(\Id_{\Sh[\CC]{\alpha}} \circ \extP{F}\)};
			
			\draw[->] (m1) to node {\(\id_{\extP{F}} \circ \extP{\eta^{T}}\)} (m2);
			\draw[<-] (m3) to node {\(\extP{\eta^{T}} \circ \id_{\extP{F}}\)} (m4);
			\draw[->] (m2) to node {\(\lambda^{\extP{T},\extP{F}}\)} (m3);
			\draw[->] (o1) to node {\(\id_{\extP{F}} \circ \psi\)} (m1);
			\draw[<-] (m4) to node {\(\psi \circ \id_{\extP{F}}\)} (o4);

			\draw[identity] (m0) -- (o1);
			\draw[identity] (m0) -- (o4);
			
			\draw[->] (m1) to node[swap] {\(\phi_{F,\Id_\CC}\)} (n1);
			\draw[->] (m2) to node {\(\phi_{F,T}\)} (n2);
			\draw[<-] (m3) to node[swap] {\(\phi^{-1}_{T,F}\)} (n3);
			\draw[->] (m4) to node {\(\phi_{\Id_\CC,F}\)} (n4);
						
			\draw[rounded corners,->] (o1) -- ++(-2,0) |- node[pos=.25] {\(\id_{\extP{F}} \circ \eta^{\extP{T}}\)} (m2);
			\draw[rounded corners,->] (o4) -- ++(2,0) |- node[pos=.25,swap] {\(\eta^{\extP{T}} \circ \id_{\extP{F}}\)} (m3);
			
			\nodeDiagMarker{0,0}{diag:extp-distlaw-eta-def-lambda};
			\nodeDiagMarker{90:1.72}{diag:extp-distlaw-eta-def-ext-lambda};	
			\nodeDiagMarker{-2.5,0}{diag:extp-distlaw-eta-def-ext-eta-r};
			\nodeDiagMarker{ 2.5,0}{diag:extp-distlaw-eta-def-ext-eta-l};
			\nodeDiagMarker{145:1.72}{diag:extp-distlaw-eta-phi-eta-r};
			\nodeDiagMarker{35:1.72}{diag:extp-distlaw-eta-phi-eta-l};
			\nodeDiagMarker{-135:1.72}{diag:extp-distlaw-eta-psi-r};
			\nodeDiagMarker{-45:1.72}{diag:extp-distlaw-eta-psi-l};
		\end{tikzpicture}
	\end{equation*}
	The pentagon (\ref{diag:extp-distlaw-eta-def-lambda}) is \eqref{eq:law-dist-mnd-end-eta} and commutes by hypothesis on $\lambda\colon FT \to TF$. 
	The square \eqref{diag:extp-distlaw-eta-def-ext-lambda} follows by definition of $\lambda^{\extP{T},\extP{F}}\colon \extP{F}\extP{T} \to \extP{T}\extP{F}$. 
	Triangles \eqref{diag:extp-distlaw-eta-def-ext-eta-r} and \eqref{diag:extp-distlaw-eta-def-ext-eta-l} commute by vertical-horizontal composition of natural transformations and by definition of $\eta^{\extP{T}}$.
	Diagrams  \eqref{diag:extp-distlaw-eta-phi-eta-r} and \eqref{diag:extp-distlaw-eta-phi-eta-r} are naturality squares of $\phi$ hence commute.
	Squares \eqref{diag:extp-distlaw-eta-psi-r} and \eqref{diag:extp-distlaw-eta-psi-l} commute by \cref{thm:extp-monoidal-functor} for they assert $\phi$ and $\psi$ coherent with right and left suitors, respectively.
	By pasting, the outer pentagon
	\begin{equation*}
		\begin{tikzpicture}[
				auto, scale=1.7, diagram font,
				baseline=(current bounding box.center)
			]	
			
			\foreach \x in {0,1,...,4}{
				\pgfmathsetmacro\a{-90-\x * 72}
				\coordinate (i\x) at (\a:1);
			}
			
			\node (n0) at (i0) {\(\extP{F}\)};
			\node (n1) at (i1) {\(\extP{F} \circ \Id_{\Sh[\CC]{\alpha}}\)};
			\node (n2) at (i2) {\(\extP{F} \circ \extP{T}\)};
			\node (n3) at (i3) {\(\extP{T} \circ \extP{F}\)};
			\node (n4) at (i4) {\(\Id_{\Sh[\CC]{\alpha}} \circ \extP{F}\)};
			
			\draw[identity] (n0) -- (n1);
			\draw[->] (n1) to node {\(\id_{\extP{F}} \circ \eta^{\extP{T}}\)} (n2);
			\draw[->] (n2) to node {\(\lambda^{\extP{T},\extP{F}}\)} (n3);
			\draw[<-] (n3) to node {\(\eta^{\extP{T}} \circ \id_{\extP{F}}\)} (n4);
			\draw[identity] (n4) -- (n0);
		\end{tikzpicture}
	\end{equation*}
	commutes. Therefore, the natural transformation $\lambda^{\extP{T},\extP{F}}$ is a distributive law of the extensions of $(T,\mu,\eta)$ over $F$.
	
	For the second part of the proof assume that $(\theta,\upsilon)$ is a distributive law morphism from $\lambda^{T,F}$ to $\lambda^{T',F'}$.
	The following diagram commutes by hypothesis on $(\theta,\upsilon)$:
	\begin{equation*}	
		\begin{tikzpicture}[
				auto, xscale=2.2, yscale=1.6, diagram font,
				baseline=(current bounding box.center)
			]	
			\node (n0) at (0,3) {\(\extP{F}\circ\extP{T}\)};
			\node (n1) at (.8,2) {\(\extP{F \circ T}\)};
			\node (n2) at (2.2,2) {\(\extP{T \circ F}\)};
			\node (n3) at (3,3) {\(\extP{T}\circ\extP{F}\)};
			\node (n4) at (0,0) {\(\extP{F'}\circ\extP{T'}\)};
			\node (n5) at (.8,1) {\(\extP{F' \circ T'}\)};
			\node (n6) at (2.2,1) {\(\extP{T' \circ F'}\)};
			\node (n7) at (3,0) {\(\extP{T'}\circ\extP{F'}\)};
			
			\draw[->] (n0) to node {\(\lambda^{\extP{T},\extP{F}}\)} (n3);
			\draw[->] (n4) to node[swap] {\(\lambda^{\extP{T'},\extP{F'}}\)}  (n7);
			
			\draw[->] (n0) to node {\(\phi_{F,T}\)} (n1);
			\draw[->] (n1) to node {\(\extP{\lambda^{T,F}}\)} (n2);
			\draw[->] (n2) to node {\(\phi^{-1}_{T,F}\)} (n3);
			\draw[->] (n4) to node[swap] {\(\phi_{F',T'}\)} (n5);
			\draw[->] (n5) to node[swap] {\(\extP{\lambda^{T',F'}}\)} (n6);
			\draw[->] (n6) to node[swap] {\(\phi^{-1}_{T',F'}\)} (n7);
			\draw[->] (n0) to node[swap] {\(\extP{\upsilon}\circ\extP{\theta}\)} (n4);
			\draw[->] (n1) to node[swap] {\(\extP{\upsilon\circ\theta}\)} (n5);
			\draw[->] (n2) to node {\(\extP{\theta\circ\upsilon}\)} (n6);
			\draw[->] (n3) to node {\(\extP{\theta}\circ\extP{\upsilon}\)} (n7);
			
		\end{tikzpicture}	
	\end{equation*}
	The outer part of this commuting diagram exhibits the pair $(\extP{\theta}, \extP{\upsilon})$ as a
	distributive law morphism from the extension of $\lambda^{T,F}$ to that of $\lambda^{T',F'}$. Functoriality conditions follow from $(\extP{-})$ being a functor and by diagram chasing. Finally, the show that the diagram in the claim commutes note that distributive law morphisms are extended as pairs of extensions.
\end{proofatend}

\subsection{Sheaf enrichment}
\label{sec:extp-sheaf-enrichment}

The pointwise extension functor enriches over $\Sh{\alpha}$ in the sense that endofunctors in its essential image are $\Sh{\alpha}$-enriched (\cf \cref{sec:sheaf-enrichment}) as stated by \cref{thm:extp-enriches-functors} below. From this result it follows that Kleisli categories and Kleisli liftings for pointwise extension of monads and endofunctors are sheaf enriched and hence admit a notion of local contractiveness.

\begin{theorem}
	\label{thm:extp-enriches-functors}
	The essential image of $(\extP{-})$ lies in the $\VEndo{\Sh{\alpha}}{\Sh[\CC]{\alpha}}$.
\end{theorem}

\begin{proofatend}
	In order to prove the thesis it suffices to show that the pointwise extension of any endofunctor $F$ over $\CC$ is $\Sh{\alpha}$-enriched \ie that each assignment $\extP{F}_{X,Y}\colon \Sh[\CC]{\alpha}(X,Y) \to \Sh[\CC]{\alpha}(\extP{F}X,\extP{F}Y)$ lies in $\Sh{\alpha}$ when hom-objects of $\Sh[\CC]{\alpha}$ are seen as sheaves of sets.
	Recall from \cref{sec:sheaf-enrichment} that for sheaves $X,Y \in \Sh[\CC]{\alpha}$, the hom-sheaf $\Sh[\CC]{\alpha}(X,Y)$ takes each stage $\beta$ to the value:
	\begin{equation*}
		\Sh[\CC]{\alpha}(X,Y)_{\beta} = \{(f,f_\beta) \mid f \in \Sh[\CC]{\alpha}(X,Y)\}
	\end{equation*}
	and each morphism $\beta \to \beta'$ to the restriction:
	\begin{equation*}
		\Sh[\CC]{\alpha}(X,Y)_{\res{\beta}{\beta'}}(f,f_\beta) = (f,f_{\beta'})
		\text{.}
	\end{equation*}
	Then, the component of $\extP{F}_{X,Y}$ at stage $\beta$, for $\beta$ a successor ordinal, is:
	\begin{equation*}
		\extP{F}_{X,Y,\beta}(f,f_\beta) = (\extP{F}f, \extP{F}f_\beta) =
		(\extP{F}f, Ff_\beta)
	\end{equation*}
	and the component at stage $\gamma$, for $\gamma$ a limit ordinal, is
	\begin{equation*}
		\extP{F}_{X,Y,\gamma}(f,f_\gamma) = (\extP{F}f, \extP{F}f_\gamma) =
		(\extP{F}f, \lim_{\beta < \gamma} Ff_\beta)
	\end{equation*}
	where  $\lim_{\beta < \gamma} Ff_\beta\colon \lim_{\beta < \gamma} FX_{\beta} \to \lim_{\beta < \gamma} FY_{\beta}$ is the mediating map for the cone $\{f_{\beta}\circ\pi_{\beta}\}_{\beta < \gamma}$. It follows from functoriality of $\extP{F}$ that these components are well-defined and satisfy the required naturality conditions as illustrated by the diagrams below:
	\begin{equation*}	
		\begin{tikzpicture}[
				auto, xscale=1.8, yscale=1.2, diagram font,
				baseline=(current bounding box.center)
			]	
			\node (n0) at (4.7,2) {\(f_\gamma\)};
			\node (n1) at (4.7,1) {\(f_{\beta'}\)};
			\node (n2) at (4.7,0) {\(f_\beta\)};
			\node (n3) at (5.7,2) {\(\lim_{\beta < \gamma} Ff_\beta\)};
			\node (n4) at (5.7,1) {\(Ff_{\beta'}\)};
			\node (n5) at (5.7,0) {\(Ff_\beta\)};
		
			\draw[|->] (n0) -- (n1);
			\draw[|->] (n0) -- (n3);
			\draw[|->] (n1) -- (n2);
			\draw[|->] (n1) -- (n4);
			\draw[|->] (n2) -- (n5);
			\draw[|->] (n3) -- (n4);
			\draw[|->] (n4) -- (n5);
			
			\node (m0) at (0,2) {\(\Sh[\CC]{\alpha}(X,Y)_\gamma\)};
			\node (m1) at (0,1) {\(\Sh[\CC]{\alpha}(X,Y)_{\beta'}\)};
			\node (m2) at (0,0) {\(\Sh[\CC]{\alpha}(X,Y)_{\beta}\)};
			\node (m3) at (2.2,2) {\(	\Sh[\CC]{\alpha}(\extP{F}X,\extP{F}Y)_{\gamma}\)};
			\node (m4) at (2.2,1) {\(	\Sh[\CC]{\alpha}(\extP{F}X,\extP{F}Y)_{\beta'}\)};
			\node (m5) at (2.2,0) {\(	\Sh[\CC]{\alpha}(\extP{F}X,\extP{F}Y)_{\beta}\)};
		
			\draw[->] (m0) to node[swap] {\(\)} (m1); %
			\draw[->] (m0) to node {\(\extP{F}_{X,Y,\gamma}\)} (m3);
			\draw[->] (m1) to node[swap] {\(\)} (m2); %
			\draw[->] (m1) to node {\(\extP{F}_{X,Y,\beta}\)} (m4);
			\draw[->] (m2) to node[swap] {\(\extP{F}_{X,Y,\beta}\)} (m5);
			\draw[->] (m3) to node {\(\Sh[\CC]{\alpha}(\extP{F}X,\extP{F}Y)_{\res{\beta'}{\gamma}}\)} (m4);
			\draw[->] (m4) to node {\(\Sh[\CC]{\alpha}(\extP{F}X,\extP{F}Y)_{\res{\beta}{\beta'}}\)} (m5);
						
		\end{tikzpicture}
	\end{equation*}
	where $\gamma$ is a limit ordinal and $\beta<\beta'$ are successor ordinals in $\gamma$.
\end{proofatend}

It follows from the above result that pointwise extensions of endofunctors are sheaf enriched, especially if they carry a monad structure. Kleisli categories of such monads share the same enrichment of their underlying category $\Sh[\CC]{\alpha}$ as stated by the following corollary.

\begin{corollary}
	\label{thm:extp-enriches-monads}
	For $(T,\mu,\eta)$ a monad, the category $\Kl{\extP{T}}$ is enriched over $\Sh{\alpha}$.
\end{corollary}

\begin{proofatend}
	Hom-objects of $\Kl{\extP{T}}$ are objects of $\Sh{\alpha}$ since the underlying category $\Sh[\CC]{\alpha}$ is enriched over $\Sh{\alpha}$.
	For $X$, $Y$, and $Z$, Kleisli composition is given as:
	\begin{equation*}	
		\begin{tikzpicture}[
				auto, xscale=6, yscale=1.2, diagram font,
				baseline=(current bounding box.center)
			]
			
			\node (m0) at (0,3) {\(\Kl{\extP{T}}(Y,Z) \times \Kl{\extP{T}}(X,Y)\)};
			\node (m3) at (0,0) {\(\Kl{\extP{T}}(X,Z)\)};

			\node (n0) at (1,3) {\(\Sh[\CC]{\alpha}(Y,\extP{T}Z) \times \Sh[\CC]{\alpha}(X,TY)\)};
			\node (n1) at (1,2) {\(\Sh[\CC]{\alpha}(\extP{T}Y,\extP{T}\extP{T}Z) \times \Sh[\CC]{\alpha}(X,TY)\)};
			\node (n2) at (1,1)  {\(\Sh[\CC]{\alpha}(X,\extP{T}\extP{T}Z)\)};
			\node (n3) at (1,0) {\(\Sh[\CC]{\alpha}(X,\extP{T}Z)\)};
						
			\draw[nat] (m0) to (n0);
			\draw[nat] (m3) to (n3);
			
			\draw[->] (m0) to node[swap] {\((-\circ_{X,Y,Z}-)\)} (m3);
			
			\draw[->] (n0) to node {\(\extP{T}_{Y,\extP{T}Z} \times \id\)} (n1);			
			\draw[->] (n1) to node {\((-\circ_{X,\extP{T}Y,\extP{T}\extP{T}Z}-)\)} (n2);
			\draw[->] (n2) to node {\((-\circ_{X,\extP{T}\extP{T}Z,\extP{T}Z}\mu_{Z})\)} (n3);			
		\end{tikzpicture}
	\end{equation*}
	It follows from \cref{thm:extp-enriches-functors,thm:c-sheaves-sheaf-enriched} that the morphism $(-\circ_{X,Y,Z}-)\colon \Kl{\extP{T}}(Y,Z) \times \Kl{\extP{T}}(X,Y) \to \Kl{\extP{T}}(X,Z)$ lies in $\Sh{\alpha}$ for it is given as the composition of morphisms in $\Sh{\alpha}$. Finally, associativity and existence of identities follow from definition of Kleisli category.
\end{proofatend}

A prerequisite of locally contractive endofunctors over $\Kl{\extP{T}}$ is to share its sheaf enrichment. This holds for Kleisli lifting of pointwise extensions (or, equivalently, for pointwise extensions of Kleisli lifting).

\begin{corollary}
	\label{thm:extp-enriches-kl-liftings}
	Any lifting $\liftKlextP{F}$ to $\Kl{\extP{T}}$ is enriched over $\Sh{\alpha}$.
\end{corollary}

\begin{proofatend}
	Let $\lambda^{\extP{T},\extP{F}}\colon \extP{F}\extP{T}\to \extP{T}\extP{F}$ be the distributive law induced by the Kleisli lifting $\liftKlextP{F}$ as per \cref{thm:kleisli-liftings-dist-laws}.
	For objects $X$ and $Y$, the functorial assignment $\liftKlextP{F}_{X,Y}\colon\Kl{\extP{T}}(X,Y)\to\Kl{\extP{T}}(\liftKlextP{F}X,\liftKlextP{F}Y)$ is given as
	\begin{equation*}	
		\begin{tikzpicture}[
				auto, xscale=4, yscale=1.2, diagram font,
				baseline=(current bounding box.center)
			]
			
			\node (m0) at (0,3) 
				{\(\Kl{\extP{T}}(X,Y)\)};
			\node (m2) at (0,1)
				{\(\Kl{\extP{T}}(\liftKlextP{F}X,\liftKlextP{F}Y)\)};

			\node (n0) at (1,3) 
				{\(\Sh[\CC]{\alpha}(X,\extP{T}Y)\)};
			\node (n1) at (1,2)
				{\(\Sh[\CC]{\alpha}(\extP{F}X,\extP{F}\extP{T}Y)\)};
			\node (n2) at (1,1) 
				{\(\Sh[\CC]{\alpha}(\extP{F}X,\extP{T}\extP{F}Y)\)};
						
			\draw[nat] (m0) -- (n0);
			\draw[nat] (m2) -- (n2);
			
			\draw[->] (m0) to node[swap] {\(\liftKlextP{F}_{X,Y}\)} (m2);
			
			\draw[->] (n0) to node {\(\extP{F}_{X,\extP{T}Y}\)} (n1);			
			\draw[->] (n1) to node {\((-\circ_{X,\extP{F}\extP{T}Y,\extP{T}\extP{F}Y}\lambda^{\extP{T},\extP{F}}_{Y})\)} (n2);			
		\end{tikzpicture}
	\end{equation*}
	and it is a morphism of sheaves in $\Sh{\alpha}$ by \cref{thm:extp-enriches-functors,thm:c-sheaves-sheaf-enriched}.
\end{proofatend}

\section{Towards guarded Kleisli (co)recursion}
\label{sec:guarded-kleisli-corecursion}
In this section we consider locally contractive endofunctors over Kleisli categories of monads obtained by pointwise extension. In particular, we are interested in Kleisli liftings of its ``guarded pointwise extension'' $\extP{F}\later$ and in the systematic derivation of these liftings from liftings of $F$ since these will be required by the constructions we introduce in \cref{sec:infinite-trace-semantics} in order to capture infinite trace semantics by finality. In \cref{sec:kleisli-lifting-guards} we show that there are settings of interest for modelling infinite traces that support this systematic derivation. In particular, we consider sheaves over $\alexT{\omega}$. In this setting, we identify a class of Kleisli liftings of $F$ that always extend to liftings of $\extP{F}\later$ and we prove that this class covers all Kleisli liftings if, and only if, the given monad is affine.

\subsection{Locally contractive Kleisli liftings}
\label{sec:locally-contractive-kleisli-liftings}

For this section let $\CD$ be a category enriched over $\Sh{\alpha}$ and let $(T,\mu,\eta)$ be a monad over $\CD$ such that its Kleisli category is enriched over $\Sh{\alpha}$---baring in mind that our prototypical example are pointwise extensions of monads to $\Sh[\CC]{\alpha}$. We are interested in locally contractive endofunctors over these Kleisli categories, especially in those that are Kleisli liftings of locally contractive endofunctors.

Recall from \cref{sec:locally-contractive-functors} that a locally contractive endofunctor over $\Kl{T}$ is any functor $F\colon \Kl{T} \to \Kl{T}$ that factors as a composition of functors enriched over $\Sh{\alpha}$:
\begin{equation*}
	\Kl{T} \xrightarrow{\nxt^{\Kl{T}}} \latercat{\Kl{T}} \longrightarrow \Kl{T}
\end{equation*}
where $\nxt^{\Kl{T}}$ is the functor induced by the point $\nxt\colon \Id_{\Sh{\alpha}} \to \later$.
Likewise, an endofunctor over $\CD$ is locally contractive if it factors as a composition of functors enriched over $\Sh{\alpha}$:
\begin{equation*}
	\CD \xrightarrow{\nxt^{\CD}} \latercat{\CD} \longrightarrow \CD
	\text{.}
\end{equation*}
Local contractiveness is only inherited by Kleisli liftings, not \viceversa.

\begin{proposition}
	\label{thm:kleisli-lifting-contractive}
	For $F$ an endofunctor over $\CD$, if $F$ is locally contractive, then any of its liftings to $\Kl{T}$ is locally contractive but not \viceversa.
\end{proposition}

\begin{proofatend}
	Let $\lambda$ be the distributive law induced by the Kleisli lifting $\liftKl{F}$ let $F$ factor as $F' \circ \nxt^{\CD}$ as per hypothesis.
	It follows from definition of Kleisli category that $\nxt^{\Kl{T}}$ is given on each hom-sheaf $\Kl{T}(X,Y) = \CD(X,TY)$, as the sheaf morphism $\nxt^{\CD}_{X,TY}\colon \CD(X,TY) \to \latercat{\CD}(X,TY)$.
	
	For any $X$ and $Y$ in $\Kl{T}$, the functorial assignment $\liftKl{F}_{X,Y}$ factors as follows:
	\begin{equation*}	
		\begin{tikzpicture}[
				auto, xscale=3.3, yscale=1.2, diagram font,
				baseline=(current bounding box.center)
			]
			
			\node (n00) at (0,4) 
				{\(\Kl{T}(X,Y)\)};
			\node (n03) at (0,1)
				{\(\Kl{T}(\liftKl{F}X,\liftKl{F}Y)\)};
			
			\node (n10) at (1,4) 
				{\(\CD(X,TY)\)};
			\node (n12) at (1,2) 
				{\(\CD(FX,FTY)\)};
			\node (n13) at (1,1)
				{\(\CD(FX,TFY)\)};

			\node (n20) at (2,4) 
				{\(\CD(X,TY)\)};
			\node (n21) at (2,3) 
				{\(\latercat{\CD}(X,TY)\)};
			\node (n22) at (2,2)
				{\(\CD(FX,FTY)\)};
			\node (n23) at (2,1) 
				{\(\CD(FX,TFY)\)};

			\node (n30) at (3,4) 
				{\(\Kl{T}(X,Y)\)};
			\node (n31) at (3,3) 
				{\(\latercat{\Kl{T}}(X,Y)\)};	
			\node (n33) at (3,1)
				{\(\Kl{T}(\liftKl{F}X,\liftKl{F}Y)\)};

			\draw[nat] (n00) -- (n10) -- (n20) -- (n30);
			\draw[nat]                   (n21) -- (n31);
			\draw[nat]          (n12) -- (n22)         ;
			\draw[nat] (n03) -- (n13) -- (n23) -- (n33);
			
			\draw[->] (n00) to node[swap] {\(\liftKl{F}_{X,Y}\)} (n03);
			
			\draw[->] (n10) to node[swap] {\(F_{X,TY}\)} (n12);
			\draw[->] (n12) to (n13);
			
			\draw[->] (n20) to node[swap] {\(\nxt^{\CD}_{X,TY}\)} (n21);
			\draw[->] (n21) to node[swap] {\(F'_{X,Y}\)} (n22);
			\draw[->] (n22) to node[swap] {\((-\circ_{X,FTY,TFY}\lambda_{Y})\)} (n23);
			
			\draw[->] (n30) to node {\(\nxt^{\Kl{T}}_{X,Y}\)} (n31);
			\draw[->] (n31) to node {\(\liftKl{F}'_{X,Y}\)} (n33);
			
		\end{tikzpicture}
	\end{equation*}
	The factorisation $\liftKl{F} = \liftKl{F}' \circ \nxt^{\Kl{T}}$ proves that the lifting $\liftKl{F}$ is locally contractive.
	Finally, note that it is not possible to identify $F'_{X,Y}$ from only $\extP{F}'_{X,Y}$. We conclude that the derivation in the other direction is not be possible.
\end{proofatend}

If the category $\CD$ comes equipped with its own instance of the modality $\later$ (\ie ${_\later}\CD(X,Y) \monoto \CD(\later X, \later Y)$) then this endofunctor provides us with a convenient way to turn any endofunctor over $\CD$ into a locally contractive one: simply compose it with $\later$. In fact, any instance of $\later$ is locally contractive and composition with locally contractive functors preserves this property. This is indeed the case in our setting of interest: pointwise extensions to $\Sh[\CC]{\alpha}$.
\begin{corollary}
	\label{thm:pointwise-kleisli-lifting-contractive}
	For $F$ an endofunctor over $\CC$, any Kleisli lifting to $\Kl{\extP{T}}$ for $\extP{F}\later$ is locally contractive.
\end{corollary}
\begin{proofatend}
	It follows from \cref{thm:extp-enriches-monads,thm:kleisli-lifting-contractive} that $\Kl{\extP{T}}$ and $\extP{F}$ are enriched over $\Sh{\alpha}$ and hence from \cref{thm:extp-enriches-kl-liftings} that $\extP{F}\later$ is locally contractive.
\end{proofatend}

\subsection{Kleisli liftings for guarded pointwise extensions}
\label{sec:kleisli-lifting-guards}

For the remaining of the section we return to our original setting: Kleisli categories of monads obtained by pointwise extension and Kleisli liftings of endofunctors obtained by ``guarding'' pointwise extensions. We investigate the derivation of liftings for $\extP{F} \later$ from liftings for $F$ and \viceversa.

The opposite derivation is always possible: the desired distributive laws are constructed from components in the image of $\Delta$ and stages associated to the so called ``double successor'' ordinals (\ie any $\beta + 2$). The derivation and its correctness are stated in \cref{thm:lifting-later-induced-dist-law} below; we will refer to distributive laws (resp.~Kleisli lifting) obtained in this way as \emph{induced} laws (resp.~lifting).

\begin{lemma}
	\label{thm:lifting-later-induced-dist-law}
	Let $\alpha > 1$ be an ordinal, $(T,\mu,\eta)$ a monad, $F$ an endofunctor, and $\xi$ a distributive law of $(\extP{T},\mu^\extP{T},\eta^\extP{T})$ over $\extP{F}\later$ on $\Sh[\CC]{\alpha}$. There is a distributive law $\lambda$ of $(T,\mu,\eta)$ over $F$ such that 
	$
		\lambda_X = \xi_{\Delta X,\beta+2}
	$
	for any $X \in \CC$ and $\beta + 1 < \alpha$.
\end{lemma}

\begin{proofatend}
	Fix an ordinal $\beta$ such that $\beta + 1 < \alpha$.
	For any object $X$ and any morphism $f \colon X \to Y$ in $\CC$, the following equalities hold:	
	\begin{align*}
		(\extP{F}\later\extP{T}\Delta X)_{\beta+2} 
		& =  %
		FTX
		&
		(\extP{F}\later\extP{T}\Delta f)_{\beta+2} 
		& =  %
		FTf
	\\
		(\extP{T}\extP{F}\later\Delta X)_{\beta+2} 
		& =  %
		TFX
		&
		(\extP{T}\extP{F}\later\Delta f)_{\beta+2} 
		& =  %
		TFf
	\end{align*}
	It follows from these equalities that the naturality square for $\xi_{\Delta X}$, $\xi_{\Delta Y}$, and $\Delta f\colon \Delta X \to \Delta Y$ at stage $\beta+2$ is exactly that for $\lambda_X$, $\lambda_Y$, and $f\colon X \to Y$. Therefore, the family $\{\lambda_X\}_{X \in \CC}$ is a natural transformation of type $FT \to TT$. 
	
	For any object $X$ and any morphism $f \colon X \to Y$ in $\CC$, the following hold:
	\begin{align*}
		(\extP{F}\later\Delta X)_{\beta+2} 
		& = 
		FX 
		&
		(\extP{F}\later\Delta f)_{\beta+2}
		& = 
		Ff
		\\
		(\extP{F}\later\extP{T}\extP{T}\Delta X)_{\beta+2} 
		& = 
		FTTX 
		&
		(\extP{F}\later\extP{T}\extP{T}\Delta f)_{\beta+2}
		& = 
		FTTf
		\\
		(\extP{T}\extP{F}\later\extP{T}\Delta X)_{\beta+2}
		& = 
		TFTX
		&
		(\extP{T}\extP{F}\later\extP{T}\Delta f)_{\beta+2} & = 
		TFTf
		\\
		(\extP{T}\extP{T}\extP{F}\later\Delta X)_{\beta+2} & = 
		TTFX
		&
		(\extP{T}\extP{T}\extP{F}\later\Delta f)_{\beta+2} & = 
		TTFf
		\\
		(\xi\circ\id_{\extP{T}})_{\Delta X,\beta + 2}
		& = 
		(\lambda \circ \id_{T})_{X}
		&
		(\id_\extP{T}\circ\xi)_{\Delta X,\beta + 2}
		& = 
		(\id_{T} \circ \lambda)_{X}
		\\
		(\id_{\extP{F}\circ\later}\circ\mu^{\extP{T}})_{\Delta X,\beta+2}
		& = 
		\id_{F} \circ \mu_X
		&
		(\mu^{\extP{T}}\circ\id_{\extP{F}\circ\later})_{\Delta X,\beta+2}
		& = 
		\mu_X \circ \id_{F}
		\\
		(\id_{\extP{F}\circ\later}\circ\eta^{\extP{T}})_{\Delta X,\beta+2}
		& = 
		\id_{F} \circ \eta_X
		&
		(\eta^{\extP{T}}\circ\id_{\extP{F}\circ\later})_{\Delta X,\beta+2}
		& = 
		\eta_X \circ \id_{F}
	\end{align*}
	Consider the diagrams asserting that $\xi$ is compatible with the structure of $(\extP{T}, \mu^{\extP{T}}, \eta^{\extP{T}})$ and, in particular, their components at stage $\beta + 2$: it follows from the equalities above that these are exactly the compatibility diagrams for $\lambda$ and the structure of $(T,\mu,\eta)$.
	
	It follows from definition of constant sheaves that the choice of the ordinal $\beta$ made at the beginning of this proof is irrelevant: indeed any restriction map $\Delta X_{\res{\beta+2}{\beta'+2}}$ is an identity.
\end{proofatend}

In general, the converse of \cref{thm:lifting-later-induced-dist-law} does not hold: it is not true that a distributive law of $(T,\mu,\eta)$ over $F$ extends to one of $(\extP{T},\mu^\extP{T},\eta^\extP{T})$ and $\extP{F}\later$. 
In order to illustrate why this is not the case, assume a distributive law $\xi$ of $(\extP{T},\mu^\extP{T},\eta^\extP{T})$ over $\extP{F}\later$ and write $\lambda$ for the distributive law induced by $\xi$. 
Stages for double successor ordinals contains the data used in the derivation of $\lambda$ and stages for limit ordinals are covered by definition of $\Sh[\CC]{\alpha}$: as a consequence, the issue at hand must arise from (first) successors of limits ordinals. In fact, for $\gamma$ a limit ordinal and $X$ a sheaf, the component $\xi_{X,\gamma+1}\colon \extP{F}\later \extP{T} X_{\gamma+1} \to \extP{F}\later X_{\gamma+1}$ is a morphism:
\begin{equation*}
	\xi_{X,\gamma+1}\colon F \lim_{\beta < \gamma} TX_\beta \to FT\lim_{\beta < \gamma} X_\beta\text{.}
\end{equation*}
It follows from the type of these components that their naturality, their compatibility with the structure of $\extP{T}$, and even their existence are non-trivial and can not be derived from $\lambda\colon FT \to TF$ without any additional information. For instance, when $F$ is the identity and $\lambda$ an automorphism for $T$, the naturality and compatibility conditions for the induced $\xi$ result to be even stronger than imposing that $T$ weakly preserve limits of $\gamma$-chains.

\subsection{Affiness and extension of Kleisli liftings}
\label{sec:kleisli-lifting-affiness}

In this section we focus on pointwise extensions to sheaves over $\alexT{\omega}$. In this setting, we identify a class of Kleisli liftings we call \emph{$\omega$-suitable} and such that the converse of \cref{thm:lifting-later-induced-dist-law} always holds. Remarkably, a monad has the property that all liftings to its Kleisli category are $\omega$-suitable if and only if it has the affiness property (\cf \cref{thm:affine-dist-law}).

\begin{definition}
	\label{def:omega-suitable-dist-law}
	A law $\lambda$ distributing a monad $(T,\mu,\eta)$ over an endofunctor $F$ is called \emph{$\omega$-suitable} whenever the diagram below commutes for any object $X$ in $\CC$:
	\begin{equation}
		\label{eq:omega-suitable-dist-law}
		\begin{tikzpicture}[
				auto, xscale=2.1, yscale=1.2, diagram font,
				baseline=(current bounding box.center)
			]	
			
			\node (n0) at (0,1) {\(FTX\)};	
			\node (n1) at (1,1) {\(TFX\)};	
			\node (n2) at (0,0) {\(F1\)};	
			\node (n3) at (1,0) {\(TF1\)};	
	
			\draw[->] (n0) to node {\(\lambda_{X}\)} (n1);
			\draw[->] (n0) to node[swap] {\(F!_{TX}\)} (n2);
			\draw[->] (n1) to node {\(TF!_{X}\)} (n3);
			\draw[->] (n2) to node[swap] {\(\eta_{F1}\)} (n3);
		\end{tikzpicture}
	\end{equation}
	Kleisli liftings are called $\omega$-suitable whenever their associated distributive laws are $\omega$-suitable.
\end{definition}

The $\omega$-suitability property provides us with the necessary information for extending distributive laws of $T$ over $F$ to distributive laws of $\extP{T}$ over $\extP{F}\later$.

\begin{theorem}
	\label{thm:omega-suitable-dist-law}
	Let $(T,\mu,\eta)$ be a monad and $F$ an endofunctor, both over $\CC$.
	The following statements, where pointwise extensions target $\Sh[\CC]{\omega}$, are true.
	\begin{itemize}
		\item For $\xi\colon \extP{F}\later\extP{T} \to \extP{T}\extP{F}\later$ a distributive law, the distributive law $\lambda\colon FT \to TF$ induced by $\xi$ is $\omega$-suitable.
		\item For $\lambda\colon FT \to TF$ an $\omega$-suitable distributive law, there is $\xi\colon \extP{F}\later\extP{T} \to \extP{T}\extP{F}\later$ a distributive law of $\extP{T}$ over $\extP{F}\later$ given on each sheaf $Y$ as:
		\begin{equation*}
			\xi_{Y,0} = \id_1 
			\qquad 
			\xi_{Y,1} = \eta_{F1}
			\qquad 
			\xi_{Y,n+2} = \lambda_{Y_{n+2}}
			\qquad
			\xi_{Y,\omega} = \rho_Y
		\end{equation*}
		where $n \in \omega$ and $\rho_Y$ is the mediating map for the cone 	$\{\xi_{Y,n}\circ \extP{F}\later\extP{T}Y_{\res{n}{\omega}}\}_{n < \omega}$. 
	\end{itemize}
\end{theorem}

\begin{proofatend}
	\looseness=-1
	Let $\xi\colon \extP{F}\later\extP{T} \to \extP{T}\extP{F}\later$ be a distributive law and write $\lambda$ for the distributive law induced by $\xi$ as per \cref{thm:lifting-later-induced-dist-law}. We proceed by showing that for each $X \in \CC$ and finite $n > 1$, diagram \eqref{eq:omega-suitable-dist-law} corresponds to the naturality square:
	\begin{equation*}
		\begin{tikzpicture}[
				auto, xscale=2.1, yscale=1.2, diagram font,
				baseline=(current bounding box.center)
			]	
			
			\node (n0) at (0,1) {\(FTX\)};	
			\node (n1) at (1,1) {\(TFX\)};	
			\node (n2) at (0,0) {\(F1\)};	
			\node (n3) at (1,0) {\(TF1\)};	

			\draw[->] (n0) to node {\(\xi_{\Delta X,n}\)} (n1);
			\draw[->] (n0) to node[swap] {\(F!_{TX}\)} (n2);
			\draw[->] (n1) to node {\(TF!_{X}\)} (n3);
			\draw[->] (n2) to node[swap] {\(\xi_{\Delta X,1}\)} (n3);
		\end{tikzpicture}
	\end{equation*}
	It follows from definition of $\lambda$ and from compatibility of $\xi$ with the unit of $\extP{T}$ that $\xi_{\Delta X,1}$ is $\eta_{F1}$. In fact, for any sheaf $Y$, the component at stage $1$ of the compatibility diagram for $\xi_Y$ and $\eta^{\extP{T}}_Y$ corresponds to the diagram below:
	\begin{equation*}
		\begin{tikzpicture}[
				auto, xscale=2.3, yscale=1, diagram font,
				baseline=(current bounding box.center)
			]	
			
			\node (n0) at (0,1) {\(1\)};	
			\node (n1) at (.5,0) {\(1\)};	
			\node (n2) at (1,1) {\(T1\)};	

			\draw[->] (n0) to node{\(\xi_{Y,1}\)} (n2);
			\draw[->] (n1) to node {\(\id_{1}\)} (n0);
			\draw[->] (n1) to node[swap] {\(\eta_{F1}\)} (n2);
		\end{tikzpicture}
	\end{equation*}
	Finally, $\xi_{\Delta X,n}$ is $\lambda_X$ by definition of $\lambda$ and by assumption on $n$. Thus, $\lambda$ is $\omega$-suitable.
	
	For the converse assume $\lambda\colon FT \to TF$ an $\omega$-suitable distributive law and let $\xi$ be the family $\{\xi_{Y,\beta}\colon \extP{F}\later\extP{T}Y_\beta \to \extP{T}\extP{F}\later Y_{\beta}\}_{Y \,\in\, \Sh[\CC]{\omega},\; \beta \,\in\, \alexT{\omega}}$ as defined above.
	First we prove that the family $\xi$ is natural in both $Y$ and $\beta$ and then that it is compatible with the structure of $\extP{T}$.
	For sheaves $Y,Y' \in \Sh[\CC]{\omega}$ and finite ordinals $n,n' > 1$, the components $\xi_{Y,n}$ and $\xi_{Y',n'}$ of $\xi$ are, by construction, $\lambda_{Y_n}$ and $\lambda_{Y'_{n'}}$, respectively, and satisfy the naturality condition for them because $\lambda$ is a natural transformation. For all sheaves $Y,Y' \in \Sh[\CC]{\omega}$ and any finite ordinal $n > 1$, the components $\xi_{Y,n}$ and $\xi_{Y',1}$ of $\xi$ are $\lambda_{Y_n}$ and $\eta_1$, respectively, and satisfy the naturality condition since the associated naturality square corresponds to diagram \eqref{eq:omega-suitable-dist-law} which commutes by $\omega$-suitability of $\lambda$. All the remaining components are at stage $0$ or $\omega$ and, by definition of morphisms in $\Sh[\CC]{\omega}$, are mediating maps. Therefore, $\xi$ is a natural transformation.
	
	For $Y$ a sheaf consider the compatibility diagrams associated with the component $\xi_Y$.
	At stage $1$ these instantiate to the diagrams in $\CC$:
	\begin{equation*}\begin{tikzpicture}[
				auto, xscale=2, yscale=1.7, diagram font,
				baseline=(current bounding box.center)]
			\node (n0) at (0,2) {\(F1\)};
			\node (n1) at (1,2) {\(TF1\)};
			\node (n2) at (2,2) {\(TTF1\)};
			\node (n3) at (0,1) {\(F1\)};
			\node (n4) at (2,1) {\(TF1\)};
			\node (n5) at (1,0) {\(F1\)};
			
			\draw[->] (n0) to node[] {\(\eta_{F1}\)} (n1);
			\draw[->] (n0) to node[swap] {\(F\id_1\)} (n3);
			\draw[->] (n1) to node[] {\(T\eta_{F1}\)} (n2);
			\draw[nat] (n1) to (n4);
			\draw[->] (n2) to node[] {\(\mu_{F1}\)} (n4);
			\draw[->] (n3) to node[swap] {\(\eta_{F1}\)} (n4);
			\draw[->] (n5) to node[] {\(F{\id_1}\)} (n3);
			\draw[->] (n5) to node[swap] {\(\eta_{F1}\)} (n4);
		\end{tikzpicture}
	\end{equation*}
	which commute by basic properties of $\eta$ and $\mu$.
	At stage $n$, for any finite $n > 1$, the compatibility diagrams for $\xi_Y$ are those for $\lambda_{Y_{n-1}}$ and commute by hypothesis on $\lambda$. Finally, diagrams at stages $0$ and $\omega$ follow from definition of pointwise extension and of morphisms in $\Sh[\CC]{\omega}$.
\end{proofatend}

A monad $(T,\mu,\eta)$ on a category with a final object $1$ is called \emph{affine} whenever its unit exhibits the isomorphism $T1 \cong 1$ \cite{kock:adm1971,lindner:adm1979,jacobs:apal1994,jacobs:cmcs2016}.\footnote{Older formulations for strong commutative monads require that components of double strengths are sections to $\langle T\pi_1,T\pi_2 \rangle$ \cite{kock:adm1971,jacobs:apal1994}.} Examples of affine monads are the non-empty powerset monad $\mathcal{P}^+$, the probability distribution monad $\mathcal{D}$, or the Giry monad $\mathcal{G}$.

\begin{lemma}
	\label{thm:affine-omega-suitable}
	Let $\lambda$ be distributive law of $(T,\mu,\eta)$ over $F$. If $T$ is affine, then $\lambda$ is $\omega$-suitable.
\end{lemma}

\begin{proofatend}
	\resetDiagMarker
	It follows from finality of $1$ in $\CC$ and the affiness of $T$ that the diagram below commutes for any object $X$ of $\CC$:
	\begin{equation}
		\label{diag:affine-omega-suitable-1}
		\stepcounter{diagrammarker}
		\tag{\roman{diagrammarker}}
		\begin{tikzpicture}[
				auto, xscale=2, yscale=1.7, diagram font,
				baseline=(current bounding box.center)
			]	
			
			\node (n0) at (0,1) {\(TX\)};	
			\node (n1) at (0,0) {\(1\)};	
			\node (n2) at (1,0) {\(T1\)};	
			\node (n3) at (1,1) {\(T1\)};	
			
			\draw[->] (n0) to node[swap] {\(!_{TX}\)} (n1);
			\draw[->] (n0) to node[] {\(T!_{X}\)} (n3);
			\draw[->] (n1) to node[swap] {\(\eta_1\)} (n2);
			\draw[->] (n3) to node[swap] {\(!_{T1}\)} (n1);
			\draw[->] (n3) to node[] {\(\id_{T1}\)} (n2);
		\end{tikzpicture}
	\end{equation}
	For $X$ an object of $\CC$, consider the following decomposition of diagram \eqref{eq:omega-suitable-dist-law}:
	\begin{equation*}
		\begin{tikzpicture}[
				auto, xscale=4, yscale=3, diagram font,
				baseline=(current bounding box.center)
			]	
			
			\node (n0) at (0,1) {\(FTX\)};	
			\node (n1) at (1,1) {\(TFX\)};	
			\node (n2) at (0,0) {\(F1\)};	
			\node (n3) at (1,0) {\(TF1\)};	
			\node (n4) at (.5,.5) {\(FT1\)};	
		
			\draw[->] (n0) to node[] {\(\lambda_X\)} (n1);
			\draw[->] (n0) to node[swap] {\(F!_{TX}\)} (n2);
			\draw[->] (n0) to node {\(FT!_{X}\)} (n4);
			\draw[->] (n1) to node[] {\(TF!_{X}\)} (n3);
			\draw[->] (n2) to node[swap] {\(\eta_1\)} (n3);
			\draw[->] (n2) to node[] {\(F\eta_1\)} (n4);
			\draw[->] (n4) to node[] {\(\lambda_1\)} (n3);
			
			\nodeDiagMarker{$(n0)!.5!(n2)!.5!(n4)$}%
				{diag:affine-omega-suitable-2};
			\nodeDiagMarker{$(n1)!.5!(n4)$}%
				{diag:affine-omega-suitable-3};
			\nodeDiagMarker{$(n2)!.5!(n3)!.5!(n4)$}%
				{diag:affine-omega-suitable-4};
				
		\end{tikzpicture}
	\end{equation*}
	Diagram \eqref{diag:affine-omega-suitable-2} commutes since it is the image of \eqref{diag:affine-omega-suitable-1}. Diagrams 	\eqref{diag:affine-omega-suitable-3} and 	\eqref{diag:affine-omega-suitable-4} follow from naturality and compatibility of $\lambda$ with $\eta$, respectively. 
\end{proofatend}

It follows from \cref{thm:omega-suitable-dist-law,thm:affine-omega-suitable} that if $(T,\mu,\eta)$ if affine, then any distributive law of $T$ over $F$ induces a distributive law of $\extP{T}$ over $\extP{F}\later$ (where pointwise extensions target $\Sh[\CC]{\omega}$). 
Remarkably, this property of distributive laws and affiness of monads are equivalent:

\begin{theorem}
	\label{thm:affine-dist-law}
	For $(T,\mu,\eta)$ a monad on $\CC$, the following statements are equivalent:
	\begin{itemize}
		\item The monad $(T,\mu,\eta)$ is affine.
		\item For any endofunctor $F$ and any distributive law $\lambda\colon FT \to TF$, there is $\xi\colon \later \extP{T} \to \extP{T}\later$ such that it is compatible with the pointwise extension of $(T,\mu,\eta)$ to $\Sh[\CC]{\omega}$ and it induces $\lambda$.
	\end{itemize}
\end{theorem}

\begin{proofatend}
	Assume that $(T,\mu,\eta)$ is an affine monad, then the implication follows from \cref{thm:omega-suitable-dist-law} and \cref{thm:affine-omega-suitable}.
	For the converse assume that for any endofunctor $F$, all distributive laws of $T$ over $F$ are induced by laws for $\extP{T}$ and $\extP{F}\later$. 
	Note that laws for $\extP{T}$ and $\later$ induce laws for $T$ and $\Id$ and these are exactly endomorphisms for the monad $T$. Thus, by assumption there is a distributive law $\xi\colon \later \extP{T} \to \extP{T}\later$ such that its induced law for $T$ and $\Id$ is the identity on $T$. In particular, consider its component for final sheaf $\Delta 1$. By construction, $\xi_{\Delta 1,n} = \id_{T1}$ for any finite successor ordinal $n$.
	It follows from naturality of $\xi$ that $\xi_{\Delta 1, 1}$ is an isomorphism since naturality of components $1$ and $n > 1$ corresponds to the following diagram:
	\begin{equation*}
		\begin{tikzpicture}[
				auto, xscale=2.1, yscale=1.2, diagram font,
				baseline=(current bounding box.center)
			]	
			
			\node (n0) at (0,1) {\(T1\)};	
			\node (n1) at (1,1) {\(T1\)};	
			\node (n2) at (0,0) {\(1\)};	
			\node (n3) at (1,0) {\(T1\)};	

			\draw[->] (n0) to node {\(\id_{T1}\)} (n1);
			\draw[->] (n0) to node[swap] {\(!_{T1}\)} (n2);
			\draw[->] (n1) to node {\(T\id_{1}\)} (n3);
			\draw[->] (n2) to node[swap] {\(\xi_{\Delta 1,1}\)} (n3);
		\end{tikzpicture}
	\end{equation*}
	It follows from compatibility of $\xi$ with $\eta^{\extP{T}}$ that the component $\xi_{\Delta 1,1}$ is $\eta_{1}$ since the associated diagram is the following:
	\begin{equation*}
		\begin{tikzpicture}[
				auto, xscale=2.2, yscale=1.2, diagram font,
				baseline=(current bounding box.center)
			]	
			
			\node (n0) at (0,1) {\(1\)};	
			\node (n1) at (.5,0) {\(1\)};	
			\node (n2) at (1,1) {\(T1\)};	

			\draw[->] (n0) to node{\(\xi_{\Delta 1,1}\)} (n2);
			\draw[->] (n1) to node {\(\id_{1}\)} (n0);
			\draw[->] (n1) to node[swap] {\(\eta_{1}\)} (n2);
		\end{tikzpicture}
	\end{equation*}
	Therefore, $\eta_1$ is an isomorphism and $(T,\mu,\eta)$ is affine.
\end{proofatend}

\phantomLabel{def:affine-part-monad}%
Recall from \cite{lindner:adm1979,jacobs:apal1994} that the \emph{affine part} of a monad $(T,\mu,\eta)$ over $\CC$ is the greatest affine submonad $T^{\mathrm{a}}$ of $T$ and that, assuming $\CC$ has enough finite limits, $T^{\mathrm{a}}$ is determined on each object $X$ by pulling back $\eta_1$ along $T_{!_X}$: 
\begin{equation*}
	\begin{tikzpicture}[
			auto, xscale=2.2, yscale=1.3, diagram font,
			baseline=(current bounding box.center)
		]	
		
		\node (n0) at (0,1) {\(T^{\mathrm{a}}X\)};	
		\node (n1) at (1,1) {\(TX_\alpha\)};	
		\node (n2) at (0,0) {\(1\)};	
		\node (n3) at (1,0) {\(T1\)};	

		\draw[->] (n0) to (n1);
		\draw[->] (n0) to (n2);
		\draw[->] (n1) to node {\(T!_{X}\)} (n3);
		\draw[->] (n2) to node[swap] {\(\eta_{1}\)} (n3);
		
		\draw (n0.south east) ++(7pt,2pt) -- ++(0,-5pt) -- ++ (-5pt,0);
	\end{tikzpicture}
\end{equation*}
(\cf \cite[Definition~4.5]{jacobs:apal1994}.)
For instance, $\Id$, $\mathcal{P}^+$, and $\mathcal{D}$ are the affine part of the writer monad $M \times \Id$, the powerset monad $\mathcal{P}$, and the generalised multiset monad $\mathcal{M}_{[0,\infty]}$, respectively.
As noted in \cite[Proposition~3.2]{cirstea:entcs2010} any law distributing a monad over an endofunctor restricts to a law distributing its affine part over the same endofunctor.

\paragraph{On alternatives to the pointwise extension}
The pointwise extension is not the only way to extend an endofunctor from $\CC$ to $\Sh[\CC]{\alpha}$. We discuss two possible alternatives detailing why they are unsatisfactory for the aims of this work.
The first approach relies on the constant sheaf adjunction $(\Delta \dashv \Gamma)\colon \Sh[\CC]{\alpha}\to \CC$ and defines the extension of an endofunctor $F$ as the composite $\Delta F \Gamma\colon \CC \to \CC$. This definition extends to a functor but not a monoidal functor since in this situation a natural isomorphism $\Id_{\Sh[\CC]{\alpha}} \cong \Delta \Gamma$ exhibits an equivalence of categories for $\Sh[\CC]{\alpha}$ and $\CC$ (for a counterexample consider sheaves of sets). However, this is not an issue: the fact that pointwise extension defines a monoidal functor allows us to prove \cref{thm:extp-monad-functor,thm:extp-dist-law-functor} from general properties of monoidal functors but these can also proven directly. In fact, the extension functor defined from $(\Delta \dashv \Gamma)$ preserves monads and lifts to the category of distributive laws. Moreover, its essential image is enriched similarly to the pointwise extension functor $(\extP{-})$. Then, why is the pointwise extension preferable to this alternative notion? It turns out that this kind of extensions pose very stringent constraints on Kleisli liftings of locally contractive endofunctors to the point of impacting their suitability with respect to the aims of this work. To understand the severity of this limitation consider a natural transformation $\xi\colon \later \Delta T \Gamma \to \Delta T \Gamma \later$ (the type of $\xi$ is exactly that of transformations associated to Kleisli liftings for $\later$). Naturality requires each component at any stage above $1$ to factor through the morphism to the final object of $\CC$ as illustrated by the naturality square below.
\begin{equation*}
	\begin{tikzpicture}[
			auto, xscale=2.2, yscale=1.3, diagram font,
			baseline=(current bounding box.center)
		]	
		
		\node (n0) at (0,1) {\(TX_\alpha\)};	
		\node (n1) at (1,1) {\(TX_\alpha\)};	
		\node (n2) at (0,0) {\(1\)};	
		\node (n3) at (1,0) {\(TX_\alpha\)};	

		\draw[->] (n0) to node {\(\id_{T1}\)} (n1);
		\draw[->] (n0) to node[swap] {\(!_{TX_\alpha}\)} (n2);
		\draw[->] (n1) to node {\(T\id_{X_\alpha}\)} (n3);
		\draw[->] (n2) to node[swap] {\(\xi_{X,\beta}\)} (n3);
	\end{tikzpicture}
\end{equation*}
Compatibility with the monad structure imposes similar constraints also on the unit and multiplication of the monad.
The second approach we discuss is usually known as \emph{right extension} and (assuming enough limits exists) associates $F$ to (the functorial part of) the right Kan extension along $\Delta$ of $\Delta F$ \ie ${Ran}_{\Delta}(\Delta F)$.
Assume, for the sake of the argument, that $\Delta$ has also a left adjoint (\eg when $\CC$ is $\Set$), then right Kan extensions are preserved by $\Delta$ and ${Ran}_{\Delta}(\Delta F) \cong \Delta \circ {Ran}_{\Delta}F$. In particular, the right extension of $\Id_\CC$ is $\Delta\circ {Ran}_{\Delta}\Id_\CC \cong \Delta \Gamma$ and at this point the argument detailed above applies.

\section{Infinite trace semantics via guarded Kleisli (co)\-re\-cur\-sion}
\label{sec:infinite-trace-semantics}

In this section we introduce a construction for capturing infinite trace semantics of systems modelled as $TF$-coalgebras via finality in a suitable category of coalgebras. 

\looseness=-1
The key observation supporting our construction is that infinite traces can be characterised by amalgamation of certain families of coherent approximations akin to how a stream is described by the infinite family of its prefixes. In general, these approximations can be thought of as observations obtained from monitoring executions for a given number of steps (the prefix length) and such observations are associated to intermediate steps of final sequences \cite{barr:jpaa1992,barr:tcs1993,adamek:jlc2002}. 

In \cref{sec:final-sequence-sheaf} we present the final sequence for an endofunctor $F$ as the unique invariant object of its guarded pointwise extension $\extP{F}\later$. In \cref{sec:guarded-coalgebras} we study coalgebras of type $\extP{F}\later$ and show that associated notion of bisimulation generalises known behavioural pseudo-ultrametrics induced by final sequences \cite{barr:tcs1993,adamek:jlc2002}.
Finally, in \cref{sec:infinite-trace-semantics} we consider the categories of coalgebras for Kleisli liftings of $\extP{F}\later$, we characterise their final objects, and provide embeddings from the category of $TF$-coalgebras.

\subsection{Final sequences as invariant objects}
\label{sec:final-sequence-sheaf}

Final sequences were introduced by \cite{barr:jpaa1992} in order to compute final coalgebras and, together with their dual structures (\ie initial sequences and initial algebras), can be thought as generalisations of Kleene's chains. These constructions have been successfully used to provide sufficient conditions for a functor to admit final (resp.~initial) invariant objects (see for example \citeauthor{barr:tcs1993} \cite{barr:tcs1993}, \citeauthor{ps:siam1982} \cite{ps:siam1982}, \citeauthor{adamek:tcs2003} \cite{ak:tcs1995,adamek:entcs2003,adamek:tcs2003,am:ic2006}, \citeauthor{worrell:entcs1999} \cite{worrell:entcs1999,worrell:tcs2005,worrell:phdthesis}, \citeauthor{bacci:phdthesis} \cite{bm:jcss2015,bacci:phdthesis}).
In this section we characterise final sequences for endofunctors as unique invariants for suitable endofunctors over categories of sheaves. These objects are proxy to all the information usually found in final coalgebras together with the sequences of observations approximating them. For instance, if final coalgebras for the given endofunctor describe infinite streams, then we obtain a sheaf that represents them by means of their prefixes.

Recall from \cite{barr:jpaa1992} that the final sequence for an endofunctor $F$ over $\CC$ is the ordinal-indexed sequence of objects $(F^\beta)_{\beta \in \Ord}$ and arrows $(f^{\beta'}_{\beta})_{\beta\leq\beta' \in \Ord}$ such that:
\begin{equation*}
	F^{\beta+1} = F (F^\beta)
	\qquad
	F^{\gamma} = \lim_{\beta < \gamma } F^\beta 
	\qquad 
	f^{\beta'+1}_{\beta+1} = F f^{\beta'}_\beta
	\qquad
	f^{\gamma}_{\beta} = \pi_\beta
\end{equation*}
where $\gamma$ is a limit ordinal (note that $0$ is considered a limit ordinal as well) and the projection $\pi_\beta\colon F^\gamma \to F^\beta$ is the $\beta$-component of the limiting cone. The final sequence for $F$ corresponds to a $\CC$-valued sheaf over the category of ordinals $\Ord$.
\begin{definition}
	\label{def:final-sequence-functor}
	For $F$ an endofunctor over $\CC$, the \emph{final sequence of $F$} is any limit-preserving functor $\fseq(F)\colon \Ord \to \CC$ such that, for all ordinals $\beta \leq \beta'$:
	\begin{itemize}
		\item $\fseq(F)(\beta + 1) = F(\fseq(F)(\beta)$;
		\item $\fseq(F)(\iota_{\beta+1,\beta+1}) = F(\fseq(F)(\iota_{\beta,\beta'}))$.
	\end{itemize}
\end{definition}
\noindent In particular, the functor $\fseq(F)$ is given on any ordinal $\beta$ and on any inclusion $\res{\beta}{\beta'}\colon \beta \to \beta'$ as follows:
\begin{equation*}
	\fseq(F)_\beta = F^\beta \qquad \fseq(F)_{\res{\beta}{\beta'}} = f^{\beta'}_\beta
	\text{.}
\end{equation*}
In the following we will be interested in the first $\alpha$ steps of the final sequence (\eg when the sequence is stable after $\alpha$ steps) and hence will restrict $\fseq(F)$ to $\alexT{\alpha}\op$. Formally, this restriction yields an object in $\Sh[\CC]{\alpha}$ and corresponds to the action on $\fseq(F)$ of the inverse image $i^\ast$, where $i$ is given by the inclusion of $\alexT{\alpha}$ into $\Ord$. 
The final sequence is said to be \emph{stable} at some ordinal $\alpha$ provided that $f_{\alpha}^{\alpha+1}$ is an isomorphism. %
For notational convenience, we will write $\fseq(F)$ instead of $i^\ast(\fseq(F))$ when $i$ is clear from the context.

\begin{example}
\label{ex:fin-sheaf-streams}
Consider the endofunctor $A \times \Id$ where $A$ is a (non-empty) set of labels and let $\alpha$ be $\omega$---the final sequence for $A \times \Id$ is stable after $\omega$. The sheaf $\fseq(A \times \Id)$ on $\alexT{\omega}$ is given as follows:
\begin{align*}
	\fseq(A \times \Id)_0 & = 1 &
	\fseq(A \times \Id)_{\res{0}{n}} & = {!}_{A^n}\\
	\fseq(A \times \Id)_{n} & = A^n &
	\fseq(A \times \Id)_{\res{n}{m}} & = A^{n} !_{A^{m-n}}\\
	\fseq(A \times \Id)_{\omega} & = \lim_{n<\omega}A^n \cong A^{\omega}	&
	\fseq(A \times \Id)_{\res{n}{\omega}} & = A^{n} !_{A^{\omega}} = \pi_n
\end{align*}
Finite words over the alphabet $A$ are the observations characterising streams \ie the abstract behaviours for $A \times \Id$-coalgebras.
\end{example}

By considering final sequences as sheaves we are able to ``internalise'' their information about how final coalgebras are identified via sequences of approximations. Besides the above direct construction, these sheaves are characterised as unique invariants (\ie final coalgebras) of guarded pointwise extensions.

\begin{lemma}
	\label{thm:extp-later-invariant}
	There is a unique $\extP{F}\later$-invariant given (up to isomorphism) by the identity on $\fseq(F)$.
\end{lemma}

\begin{proofatend}
	Let $\alpha$ be an ordinal.	It follows from \cref{thm:extp-enriches-functors,thm:locally-contractive-ops} that $\extP{F}$ is enriched over $\Sh{\alpha}$ and that $\extP{F}\later$ is locally contractive. 
	The category $\Sh[\CC]{\alpha}$ has limits of $\alpha$-sequences since, by hypothesis, we have $(\Delta \dashv \Gamma)$ and limits of $\alpha$-sequences in $\CC$.	
	It follows from \cref{thm:sh-contractively-complete} that the endofunctor $\extP{F}\later$ has a unique (up to isomorphism) invariant. Therefore, to prove the claim it suffices to show that $\id_{\fseq(F)}$ is an $\extP{F}\later$-(co)algebra.
	On successors ordinals we have that:
	\begin{equation*}
		\fseq(F)_{\beta+1} = F^{\beta+1} = FF^{\beta} = (\extP{F}\later \fseq(F))_{\beta+1}
	\end{equation*}
	and on limit ones that:
	\begin{equation*}
		\fseq(F)_\gamma = \lim_{\beta < \gamma} \fseq(F)_\beta \stackrel{\ddagger}{=} \lim_{\beta+1 < \gamma} \fseq(F)_{\beta+1} = \lim_{\beta < \gamma} (\extP{F}\later\fseq(F))_\beta = (\extP{F}\later \fseq(F))_{\gamma}
	\end{equation*}
	where $(\ddagger)$ follows by restriction to a family of successor ordinals covering $\gamma$.
\end{proofatend}

\subsection{Guarded coalgebras}
\label{sec:guarded-coalgebras}

For an endofunctor $F$, we refer to coalgebras of type $\extP{F}\later$ as \emph{guarded}. Intuitively, the modality $\later$ guarding $\extP{F}$ forces transitions at any successor stage $\beta+1$ to have targets at their predecessor stage $\beta$:
\begin{equation*}
	X_{\beta+1} \xrightarrow{h_{\beta+1}} FX_\beta 
\end{equation*}
whereas transitions at stages that are limit ordinals have targets at the same stage and are obtained as mediating maps.

For instance, take $F$ as the endofunctor $A \times \Id$ and $\alpha$ as $\omega$ since, as discussed in \cref{ex:fin-sheaf-streams}, the final sequence for $A \times \Id$ is stable at $\omega$. Then, $\extP{F}\later$ is the endofunctor $\Delta A \times \later$ over $\Sh{\omega}$. Let $h\colon X \to \Delta A \times \later X$ be a guarded coalgebra. The component at stage $0$ of $h$ is determined by the structure of sheaves and is the identity on the singleton $1$; hence the only element inhabiting this stage can be seen as a sink state $\bottom$. This interpretation for $h_0$ and $\bottom$ is fostered by looking at the other components of $h$. At stage $1$, $\later X$ takes value $X_0 = 1$ and hence all transitions described by $h_1 \colon X_1 \to A \times X_0$ necessarily end in the sink $\bottom$ which essentially means they terminate producing a label ($A \times X_0 \cong A$). In general, transitions described by $h_{n+2}\colon X_{n+2} \to A \times X_{n+1}$ start at stage $n+2$ and end at stage $n+1$, those described by $h_{n+1}\colon X_{n+1} \to A \times X_n$ go from $n+1$ to $n$, and so on until $\bottom$ is reached after $n+2$ steps. At stage $\omega$, $X$ and $h$ are defined by amalgamation from the underlying stages:
\begin{equation*}
	h_{\omega}(x) = (a,x') \iff \forall n < \omega\, (h_{n+1}\circ X_{\res{n+1}{\omega}})(x) = (a,X_{\res{n}{\omega}}(x'))
	\text{.}
\end{equation*}
Computations described by this component of $h$ never leave stage $\omega$ and each of their countably many steps projects, coherently with restriction maps, to a step at stage $n$ for any $n < \omega$. It follows that $h$ outputs streams and words forming their prefixes.

This example fosters the intuition of stages as describing the ``number of available steps'' or the ``observations length''. Form this perspective, the component at stage $\beta$ of the final semantics map describes behaviours distinguishable by means of observations at stage $\beta$ that is, baring with the above intuition, ``by considering executions up to $\beta$-many\footnote{ %
	In general $\beta$ is the index of a step in the final sequence and not an actual length, however the two coincide for sequences stable at $\omega$ like those arising from the examples considered in this section.
} transition steps''. This perspective generalises ideas from \cite{barr:tcs1993,adamek:jlc2002} where \citeauthor{barr:tcs1993} and \citeauthor{adamek:jlc2002} observed how final sequences for $\omega$-continuous endofunctors over $\Set$ determine a pseudo-ultrametric on their final coalgebra carrier (and hence on each coalgebra carrier). In particular, for $(X,h)$ an $(A \times \Id)$-coalgebra, the distance of two states $x$ and $x'$ in $X$ is defined as $2^{-n}$ where $n$ is the length of the longest prefix shared by the streams generated from $x$ and $x'$:
\begin{equation}
	\label{eq:stream-distance}
	d(x,x') = \inf \left\{
		2^{-n} \,\middle\vert\,
		\begin{array}{l}
		\exists (x_i  \in X)_{i < n},\ (x'_i  \in X)_{i < n},\ (a_i \in A)_{i < n}  \text{ such that }
		x_0 = x,\ x'_0 = x',\\ \text{and } \forall i < n-1
		h(x_i) = (a_i,x_{i+1}) \text{ and } h(x'_i) = (a_i,x'_{i+1})
		\end{array}
	\right\}
\end{equation}
Thus, $d(x,x') = 0$ if and only if $x$ and $x'$ are behaviourally equivalent.

Coalgebras of type $\extP{F}\later$ and their bisimulations rephrase the above situation in the language of sheaves: these structures localise the information contained in the pseudo-ultrametric \eqref{eq:stream-distance} by restriction to the values associated to each $n$ or, equivalently, to each step of the final sequence for $F = A \times \Id$. In this setting, a bisimulation is a span $X \from R \to X'$ of sheaves making the usual diagram commute and can be understood (without loss of generality) as a decreasing $\omega$-indexed sequence of relations $R_0 \supseteq R_1 \supseteq \dots$ such that:
\begin{equation*}
	x \mathrel{R_n} x' \implies d(x,x') \leq 2^{-n}\text{.}
\end{equation*}
From this perspective, a bisimulation at stage $\beta$ captures observational equivalence where observations are restricted to those described by the $\beta$-step of the final sequence. Therefore, guarded coalgebras and their bisimulations are a (conservative) generalisation of \citeauthor{barr:tcs1993}'s ideas to arbitrary endofunctors (albeit a metric cannot be defined in general, \eg when $\alexT{\alpha}$ is not metrizable).

To conclude this section, we show that all coalgebras are guarded in the sense that $F$-coalgebras form a subcategory of $\Coalg{\extP{F}\later}$.
Recall from \cref{sec:locally-contractive-functors} that $\later$ is a well-pointed endofunctor and its point is $\nxt\colon \Id \to \later$.
This natural transformation induces the functor between coalgebra categories:
\begin{equation*}
	\Coalg{\id_{\extP{F}}\circ\nxt}\colon \Coalg{\extP{F}} \to \Coalg{\extP{F}\later}
\end{equation*}
given, on each coalgebra $(X,h)$ and homomorphism $f$ by the assignments:
\begin{equation*}
	(X,h) \mapsto (X, \extP{F}(\nxt_{X}) \circ h)
	\qquad
	f \mapsto f
	\text{.}
\end{equation*}
Intuitively, this functor uses restriction morphisms to ``guard transition targets'' as clear from unfolding the definition of $(\extP{F}(\nxt_{X}) \circ h)_{\beta+1}$:
\begin{equation*}
 X_{\beta+1} \xrightarrow{h_{\beta+1}} FX_{\beta+1} \xrightarrow{FX_{\res{\beta}{\beta+1}}} FX_{\beta}
 \text{.}
\end{equation*}
We remark that $\Coalg{\id_F \circ \nxt}$ is a lifting of the identity on $\Sh[\CC]{\alpha}$ along the forgetful functors for $\Coalg{\extP{F}}$ and $\Coalg{\extP{F}\later}$ since it acts as the identity coalgebra homomorphisms.
We write $(-)^\later$ for $\Coalg{\id_{\extP{F}}\circ\nxt}$.

Recall from \cref{sec:extp-functor} that the constant sheaf functor $\Delta$ lifts to categories of coalgebras and hence, by composition with $(-)^\later$, we have a functor $\Delta^\later$ turning every $F$-coalgebra into a guarded coalgebra while acting as $\Delta$ on their carrier. 
\begin{equation}	
	\label{eq:lift-delta-guarded-coalg}
	\begin{tikzpicture}[
			auto, xscale=2.1, yscale=1.2, diagram font,
			baseline=(current bounding box.center)
		]	
		\node[] (n0) at (0,0) {\(\CC\)};	
		\node[] (n1) at (1,0) {\(\Sh[\CC]{\alpha}\)};	
		\node[] (n2) at (0,1) {\(\Coalg{F}\)};	
		\node[] (n3) at (1,1) {\(\Coalg{\extP{F}\later}\)};	
		
		\draw[->] (n0) to node[swap] {\(\Delta\)} (n1);
	 	\draw[<-] (n0) -- (n2);
	 	\draw[<-] (n1) -- (n3);
	 	\draw[->] (n2) to node {\(\Delta^\later\)} (n3);
		
	\end{tikzpicture}
\end{equation}
Moreover, this functor is an inclusion whenever $\alpha$ is greater than $1$ since $\Delta X_{\iota_{\beta,\beta'}}$ is $\id_X$ for any $\beta'\geq \beta > 0$.
It follows from \cref{thm:extp-final-invariants,thm:extp-later-invariant} that $\Delta^\later$ takes final coalgebras to final (guarded) coalgebras. The functor $\Delta^\later$ exhibits $\Coalg{F}$ as a subcategory of $\Coalg{\extP{F}\later}$ and, under mild assumptions akin to \cref{thm:extp-delta-gamma-coalg}, as a coreflective subcategory.

\begin{proposition}
	\label{thm:extp-delta-gamma-guarded-coalg}
	For $\alpha$ an infinite limit ordinal, if $\varrho\colon F\Gamma \to \Gamma\extP{F}$ from \eqref{eq:extp-lift-along-gamma} is a retraction then the constant sheaf adjunction $(\Delta \dashv \Gamma)\colon \Sh[\CC]{\alpha} \to \CC$ lifts along the forgetful functors for $\Coalg{F}$ and $\Coalg{\extP{F}\later}$. Moreover, the lifted adjunction is a coreflection.
	\begin{equation*}	
		\begin{tikzpicture}[
				auto, xscale=2.1, yscale=1.2, diagram font,
				baseline=(current bounding box.center)
			]	
			\node[] (n0) at (0,0) {\(\CC\)};	
			\node[] (n1) at (1,0) {\(\Sh[\CC]{\alpha}\)};	
			\node[] (n2) at (0,1) {\(\Coalg{F}\)};	
			\node[] (n3) at (1,1) {\(\Coalg{\extP{F}\later}\)};	
			
			\draw[->,bend left] (n0) to node[pos=.6] {\(\Delta\)} (n1);
		 	\draw[->,bend left] (n1) to node[pos=.4] {\(\Gamma\)} (n0);
		 	\node[rotate=90] at ($(n0)!.5!(n1)$) {\(\vdash\)};
		 	\draw[<-] (n0) -- (n2);
		 	\draw[<-] (n1) -- (n3);
		 	\draw[->,bend left] (n2) to node {\(\Delta^\later\)}(n3);
		 	\draw[->,bend left] (n3) to (n2);
		 	\node[rotate=90] at ($(n2)!.5!(n3)$) {\(\vdash\)};
			
		\end{tikzpicture}
	\end{equation*}
\end{proposition}

\begin{proofatend}
	Assume $\varrho\colon F\Gamma \to \Gamma \extP{F}$ is a retraction and let $\varsigma\colon \Gamma \extP{F} \to F \Gamma$ be any of its sections. 
	Akin to \cref{thm:extp-delta-gamma-coalg}, the statement can be shown to follow from \cite[Theorem~2.5]{kkw:cmcs2014} by providing $\vartheta^\later\colon \Delta F \to \extP{F} \later \Delta$ and $\varsigma^\later\colon \Gamma \extP{F} \later \to F \Gamma$ such that the necessary diagrams commute. 
	
	Note that the natural transformation $\Gamma\extP{F}\nxt\colon \Gamma\extP{F} \to \Gamma\extP{F}\later$ has an inverse for it is given, on each sheaf $X$, by the equality
	\begin{equation*}
		\Gamma \extP{F} \later X =
		\extP{F} \later X_\alpha = 
		\lim_{\beta+1 < \alpha} \extP{F}\later X_{\beta+1} =
		\lim_{\beta+1 < \alpha} FX_\beta = 
		\extP{F} X_\alpha = 
		\Gamma \extP{F} X
	\end{equation*}
	where the limits are restricted to the base of successor ordinals.	%
	Define the natural transformations $\vartheta^\later\colon \Delta F \to \extP{F} \later \Delta$ and $\varsigma^\later\colon \Gamma \extP{F} \later \to F \Gamma$ as
	\begin{equation*}
		\Delta  F 
		\xrightarrow{\vartheta} 
		\extP{F} \Delta 
		\xrightarrow{\extP{F}\nxt_{\Delta}} 
		\extP{F} \later \Delta
		\qquad\text{and}\qquad
		\Gamma \extP{F} \later
		\xrightarrow{(\Gamma\extP{F}\nxt)^{-1}} 
		\Gamma \extP{F}
		\xrightarrow{\varsigma} 
		F \Gamma
	\end{equation*}
	where $\vartheta\colon \Delta F \to \extP{F} \Delta$ and $\varsigma\colon \Gamma \extP{F} \to F \Gamma$ are the equality \eqref{eq:extp-extension-along-delta} and a section for $\varrho$ as per  \eqref{eq:extp-lift-along-gamma}. 
	In order to prove the necessary diagrams commute decompose them as follows:
	\resetDiagMarker
	\begin{equation*}	
		\begin{tikzpicture}[
				auto, xscale=3, yscale=2, diagram font,
				baseline=(current bounding box.center)
			]	

			\node[] (n0) at (0,2) {\(F\)};	
			\node[] (n1) at (2,2) {\(F\circ \Gamma \circ \Delta\)};	
			\node[] (n2) at (0,0) {\(\Gamma\circ \Delta \circ F\)};	
			\node[] (n3) at (2,0) {\(\Gamma\circ \extP{F} \circ \later \circ \Delta\)};
			\node[] (n4) at (1,1) {\(\Gamma\circ \extP{F} \circ \Delta\)};
			
			\draw[->] (n0) to node {\(\id_F \circ \eta\)} (n1);
			\draw[->] (n0) to node[swap] {\(\eta \circ \id_F\)} (n2);
			\draw[->] (n1) to node[swap] {\(\varsigma \circ \id_\Delta\)} (n4);
			\draw[->] (n1) to node {\(\varsigma^\later \circ \id_\Delta\)} (n3);
			\draw[->] (n2) to node {\(\id_\Gamma \circ \vartheta\)} (n4);
			\draw[->] (n2) to node[swap] {\(\id_\Gamma \circ \vartheta^\later\)} (n3);
			\draw[->] (n4) to node[swap] {\(\id_{\Gamma \circ \extP{F}} \circ \nxt \circ \id_\Delta\)} (n3);
						
			\nodeDiagMarker{$(n0)!.5!(n4)$}%
				{diag:extp-delta-gamma-guarded-coalg-10};
			\nodeDiagMarker{$(n2)!.5!(n3)!.6!(n4)$}%
				{diag:extp-delta-gamma-guarded-coalg-11};
			\nodeDiagMarker{$(n1)!.5!(n3)!.3!(n4)$}%
				{diag:extp-delta-gamma-guarded-coalg-12};
		\end{tikzpicture}
	\end{equation*}
	\begin{equation*}	
		\begin{tikzpicture}[
				auto, xscale=2.6, yscale=1.6, diagram font,
				baseline=(current bounding box.center)
			]	
			
			\node[] (m0) at (0,0) {\(\extP{F}\circ\later\circ \Delta \circ \Gamma\)};	
			\node[] (m1) at (3,0) {\(\Delta\circ F\circ \Gamma\)};	
			\node[] (m2) at (0,3) {\(\extP{F}\circ \later\)};	
			\node[] (m3) at (3,3) {\(\Delta\circ \Gamma\circ \extP{F}\circ \later\)};
			
			\node[] (n0) at (1,1) {\(\extP{F}\circ \Delta \circ \Gamma\)};	
			\node[] (n1) at (2,1) {\(\Delta\circ F\circ \Gamma\)};	
			\node[] (n2) at (1,2) {\(\extP{F}\)};	
			\node[] (n3) at (2,2) {\(\Delta\circ \Gamma\circ \extP{F}\)};	
			
			\draw[<-] (n0) to node[swap] {\(\vartheta \circ \id_\Gamma\)} (n1);
			\draw[<-] (n1) to node[swap] {\(\id_\Delta \circ \varsigma\)} (n3);
			\draw[<-] (n2) to node[swap] {\(\id_{\extP{F}} \circ \varepsilon\)} (n0);
			\draw[<-] (n2) to node {\(\varepsilon \circ \id_{\extP{F}}\)} (n3);

			\draw[<-] (m0) to node[swap] {\(\vartheta^\later \circ \id_\Gamma\)} (m1);
			\draw[<-] (m1) to node[swap] {\(\id_\Delta \circ \varsigma^\later\)} (m3);
			\draw[<-] (m2) to node[swap] {\(\id_{\extP{F}\circ\later} \circ \varepsilon\)} (m0);
			\draw[<-] (m2) to node {\(\varepsilon \circ \id_{\extP{F}\circ\later}\)} (m3);

			\draw[<-] (m0) to node[swap] {\(\id_{\extP{F}} \circ \nxt \circ \id_{\Delta \circ \Gamma}\)} (n0);
			\draw[nat] (m1) to (n1);
			\draw[<-] (m2) to node {\(\id_{\extP{F}} \circ \nxt\)} (n2);
			\draw[<-] (m3) to node[swap] {\(\id_{\Delta \circ \Gamma \circ \extP{F}} \circ \nxt\)} (n3);
			
			\nodeDiagMarker{$(n0)!.5!(n3)$}%
				{diag:extp-delta-gamma-guarded-coalg-20};
			\nodeDiagMarker{$(m0)!.6!(m1)!.5!(n1)$}%
				{diag:extp-delta-gamma-guarded-coalg-21};
			\nodeDiagMarker{$(m3)!.6!(m1)!.5!(n1)$}%
				{diag:extp-delta-gamma-guarded-coalg-22};
		\end{tikzpicture}
	\end{equation*}
	Both diagrams commute:	
	\eqref{diag:extp-delta-gamma-guarded-coalg-10} and 	\eqref{diag:extp-delta-gamma-guarded-coalg-20} are shown to commute in the proof of \cref{thm:extp-delta-gamma-coalg}; 	\eqref{diag:extp-delta-gamma-guarded-coalg-11} and \eqref{diag:extp-delta-gamma-guarded-coalg-21} define $\vartheta^\later$; 
	\eqref{diag:extp-delta-gamma-guarded-coalg-12} and	\eqref{diag:extp-delta-gamma-guarded-coalg-22} define $\varsigma^\later$; the remaining squares follow by naturality of $\varepsilon$ e $\nxt$.	
	It follows from \cite[Theorem~2.5]{kkw:cmcs2014} that the desired lifting exists. In particular, the lifting of $\Gamma$, is given on each $\extP{F}$-coalgebra $(X,h)$ and homomorphism $f$, by the assignments 
	\begin{equation*}
		(X,h) \mapsto (\Gamma X,\varsigma^\later_X \circ \Gamma h)
		\qquad 
		f \mapsto \Gamma f
	\end{equation*}	
	and the lifting of $\Delta$ given on each $F$-coalgebra $(Y,k)$ and homomorphism $g$, by the assignments 
	\begin{equation*}
		(Y,k) \mapsto (\Delta Y, \vartheta^\later_Y \circ \Delta k)
		\qquad 
		g \mapsto \Delta g
		\text{.}
	\end{equation*}
	The latter is exactly the inclusion functor $\Delta^\later = \Coalg{\id_{\extP{F}}\circ \nxt}$ since $\vartheta^\later$ is defined as $(\id_{\extP{F}}\circ \nxt) \bullet \vartheta$ and $\vartheta$ as the equality $\Delta F = \extP{F}\Delta$.
\end{proofatend}

\subsection{Infinite trace semantics}
\label{sec:infinite-trace-semantics-thm}

\looseness=-1
In this section we combine guarded coalgebras with extensions to Kleisli categories in order to capture infinite trace semantics via final semantics. Intuitively, the former provides us with the tools for collecting observations into coherent families whereas the latter offers us the setting where to abstract the effects modelled by the branching type \ie ensure observations come from the linear semantics of systems under scrutiny.
In practice, for systems modelled as $TF$-coalgebras, we consider coalgebras for Kleisli liftings of $\extP{F}\later$ where the pointwise extension targets sheaves on an ordinal $\alpha$ large enough for the final sequence of $F$ to stabilise.

Before we discuss $\liftKl{\extP{F}\later}$-coalgebras in general let us illustrate the construction in the case of non-deterministic labelled transition systems. 
To this end, take $T$, $F$, $\lambda$, and $\alpha$ as follows:
\begin{itemize}
\item the affine and commutative monad $\mathcal{P}^+$ (the double strength of $\mathcal{P}$ readily restricts to its affine part);
\item the polynomial functor $A \times \Id$ (for $A$ non-empty);
\item the distributive law $\lambda\colon A \times \mathcal{P}^+ \to \mathcal{P}^+(A \times \Id)$ associated to the canonical Kleisli lifting of $A \times \Id$ to $\Kl{\mathcal{P^+}}$; 
\item the first infinite ordinal $\omega$ (the final sequence for $A \times \Id$ stabilises at $\omega$).
\end{itemize}
Since $\mathcal{P}^+$ is affine, we can apply \cref{thm:omega-suitable-dist-law,thm:affine-dist-law} to $\lambda$ obtaining the distributive law of $\extP{\mathcal{P}^+}$ over $\extP{F}\later$:
\begin{equation}
	\label{eq:dist-law-guarded-lts}
	\extP{A \times \Id} \circ \later \circ \mathcal{P}^+
	\xrightarrow{\id_{\extP{A \times \Id}} \,\circ\, \theta}
	\extP{A \times \Id} \circ \mathcal{P}^+ \circ \later
	\xrightarrow{\extP{\lambda} \,\circ\, \id_{\later}}
	\mathcal{P}^+ \circ \extP{A \times \Id} \circ \later
	\text{.}
\end{equation}
The fact that this distributive law factors through $\extP{A \times \Id} \circ \mathcal{P}^+ \circ \later$ corresponds to its associated Kleisli lifting of $\extP{F}\later$ being the composition of Kleisli liftings of $\extP{F}$ and $\later$ given by the extension of $\lambda$ and the distributive law $\xi$ constructed in the proof of \cref{thm:affine-dist-law}, respectively. We remark that this strategy is the equivalent for Kleisli liftings of the constructions  for distributive laws presented in \cref{sec:guarded-kleisli-corecursion}.
The distributive law \eqref{eq:dist-law-guarded-lts} acts essentially as $\lambda$ since its component for a sheaf $X$ is the arrow given at stage $1$ as the function
\begin{equation*}	
	\begin{tikzpicture}[
			auto, xscale=2.3, yscale=1.8, diagram font,
			baseline=(current bounding box.center)
		]	
		\node (n0) at (0,0) {\(\extP{\mathcal{P}^+}(\Delta A \times \later X)_{1} \)};
		\node (n1) at (0,1) {\((\Delta A \times \later \extP{\mathcal{P}^+} X)_{1}\)};
		\node (n2) at (0,2) {\((\Delta A \times \later \extP{\mathcal{P}^+} X)_{1}\)};
		\node (m0) at (1,0) {\(\mathcal{P}^+A\)};
		\node (m1) at (1,1) {\(A\)};
		\node (m2) at (1,2) {\(A\)};

		\draw[nat] (n0) to (m0);
		\draw[nat] (n1) to (m1);
		\draw[nat] (n2) to (m2);
		
		\draw[->] (n1) to node[swap] {\((\extP{\lambda} \circ \id_{\later})_{X,1}\)} (n0);
		\draw[->] (n2) to node[swap] {\((\id_{\extP{A \times \Id}} \circ \theta)_{X,1}\)} (n1);
		\draw[->] (m1) to node {\(\eta_A\)} (m0);
		\draw[->] (m2) to node {\(\id_A\)} (m1);

	\end{tikzpicture}
\end{equation*}
and at stage $n+2$ as the function
\begin{equation*}	
	\begin{tikzpicture}[
			auto, xscale=3.4, yscale=1.8, diagram font,
			baseline=(current bounding box.center)
		]	
		\node (n0) at (0,0) {\(\extP{\mathcal{P}^+}(\Delta A \times \later X)_{n+2}\)};
		\node (n1) at (0,1) {\((\Delta A \times \later \extP{\mathcal{P}^+} X)_{n+2}\)};
		\node (n2) at (0,2) {\((\Delta A \times \later \extP{\mathcal{P}^+} X)_{n+2}\)};
		\node (m0) at (1,0) {\(\mathcal{P}^+(A \times X_{n+1})\)};
		\node (m1) at (1,1) {\(A \times \mathcal{P}^+ X_{n+1}\)};
		\node (m2) at (1,2) {\(A \times \mathcal{P}^+ X_{n+1}\)};

		\draw[nat] (n0) to (m0);
		\draw[nat] (n1) to (m1);
		\draw[nat] (n2) to (m2);
		
		\draw[->] (n1) to node[swap] {\((\extP{\lambda} \circ \id_{\later})_{X,n+2}\)} (n0);
		\draw[->] (n2) to node[swap] {\((\id_{\extP{A \times \Id}} \circ \theta)_{X,n+2}\)} (n1);
		\draw[->] (m1) to node {\(\lambda_{X_{n+1}}\)} (m0);
		\draw[->] (m2) to node {\(\id_{A \times \mathcal{P}^+ X_{n+1}}\)} (m1);

	\end{tikzpicture}
\end{equation*}

\looseness=-1
Coalgebras for endofunctors like these are guarded coalgebras: this means that the executions they describe are intertwined with stages whose ordinal number represents the number of steps available to the computation. Akin to \cref{sec:guarded-coalgebras}, consider the components of a coalgebra $h\colon X \to \extP{\mathcal{P}^+}(\Delta A \times \later X)$: $h_0$ has type $1 \to 1$ but since $\mathcal{P}^+$ is affine this is an isomorphism which means the computation it describes cannot evolve in any meaningful way; $h_1$ has type $X_1 \to \mathcal{P}^+\Delta A$ and the computations it describes non-deterministically output a label before reaching the sink at stage $0$; $h_{n}$ for $n > 1$ has type $X_{n} \to \mathcal{P}^+(\Delta A \times X_{n-1})$ and the computations it describes non-deterministically output a label before reaching a state at stage $n-1$. The fundamental difference with respect to the situation discussed in \cref{sec:guarded-coalgebras} is that steps are now concatenated by means of Kleisli composition hence abstracting from non-deterministic branching. In order to illustrate this difference and elucidate the key r\^ole played by the Kleisli category let us consider sequences of steps. Sequences of two steps in the system modelled by $h$ are described by the composite 
\begin{equation*}
	h' \defeq \liftKl{\Delta A \times \later}\left(h\right) \circ h
\end{equation*}
\ie the arrow $h'\colon X \to \extP{\mathcal{P}^+}\left(\Delta A \times \later \left(\Delta A \times \later X\right)\right)$ in $\Sh{\omega}$ defined as:
\begin{equation*}
	h' 
	\defeq 
	\extP{\mu}_{X} 
	\circ 
	\extP{\mathcal{P}^+}\left(\extP{\lambda}_{\later \left(\Delta A \times \later X\right)}\right) 
	\circ
	\extP{\mathcal{P}^+}\left(\Delta A \times \later\theta_{\Delta A \times X}\right) 
	\circ
	\extP{\mathcal{P}^+}\left(\Delta A \times \later h\right) \circ h
	\text{.}
\end{equation*}
At stage $1$, $h'$ equals to $h$:
\begin{equation*}
	a \in h'_{1}(x) \iff a \in h_1(x)
\end{equation*}
since the outermost occurrence of $\later$ in the behavioural functor takes value $1$ at this stage. 
We encounter the first difference at stage $2$:
\begin{equation*}
	(a,a') \in h'_{2}(x) \iff \exists x' \in X_1 \text{ s.t. } (a,x') \in h_{2}(x) \land a' \in h_{1}(x')
	\text{.}
\end{equation*}
At this stage the outermost occurrence of $\later$ takes the value of its argument at stage $1$ and hence the innermost occurrence takes value at stage $0$ meaning that sequences always end in the sink $\bottom$. Because of Kleisli composition, intermediate states ($x'$ above) are stripped from the outcome. Components at greater stages behave similarly except for the ending state not being the sink. In particular, at any stage $n+3$ we have that:
\begin{equation*}
	(a,a',x'') \in h'_{n+3}(x) \iff \exists x' \in X_{n+2} ((a,x') \in h_{n+3}(x) \land (a',x'') \in h_{n+2}(x'))
	\text{.}
\end{equation*}
The same considerations apply to sequences of arbitrary length: observations at stage $n$ are partial traces of length\footnote{
	In presence of explicit termination, as in the case for $F = A \times \id + 1$, length of executions at stage $n$ is \emph{at most} $n$.
} $n$ and partial traces observed at different stages abide restriction maps as illustrated by the schema:
\begin{equation*}	
	\begin{tikzpicture}[
			auto, xscale=1.4, yscale=1.2, diagram font,
			baseline=(current bounding box.center)
		]
		
		\node[anchor=east,black!80] at (-.2,1) {\(n+1 \vdash\)};	
		\node[anchor=east,black!80] at (-.2,0) {\(n \vdash\)};	
		
		\node (n0) at (0,1) {\(x_1\)};
		\node (n1) at (1,1) {\(x_2\)};
		\node (n2) at (2,1) {\(\cdots\)};
		\node (n3) at (3,1) {\(x_{n}\)};
		\node (n4) at (4,1) {\(x_{n+1}\)};
		\node (n5) at (5,1) {\(\bottom\)};
		\node (n6) at (6,1) {\(\cdots\)};

		\node (m0) at (0,0) {\(x_1|_n\)};
		\node (m1) at (1,0) {\(x_2|_n\)};
		\node (m2) at (2,0) {\(\cdots\)};
		\node (m3) at (3,0) {\(x_{n}|_n\)};
		\node (m4) at (4,0) {\(\bottom\)};
		\node (m5) at (5,0) {\(\bottom\)};
		\node (m6) at (6,0) {\(\cdots\)};
		
		\draw[->] (n0) to node {\(a_1\)} (n1);
		\draw[->] (n1) to node {\(a_2\)} (n2);
		\draw[->] (n2) to node {\(a_{n-1}\)} (n3);
		\draw[->] (n3) to node {\(a_{n}\)} (n4);
		\draw[->] (n4) to node {\(a_{n+1}\)} (n5);
		\draw[->] (n5) to node {\(\)} (n6);
		
		\draw[->] (m0) to node {\(a_1\)} (m1);
		\draw[->] (m1) to node {\(a_2\)} (m2);
		\draw[->] (m2) to node {\(a_{n-1}\)} (m3);
		\draw[->] (m3) to node {\(a_{n}\)} (m4);
		\draw[->] (m4) to node {\(\)} (m5);
		\draw[->] (m5) to node {\(\)} (m6);
		
		\draw[|->] (n0) to node {\(\)} (m0);
		\draw[|->] (n1) to node {\(\)} (m1);
		\draw[|->] (n3) to node {\(\)} (m3);
		\draw[|->] (n4) to node {\(\)} (m4);
		\draw[|->] (n5) to node {\(\)} (m5);
	\end{tikzpicture}
\end{equation*}
where vertical arrows are mappings induced by the restriction function $X_{\res{n}{n+1}}$ and horizontal arrows are transitions described by $h$ at stages $n$ and $n+1$.

Stage $\omega$ is defined by amalgamation and the associated observations are $\omega$-sequences. It follows from definition of sheaves and their morphisms that observations made at stage $\omega$ restrict, for each finite ordinal $n$, to observations at stage $n$. Symmetrically, a family with an observation for each stage $n < \omega$ that is coherent with respect to restriction maps induces an observation at stage $\omega$. We conclude that since observations at stage $n < \omega$ are (partial) trace of length $n$ observations at stage $\omega$ are necessarily infinite traces.

Consider the diagram asserting that $f\colon X \to Y$ extends to some morphism of $\liftKl{\Delta A \times \later}$-coalgebras $f\colon (X,h) \to (Y,k)$ and in particular its unfolding in $\Sh{\omega}$:
\begin{equation*}	
	\begin{tikzpicture}[
			auto, xscale=3.5, yscale=-1.5, rotate=90, diagram font,
			baseline=(current bounding box.center)
		]	
		\node (n0) at (0,2) {\(X\)};	
		\node (n1) at (0,0) {\(\extP{\mathcal{P}^+}(\Delta A \times \later X)\)};	
		\node (n2) at (4,2) {\(\extP{\mathcal{P}^+}Y\)};	
		\node (n3) at (4,0) {\(\extP{\mathcal{P}^+}(\Delta A \times \later Y)\)};	

		\node (m0) at (4,1.2) {\(\extP{\mathcal{P}^+}\extP{\mathcal{P}^+}(\Delta A \times \later Y)\)};
		\node (m1) at (1,0) {\(\extP{\mathcal{P}^+}(\Delta A \times \later \extP{\mathcal{P}^+}Y)\)};
		\node (m2) at (2,0) {\(\extP{\mathcal{P}^+}(\Delta A \times \extP{\mathcal{P}^+}\later Y)\)};
		\node (m3) at (3,0) {\(\extP{\mathcal{P}^+}\extP{\mathcal{P}^+}(\Delta A \times \later Y)\)};

		\draw[->] (n0) to node {\(h\)} (n1);
		\draw[->] (n2) to node[swap] {\(\extP{\mathcal{P}^+}k\)} (m0);
		\draw[->] (m0) to node[swap] {\(\extP{\mu}_{\Delta A \times \later Y}\)} (n3);
		
		\draw[->] (n0) to node[swap] {\(f\)} (n2);
		\draw[->] (n1) to node {\(\extP{\mathcal{P}^+}(\Delta A \times \later f))\)} (m1);
		\draw[->] (m1) to node {\(\extP{\mathcal{P}^+}(\Delta A \times \theta_Y)\)} (m2);
		\draw[->] (m2) to node {\(\extP{\mathcal{P}^+}\extP{\lambda}_{\later Y}\)} (m3);
		\draw[->] (m3) to node {\(\extP{\mu}_{\Delta A \times \later Y}\)} (n3);
				
	\end{tikzpicture}
\end{equation*}
In order to show that the above diagram commutes, it suffices to show that it commutes when restricted to successor ordinals in $\omega$ and hence only two cases need to be checked; the first corresponds to stage $1$ and the second to all the remaining $1 < n < \omega$. The component at stage $1$ commutes if, and only if, for any label $a \in A$ and any state $x \in X_1$ it holds that:
\begin{equation*}
	a \in h_1(x) \iff \exists y \in f_1(x) (a \in k_1(y))
	\text{.}
\end{equation*}
A component at stage $n > 1$ commutes if, and only if, for any $a \in A$, $x \in X_{n}$, and $x' \in X_{n-1}$, it holds that:
\begin{equation*}
	(a,x') \in h_{n}(x) \iff \forall y' \in f_{n-1}(x') \exists y \in f_{n}(x) ((a,y') \in k_{n}(y))
	\text{.}
\end{equation*}
Therefore, $\liftKl{\extP{F}\later}$-coalgebra homomorphisms, like $\extP{F}\later$-coalgebra homomorphisms, tie execution steps to stages and, like $\extP{F}$-coalgebra homomorphisms, they abstract from branching.

The example above suggests that final $\liftKl{\Delta A \times \later}$-coalgebras capture infinite trace semantics for labelled transition systems (without implicit termination). Indeed, this is the case and the same result holds for arbitrary systems modelled by $TF$-coalgebras---provided that the sequence for $F$ stabilises %
and that $\extP{F}\later$ %
has a Kleisli lifting. Under these assumptions, final $\liftKl{\extP{F}\later}$-coalgebras are images through the canonical inclusion $K\colon \CC \to \Kl{T}$ of final $\extP{F}\later$-coalgebras and the latter characterise, by construction, final $F$-coalgebras (\cf \cref{thm:extp-later-invariant,thm:extp-delta-gamma-guarded-coalg}). Therefore, 
final semantics for $\liftKl{\extP{F}\later}$-coalgebras captures (possibly) infinite trace semantics (\cf \cref{def:trace-as-linear}) for systems modelled by $TF$-coalgebras. Formally:

\begin{theorem}
	\label{thm:liftkl-extp-later-invariant}
	Let $(T,\mu,\eta)$ be a monad and $F$ an endofunctor, both over some category $\CC$. Let $\alpha$ be an ordinal such that the adjunction $(\Delta \dashv \Gamma)\colon \Sh[\CC]{\alpha} \to \CC$ is defined and the final sequence for $F$ is stable at $\alpha$. For $\liftKl{\extP{F}\later}$ Kleisli lifting of $\extP{F}\later$, there is a unique $\liftKl{\extP{F}\later}$-invariant and it is the identity on $\fseq(F)$.
\end{theorem}

\begin{proofatend}
	It follows from \cref{thm:extp-later-invariant} that the identity on $\fseq(F) \in \Sh[\CC]{\alpha}$ exhibits the unique $\extP{F}\later$-invariant. Because the canonical inclusion $K\colon \Sh[\CC]{\alpha} \to \Kl{\extP{T}}$ transports initial invariants to initial invariants of Kleisli liftings (see \cref{thm:kleisli-initial-invariants}), ${K\left(\id_{\fseq(F)}\right)} = \eta_{\fseq(F)}$ is the initial invariant of $\liftKl{\extP{F}\later}$.
	We conclude by noting that by \cref{thm:kleisli-lifting-contractive} $\liftKl{\extP{F}\later}$ is locally contractive and thus it follows from  \cref{thm:locally-contractive-unique} that $\liftKl{\extP{F}\later}$ has a unique invariant \ie $\id_{\fseq(F)} \in \Kl{\extP{T}}$. 
\end{proofatend}

\begin{corollary}
	Let $(T,\mu,\eta)$ be a monad and $F$ an endofunctor, both over some category $\CC$. Let $\alpha$ be an ordinal such that the adjunction $(\Delta \dashv \Gamma)\colon \Sh[\CC]{\alpha} \to \CC$ is defined and the final sequence for $F$ is stable at $\alpha$. The final $\liftKl{\extP{F}\later}$-coalgebra captures infinite trace semantics.
\end{corollary}

We conclude the section by noting that there is a functor associating $TF$-coalgebras to $\liftKl{\extP{F}\later}$-coalgebras while acting as the constant sheaf functor $\Delta$ on carriers and as $K$ on homomorphisms.
In fact, there is a lifting of $K \Delta$ given by composition of \eqref{eq:lift-delta-guarded-coalg} and \eqref{eq:lift-k-coalg} as the commuting diagram below illustrates.
\begin{equation*}	
	\begin{tikzpicture}[
			auto, xscale=2.4, yscale=1.2, diagram font,
			baseline=(current bounding box.center)
		]	
		\node[] (n0) at (0,0) {\(\CC\)};	
		\node[] (n1) at (1,0) {\(\Sh[\CC]{\alpha}\)};	
		\node[] (n2) at (2,0) {\(\Kl{\extP{T}}\)};	
		\node[] (n3) at (0,1) {\(\Coalg{TF}\)};	
		\node[] (n4) at (1,1) {\(\Coalg{\extP{T}\extP{F}\later}\)};	
		\node[] (n5) at (2,1) {\(\Coalg{\liftKl{\extP{F}\later}}\)};	
		
		\draw[->] (n0) to node[swap] {\(\Delta\)} (n1);
		\draw[->] (n1) to node[swap] {\(K\)} (n2);
		\draw[->] (n3) -- (n4);
		\draw[->] (n4) -- (n5);
		\draw[<-] (n0) -- (n3);
	 	\draw[<-] (n1) -- (n4);
	 	\draw[<-] (n2) -- (n5);
		
	\end{tikzpicture}
\end{equation*}
If $\alpha > 1$, then this functor is an inclusion for the category of $TF$-coalgebras into that of $\liftKl{\extP{F}\later}$-coalgebras.
In particular, it acts on any $F$-coalgebra $(X,h)$ as $\Delta^\later$ and on any $F$-coalgebra homomorphism $f\colon (X,h) \to (Y,k)$ as $K \Delta$:
\begin{equation*}
	(X,h) \mapsto (\Delta X, \extP{TF}(\nxt_{\Delta X}) \circ \Delta h)
	\qquad
	f \mapsto \eta^\extP{T}_{\extP{F}\Delta Y} \circ \Delta f
	\text{.}
\end{equation*}

Thanks to this inclusion we are able to define the infinite trace semantics of a $TF$-coalgebra $(X,h)$ as the unique coalgebra homomorphism from $(\Delta X, \Delta^\later h)$ to the final $\liftKl{\extP{F}\later}$-coalgebra. 
By definition unfolding, this morphism is the unique arrow $!_{\Delta^\later h}$ that makes the following diagram in $\Sh[\CC]{\alpha}$ commute:
\begin{equation*}	
	\begin{tikzpicture}[
			auto, xscale=3.5, yscale=-1.5, rotate=90,diagram font,
			baseline=(current bounding box.center)
		]	
		\node (n0) at (0,2) {\(\Delta X\)};	
		\node (n1) at (0,0) {\(\extP{T}\extP{F}\later \Delta X\)};	
		\node (n2) at (4,2) {\(\extP{T}\fseq(F)\)};	
		\node (n3) at (4,0) {\(\extP{T}\extP{F} \later \fseq(F)\)};	

		\node (m0) at (0,1) {\(\extP{T}\extP{F} \Delta X\)};
		\node (m1) at (1,0) {\(\extP{T}\extP{F} \later \extP{T}\fseq(F)\)};
		\node (m2) at (2,0) {\(\extP{T}\extP{F} \extP{T}\later \fseq(F)\)};
		\node (m3) at (3,0) {\(\extP{T}\extP{T}\extP{F} \later \fseq(F)\)};

		\draw[->] (n0) to node {\(\Delta^\later h\)} (m0);
		\draw[->] (m0) to node {\(\extP{T}\extP{F}\nxt_{\Delta X}\)} (n1);
		\draw[nat] (n2) to node[swap] {\(\extP{T}\ffix \extP{F}\later\)} (n3);
		
		\draw[dashed,->] (n0) to node[swap] {\(!_{\Delta^\later h}\)} (n2);
		\draw[->] (n1) to node {\(\extP{T}\extP{F} \later !_{\Delta^\later h}\)} (m1);
		\draw[->] (m1) to node {\(\extP{T}\extP{F} \theta_{\fseq(F)}\)} (m2);
		\draw[->] (m2) to node {\(\extP{T}\extP{\lambda}_{\later \fseq(F)}\)} (m3);
		\draw[->] (m3) to node {\(\extP{\mu}_{\extP{F} \later \fseq(F)}\)} (n3);
				
	\end{tikzpicture}
\end{equation*}

Let $T$ be an affine monad, $F$ an endofunctor whose final sequence is stable at $\omega$, and $\lambda$ a distributive law for them. Given a $TF$-coalgebra $(X,h)$, ${!}_{\Delta^\later h}$ is the unique morphism of sheaves such that:
\begin{equation*}
	{!}_{\Delta^\later h,1} = TF!_X \circ h
	\qquad
	{!}_{\Delta^\later h,n} = \mu_{F^{n}} \circ T\lambda_{F^{n-1}} \circ F{!}_{\Delta^\later h,n-1} \circ h
\end{equation*}
where $1 < n < \omega$. With reference to our initial example on non-deterministic labelled transition systems, when $(X,h)$ is a $\mathcal{P}^+(A \times \Id)$-coalgebra, ${!}_{\Delta^\later h}$ is the unique morphism of sheaves such that:
\begin{equation*}
	{!}_{\Delta^\later h,1} = \mathcal{P}^+(A \times !_X) \circ h
	\qquad
	{!}_{\Delta^\later h,n} = \mu_{A^{n}} \circ \mathcal{P}^+\lambda_{A^{n-1}} \circ \mathcal{P}^+{!}_{\Delta^\later h,n-1} \circ h
\end{equation*}
where $1 < n < \omega$. The first equation corresponds to the double implication
\begin{equation*}
	a \in {!}_{\Delta^\later h,1}(x) 
	\iff
	\exists x' \in X \text{ s.t. } (a,x') \in h(x)
\end{equation*}
whereas the second to:
\begin{equation*}
	(a_1,a_2, \dots ,a_{n+1}) \in {!}_{\Delta^\later h,n+1}(x)
	\iff 
	\exists x' \in X \text{ s.t. } (a_1,x') \in h(x) \\
	\land	(a_2, \dots ,a_{n}) \in {!}_{\Delta^\later h,n}(x')
	\text{.}
\end{equation*}
In other words, a state $x$ is assigned by ${!}_{\Delta^\later h,n}$ to the set of its (partial) traces of length $n$.
Restriction from stage $n+1$ to $n$ corresponds to the implication:
\begin{equation*}
	(a_1, \dots ,a_{n},a_{n+1}) \in {!}_{\Delta^\later h,n+1}(x)
	\implies 
	(a_1, \dots ,a_{n}) \in {!}_{\Delta^\later h,n}(x)
\end{equation*}
and amalgamation to the double implication:
\begin{equation*}
	(a_1,a_2, \dots ) \in {!}_{\Delta^\later h,\omega}(x)
	\iff \forall n < \omega
	(a_1,a_2, \dots ,a_{n}) \in {!}_{\Delta^\later h,n}(x)
\end{equation*}
In other words, ${!}_{\Delta^\later h,\omega}$ captures infinite trace semantics.

\section{Concluding remarks}
\label{sec:trace-semantics-remarks}

In this work we presented a general coalgebraic account of infinite trace semantics covering several systems such as non-deterministic, discrete and continuous probabilistic labelled transition systems.
Many authors and works have investigated infinite trace semantics under the lens the theory of coalgebras; we mention \cite{jacobs:cmcs2004,hjs:entcs2006,hjs:lmcs2007,kk:lmcs2013,hu:calco2015,hsu:concur2016} and \cite{cirstea:entcs2010} which is perhaps the closest to ours. 
The main improvements with respect to related works introduced in this work are summarised below:
\begin{itemize}
	\item infinite trace semantics coincides with final semantics in a suitable category of coalgebras;
	\item monads modelling the branching type considered need not to induce enriched Kleisli categories;
	\item the final sequence for the functor modelling linear behaviours can stabilise after $\omega$.
\end{itemize}
In retrospective, the main motivation behind this work is that related works capture infinite trace semantics by means of certain maps into weakly final coalgebras \cite{hjs:lmcs2007,cirstea:entcs2010,hu:calco2015,hsu:concur2016}. 

As a further contribution, we proved in \cref{sec:guarded-coalgebras} that final semantics for guarded coalgebras provides a conservative generalisation of certain behavioural pseudo-metrics due to \citeauthor{barr:tcs1993} and \citeauthor{adamek:jlc2002} \cite{barr:tcs1993,adamek:jlc2002}. In comparison to the rich theory of behavioural metrics \cite{bbkk:fsttcs2014,bbkk:calco2015,breugel:ipl2012,breugel:tcs2001,bw:tcs2005} this result is preliminary, only simple behavioural metrics were considered, nonetheless it calls for future investigations. 

The only non-trivial piece of information required in order to apply our construction are Kleisli liftings for guarded pointwise extensions of behavioural endofunctors. A strategy for obtaining this data is to extend Kleisli liftings for ``unguarded'' behavioural functors (in practice, to start from $\lambda\colon FT \to TF$) and then compose them with liftings for $\later$. Although this path is not always available, we identified mild assumptions that are sufficient for the strategy to succeed: if the construction is done in the context of sheaves on $\alexT{\omega}$, then existence of these liftings is equivalent to affiness of $T$. Affine monads were considered also in \cite{cirstea:entcs2010} where \citeauthor{cirstea:entcs2010} identified in this property a sufficient condition for obtaining maps to the weakly final coalgebra of infinite traces. We remark that our is an equivalence result and holds whenever the final sequence for $F$ stabilises at $\omega$ \ie the same assumption made in \loccit. Affiness and stabilisation at $\omega$ are met by monads and functors used in the modelling of several systems of interest, especially those considered in \cite{hjs:lmcs2007,cirstea:entcs2010,hu:calco2015,hsu:concur2016}: non-deterministic, discrete and continuous probabilistic labelled transition systems.

\bibliography{biblio}


\begin{thebibliography}{85}


\ifx \showCODEN    \undefined \def \showCODEN     #1{\unskip}     \fi
\ifx \showISBNx    \undefined \def \showISBNx     #1{\unskip}     \fi
\ifx \showISBNxiii \undefined \def \showISBNxiii  #1{\unskip}     \fi
\ifx \showISSN     \undefined \def \showISSN      #1{\unskip}     \fi
\ifx \showLCCN     \undefined \def \showLCCN      #1{\unskip}     \fi
\ifx \shownote     \undefined \def \shownote      #1{#1}          \fi
\ifx \showarticletitle \undefined \def \showarticletitle #1{#1}   \fi
\ifx \showURL      \undefined \def \showURL       {\relax}        \fi
\providecommand\bibfield[2]{#2}
\providecommand\bibinfo[2]{#2}
\providecommand\natexlab[1]{#1}
\providecommand\showeprint[2][]{arXiv:#2}

\bibitem[Ad{\'{a}}mek(2002)]%
        {adamek:jlc2002}
\bibfield{author}{\bibinfo{person}{Jir{\'{\i}} Ad{\'{a}}mek}.} \bibinfo{year}{2002}\natexlab{}.
\newblock \showarticletitle{Final Coalgebras are Ideal Completions of Initial Algebras}.
\newblock \bibinfo{journal}{\emph{Journal of Logic and Computation}} \bibinfo{volume}{12}, \bibinfo{number}{2} (\bibinfo{year}{2002}), \bibinfo{pages}{217--242}.
\newblock


\bibitem[Ad{\'{a}}mek(2003a)]%
        {adamek:entcs2003}
\bibfield{author}{\bibinfo{person}{Jir{\'{\i}} Ad{\'{a}}mek}.} \bibinfo{year}{2003}\natexlab{a}.
\newblock \showarticletitle{On a Description of Terminal Coalgebras and Iterative Theories}.
\newblock \bibinfo{journal}{\emph{Electronic Notes in Theoretical Computer Science}} \bibinfo{volume}{82}, \bibinfo{number}{1} (\bibinfo{year}{2003}), \bibinfo{pages}{1--16}.
\newblock


\bibitem[Ad{\'{a}}mek(2003b)]%
        {adamek:tcs2003}
\bibfield{author}{\bibinfo{person}{Jir{\'{\i}} Ad{\'{a}}mek}.} \bibinfo{year}{2003}\natexlab{b}.
\newblock \showarticletitle{On final coalgebras of continuous functors}.
\newblock \bibinfo{journal}{\emph{Theoretical Computer Science}} \bibinfo{volume}{294}, \bibinfo{number}{1/2} (\bibinfo{year}{2003}), \bibinfo{pages}{3--29}.
\newblock


\bibitem[Ad{\'{a}}mek and Koubek(1995)]%
        {ak:tcs1995}
\bibfield{author}{\bibinfo{person}{Jir{\'{\i}} Ad{\'{a}}mek} {and} \bibinfo{person}{V{\'{a}}clav Koubek}.} \bibinfo{year}{1995}\natexlab{}.
\newblock \showarticletitle{On the Greatest Fixed Point of a Set Functor}.
\newblock \bibinfo{journal}{\emph{Theoretical Computer Science}} \bibinfo{volume}{150}, \bibinfo{number}{1} (\bibinfo{year}{1995}), \bibinfo{pages}{57--75}.
\newblock


\bibitem[Ad{\'{a}}mek and Milius(2006)]%
        {am:ic2006}
\bibfield{author}{\bibinfo{person}{Jir{\'{\i}} Ad{\'{a}}mek} {and} \bibinfo{person}{Stefan Milius}.} \bibinfo{year}{2006}\natexlab{}.
\newblock \showarticletitle{Terminal coalgebras and free iterative theories}.
\newblock \bibinfo{journal}{\emph{Information and Computation}} \bibinfo{volume}{204}, \bibinfo{number}{7} (\bibinfo{year}{2006}), \bibinfo{pages}{1139--1172}.
\newblock


\bibitem[Atkey and McBride(2013)]%
        {am:icfp2013}
\bibfield{author}{\bibinfo{person}{Robert Atkey} {and} \bibinfo{person}{Conor McBride}.} \bibinfo{year}{2013}\natexlab{}.
\newblock \showarticletitle{Productive coprogramming with guarded recursion}. In \bibinfo{booktitle}{\emph{{ICFP}}}. \bibinfo{publisher}{{ACM}}, \bibinfo{pages}{197--208}.
\newblock


\bibitem[Axelsen and Kaarsgaard(2016)]%
        {ak:fossacs2016}
\bibfield{author}{\bibinfo{person}{Holger~Bock Axelsen} {and} \bibinfo{person}{Robin Kaarsgaard}.} \bibinfo{year}{2016}\natexlab{}.
\newblock \showarticletitle{Join Inverse Categories as Models of Reversible Recursion}. In \bibinfo{booktitle}{\emph{FoSSaCS}} \emph{(\bibinfo{series}{Lecture Notes in Computer Science}, Vol.~\bibinfo{volume}{9634})}. \bibinfo{publisher}{Springer}, \bibinfo{pages}{73--90}.
\newblock


\bibitem[Bacci(2012)]%
        {bacci:phdthesis}
\bibfield{author}{\bibinfo{person}{Giorgio Bacci}.} \bibinfo{year}{2012}\natexlab{}.
\newblock \emph{\bibinfo{title}{Generalized labelled Markov processes, coalgebraically}}.
\newblock \bibinfo{thesistype}{Ph.\,D. Dissertation}. \bibinfo{school}{Department of Mathematics and Computer Science, University of Udine}.
\newblock


\bibitem[Bacci and Miculan(2015)]%
        {bm:jcss2015}
\bibfield{author}{\bibinfo{person}{Giorgio Bacci} {and} \bibinfo{person}{Marino Miculan}.} \bibinfo{year}{2015}\natexlab{}.
\newblock \showarticletitle{Structural operational semantics for continuous state stochastic transition systems}.
\newblock \bibinfo{journal}{\emph{J. Comput. System Sci.}} \bibinfo{volume}{81}, \bibinfo{number}{5} (\bibinfo{year}{2015}), \bibinfo{pages}{834--858}.
\newblock
\href{https://doi.org/10.1016/j.jcss.2014.12.003}{doi:\nolinkurl{10.1016/j.jcss.2014.12.003}}


\bibitem[Baldan et~al\mbox{.}(2014)]%
        {bbkk:fsttcs2014}
\bibfield{author}{\bibinfo{person}{Paolo Baldan}, \bibinfo{person}{Filippo Bonchi}, \bibinfo{person}{Henning Kerstan}, {and} \bibinfo{person}{Barbara K{\"{o}}nig}.} \bibinfo{year}{2014}\natexlab{}.
\newblock \showarticletitle{Behavioral Metrics via Functor Lifting}. In \bibinfo{booktitle}{\emph{{FSTTCS}}} \emph{(\bibinfo{series}{LIPIcs}, Vol.~\bibinfo{volume}{29})}. \bibinfo{publisher}{Schloss Dagstuhl - Leibniz-Zentrum fuer Informatik}, \bibinfo{pages}{403--415}.
\newblock


\bibitem[Baldan et~al\mbox{.}(2015)]%
        {bbkk:calco2015}
\bibfield{author}{\bibinfo{person}{Paolo Baldan}, \bibinfo{person}{Filippo Bonchi}, \bibinfo{person}{Henning Kerstan}, {and} \bibinfo{person}{Barbara K{\"{o}}nig}.} \bibinfo{year}{2015}\natexlab{}.
\newblock \showarticletitle{Towards Trace Metrics via Functor Lifting}. In \bibinfo{booktitle}{\emph{{CALCO}}} \emph{(\bibinfo{series}{LIPIcs}, Vol.~\bibinfo{volume}{35})}. \bibinfo{publisher}{Schloss Dagstuhl - Leibniz-Zentrum fuer Informatik}, \bibinfo{pages}{35--49}.
\newblock


\bibitem[Barr(1992)]%
        {barr:jpaa1992}
\bibfield{author}{\bibinfo{person}{Michael Barr}.} \bibinfo{year}{1992}\natexlab{}.
\newblock \showarticletitle{Algebraically compact functors}.
\newblock \bibinfo{journal}{\emph{Journal of Pure and Applied Algebra}} \bibinfo{volume}{82}, \bibinfo{number}{3} (\bibinfo{year}{1992}), \bibinfo{pages}{211--231}.
\newblock


\bibitem[Barr(1993)]%
        {barr:tcs1993}
\bibfield{author}{\bibinfo{person}{Michael Barr}.} \bibinfo{year}{1993}\natexlab{}.
\newblock \showarticletitle{Terminal Coalgebras in Well-Founded Set Theory}.
\newblock \bibinfo{journal}{\emph{Theoretical Computer Science}} \bibinfo{volume}{114}, \bibinfo{number}{2} (\bibinfo{year}{1993}), \bibinfo{pages}{299--315}.
\newblock


\bibitem[Barr et~al\mbox{.}(1971)]%
        {bgv:sheaves}
\bibfield{author}{\bibinfo{person}{Michael Barr}, \bibinfo{person}{Pierre~Antoine Grillet}, {and} \bibinfo{person}{Donovan~H. Van~Osdol}.} \bibinfo{year}{1971}\natexlab{}.
\newblock \bibinfo{booktitle}{\emph{Exact categories and categories of sheaves}}.
\newblock \bibinfo{publisher}{Springer-Verlag}, \bibinfo{address}{Berlin, New York}.
\newblock
\showISBNx{0-387-05678-5}


\bibitem[Benton et~al\mbox{.}(2000)]%
        {bhm:appsem2000}
\bibfield{author}{\bibinfo{person}{Nick Benton}, \bibinfo{person}{John Hughes}, {and} \bibinfo{person}{Eugenio Moggi}.} \bibinfo{year}{2000}\natexlab{}.
\newblock \showarticletitle{Monads and Effects}. In \bibinfo{booktitle}{\emph{{APPSEM}}} \emph{(\bibinfo{series}{Lecture Notes in Computer Science}, Vol.~\bibinfo{volume}{2395})}. \bibinfo{publisher}{Springer}, \bibinfo{pages}{42--122}.
\newblock


\bibitem[Birkedal et~al\mbox{.}(2016)]%
        {bbcgsv:csl2016}
\bibfield{author}{\bibinfo{person}{Lars Birkedal}, \bibinfo{person}{Ales Bizjak}, \bibinfo{person}{Ranald Clouston}, \bibinfo{person}{Hans~Bugge Grathwohl}, \bibinfo{person}{Bas Spitters}, {and} \bibinfo{person}{Andrea Vezzosi}.} \bibinfo{year}{2016}\natexlab{}.
\newblock \showarticletitle{Guarded Cubical Type Theory: Path Equality for Guarded Recursion}. In \bibinfo{booktitle}{\emph{{CSL}}} \emph{(\bibinfo{series}{LIPIcs}, Vol.~\bibinfo{volume}{62})}. \bibinfo{publisher}{Schloss Dagstuhl - Leibniz-Zentrum fuer Informatik}, \bibinfo{pages}{23:1--23:17}.
\newblock


\bibitem[Birkedal et~al\mbox{.}(2013)]%
        {bbs:lmcs2013}
\bibfield{author}{\bibinfo{person}{Lars Birkedal}, \bibinfo{person}{Ales Bizjak}, {and} \bibinfo{person}{Jan Schwinghammer}.} \bibinfo{year}{2013}\natexlab{}.
\newblock \showarticletitle{Step-Indexed Relational Reasoning for Countable Nondeterminism}.
\newblock \bibinfo{journal}{\emph{Logical Methods in Computer Science}} \bibinfo{volume}{9}, \bibinfo{number}{4} (\bibinfo{year}{2013}).
\newblock


\bibitem[Birkedal et~al\mbox{.}(2012)]%
        {bmss:lmcs2012}
\bibfield{author}{\bibinfo{person}{Lars Birkedal}, \bibinfo{person}{Rasmus~Ejlers M{\o}gelberg}, \bibinfo{person}{Jan Schwinghammer}, {and} \bibinfo{person}{Kristian St{\o}vring}.} \bibinfo{year}{2012}\natexlab{}.
\newblock \showarticletitle{First steps in synthetic guarded domain theory: step-indexing in the topos of trees}.
\newblock \bibinfo{journal}{\emph{Logical Methods in Computer Science}} \bibinfo{volume}{8}, \bibinfo{number}{4} (\bibinfo{year}{2012}).
\newblock
\href{https://doi.org/10.2168/LMCS-8(4:1)2012}{doi:\nolinkurl{10.2168/LMCS-8(4:1)2012}}


\bibitem[Bizjak et~al\mbox{.}(2014)]%
        {bbm:tlca2014}
\bibfield{author}{\bibinfo{person}{Ales Bizjak}, \bibinfo{person}{Lars Birkedal}, {and} \bibinfo{person}{Marino Miculan}.} \bibinfo{year}{2014}\natexlab{}.
\newblock \showarticletitle{A Model of Countable Nondeterminism in Guarded Type Theory}. In \bibinfo{booktitle}{\emph{{RTA-TLCA}}} \emph{(\bibinfo{series}{Lecture Notes in Computer Science}, Vol.~\bibinfo{volume}{8560})}. \bibinfo{publisher}{Springer}, \bibinfo{pages}{108--123}.
\newblock


\bibitem[Brengos et~al\mbox{.}(2015)]%
        {bmp:jlamp2015}
\bibfield{author}{\bibinfo{person}{Tomasz Brengos}, \bibinfo{person}{Marino Miculan}, {and} \bibinfo{person}{Marco Peressotti}.} \bibinfo{year}{2015}\natexlab{}.
\newblock \showarticletitle{Behavioural equivalences for coalgebras with unobservable moves}.
\newblock \bibinfo{journal}{\emph{Journal of Logical and Algebraic Methods in Programming}} \bibinfo{volume}{84}, \bibinfo{number}{6} (\bibinfo{year}{2015}), \bibinfo{pages}{826--852}.
\newblock


\bibitem[Brengos and Peressotti(2016)]%
        {bp:concur2016}
\bibfield{author}{\bibinfo{person}{Tomasz Brengos} {and} \bibinfo{person}{Marco Peressotti}.} \bibinfo{year}{2016}\natexlab{}.
\newblock \showarticletitle{A uniform framework for timed automata}. In \bibinfo{booktitle}{\emph{{CONCUR}}} \emph{(\bibinfo{series}{LIPIcs}, Vol.~\bibinfo{volume}{59})}. \bibinfo{publisher}{Schloss Dagstuhl - Leibniz-Zentrum fuer Informatik}, \bibinfo{pages}{26:1--26:14}.
\newblock


\bibitem[Brengos and Peressotti(2019)]%
        {bp:lmcs2019}
\bibfield{author}{\bibinfo{person}{Tomasz Brengos} {and} \bibinfo{person}{Marco Peressotti}.} \bibinfo{year}{2019}\natexlab{}.
\newblock \showarticletitle{Behavioural equivalences for timed systems}.
\newblock \bibinfo{journal}{\emph{Logical Methods in Computer Science}} \bibinfo{volume}{15}, \bibinfo{number}{1} (\bibinfo{year}{2019}).
\newblock
\href{https://doi.org/10.23638/LMCS-15(1:17)2019}{doi:\nolinkurl{10.23638/LMCS-15(1:17)2019}}


\bibitem[Capretta(2011)]%
        {capretta:tcs2011}
\bibfield{author}{\bibinfo{person}{Venanzio Capretta}.} \bibinfo{year}{2011}\natexlab{}.
\newblock \showarticletitle{Coalgebras in functional programming and type theory}.
\newblock \bibinfo{journal}{\emph{Theoretical Computer Science}} \bibinfo{volume}{412}, \bibinfo{number}{38} (\bibinfo{year}{2011}), \bibinfo{pages}{5006--5024}.
\newblock
\showISSN{0304-3975}
\href{https://doi.org/10.1016/j.tcs.2011.04.024}{doi:\nolinkurl{10.1016/j.tcs.2011.04.024}}


\bibitem[C{\^{i}}rstea(2010)]%
        {cirstea:entcs2010}
\bibfield{author}{\bibinfo{person}{Corina C{\^{i}}rstea}.} \bibinfo{year}{2010}\natexlab{}.
\newblock \showarticletitle{Generic Infinite Traces and Path-Based Coalgebraic Temporal Logics}.
\newblock \bibinfo{journal}{\emph{Electronic Notes in Theoretical Computer Science}} \bibinfo{volume}{264}, \bibinfo{number}{2} (\bibinfo{year}{2010}), \bibinfo{pages}{83--103}.
\newblock


\bibitem[Desharnais et~al\mbox{.}(2002)]%
        {prakash:ic2002}
\bibfield{author}{\bibinfo{person}{Josee Desharnais}, \bibinfo{person}{Abbas Edalat}, {and} \bibinfo{person}{Prakash Panangaden}.} \bibinfo{year}{2002}\natexlab{}.
\newblock \showarticletitle{Bisimulation for Labelled Markov Processes}.
\newblock \bibinfo{journal}{\emph{Information and Computation}} \bibinfo{volume}{179}, \bibinfo{number}{2} (\bibinfo{year}{2002}), \bibinfo{pages}{163--193}.
\newblock
\href{https://doi.org/10.1006/inco.2001.2962}{doi:\nolinkurl{10.1006/inco.2001.2962}}


\bibitem[Diekert et~al\mbox{.}(1995)]%
        {dr:traces}
\bibfield{author}{\bibinfo{person}{Volker Diekert}, \bibinfo{person}{Grzegorz Rozenberg}, {and} \bibinfo{person}{G Rozenburg}.} \bibinfo{year}{1995}\natexlab{}.
\newblock \bibinfo{booktitle}{\emph{The book of traces}}. Vol.~\bibinfo{volume}{15}.
\newblock \bibinfo{publisher}{World Scientific}.
\newblock


\bibitem[Doberkat(2009)]%
        {doberkat:stoclogic}
\bibfield{author}{\bibinfo{person}{Ernst{-}Erich Doberkat}.} \bibinfo{year}{2009}\natexlab{}.
\newblock \bibinfo{booktitle}{\emph{Stochastic Coalgebraic Logic}}.
\newblock \bibinfo{publisher}{Springer}.
\newblock
\showISBNx{978-3-642-02994-3}
\href{https://doi.org/10.1007/978-3-642-02995-0}{doi:\nolinkurl{10.1007/978-3-642-02995-0}}


\bibitem[Freyd(1991)]%
        {freyd:ct1991}
\bibfield{author}{\bibinfo{person}{Peter~J. Freyd}.} \bibinfo{year}{1991}\natexlab{}.
\newblock \showarticletitle{Algebraically complete categories}. In \bibinfo{booktitle}{\emph{Category Theory}} \emph{(\bibinfo{series}{Lecture Notes in Mathematics}, Vol.~\bibinfo{volume}{1488})}, \bibfield{editor}{\bibinfo{person}{Aurelio Carboni}, \bibinfo{person}{MariaCristina Pedicchio}, {and} \bibinfo{person}{Guiseppe Rosolini}} (Eds.). \bibinfo{publisher}{Springer Berlin Heidelberg}, \bibinfo{pages}{95--104}.
\newblock
\showISBNx{978-3-540-54706-8}
\href{https://doi.org/10.1007/BFb0084215}{doi:\nolinkurl{10.1007/BFb0084215}}


\bibitem[Freyd(1992)]%
        {freyd:lmms1992}
\bibfield{author}{\bibinfo{person}{Peter~J. Freyd}.} \bibinfo{year}{1992}\natexlab{}.
\newblock \showarticletitle{Remarks on algebraically compact categories}. In \bibinfo{booktitle}{\emph{Applications of Categories in Computer Science: Proceedings of the London Mathematical Society Symposium}}, Vol.~\bibinfo{volume}{177}. Cambridge University Press, \bibinfo{pages}{95}.
\newblock


\bibitem[Gianantonio and Miculan(2004)]%
        {dgm:fossacs2004}
\bibfield{author}{\bibinfo{person}{Pietro~Di Gianantonio} {and} \bibinfo{person}{Marino Miculan}.} \bibinfo{year}{2004}\natexlab{}.
\newblock \showarticletitle{Unifying Recursive and Co-recursive Definitions in Sheaf Categories}. In \bibinfo{booktitle}{\emph{Proc.~{FOSSACS}}} \emph{(\bibinfo{series}{Lecture Notes in Computer Science}, Vol.~\bibinfo{volume}{2987})}, \bibfield{editor}{\bibinfo{person}{Igor Walukiewicz}} (Ed.). \bibinfo{publisher}{Springer}, \bibinfo{pages}{136--150}.
\newblock
\href{https://doi.org/10.1007/978-3-540-24727-2_11}{doi:\nolinkurl{10.1007/978-3-540-24727-2_11}}


\bibitem[Gray(1965)]%
        {gray:top1965}
\bibfield{author}{\bibinfo{person}{John~W. Gray}.} \bibinfo{year}{1965}\natexlab{}.
\newblock \showarticletitle{Sheaves with values in a category}.
\newblock \bibinfo{journal}{\emph{Topology}} \bibinfo{volume}{3}, \bibinfo{number}{1} (\bibinfo{year}{1965}), \bibinfo{pages}{1--18}.
\newblock
\showISSN{0040-9383}
\href{https://doi.org/10.1016/0040-9383(65)90066-2}{doi:\nolinkurl{10.1016/0040-9383(65)90066-2}}


\bibitem[Haghverdi(2011)]%
        {hs:tcs2011}
\bibfield{author}{\bibinfo{person}{Esfandiar Haghverdi}.} \bibinfo{year}{2011}\natexlab{}.
\newblock \showarticletitle{Towards a geometry of recursion}.
\newblock \bibinfo{journal}{\emph{Theoretical Computer Science}} \bibinfo{volume}{412}, \bibinfo{number}{20} (\bibinfo{year}{2011}), \bibinfo{pages}{2015--2028}.
\newblock


\bibitem[Haghverdi and Scott(2010)]%
        {hs:mscs2010}
\bibfield{author}{\bibinfo{person}{Esfandiar Haghverdi} {and} \bibinfo{person}{Philip~J. Scott}.} \bibinfo{year}{2010}\natexlab{}.
\newblock \showarticletitle{Towards a typed Geometry of Interaction}.
\newblock \bibinfo{journal}{\emph{Mathematical Structures in Computer Science}} \bibinfo{volume}{20}, \bibinfo{number}{3} (\bibinfo{year}{2010}), \bibinfo{pages}{473--521}.
\newblock


\bibitem[Hasuo et~al\mbox{.}(2006)]%
        {hjs:entcs2006}
\bibfield{author}{\bibinfo{person}{Ichiro Hasuo}, \bibinfo{person}{Bart Jacobs}, {and} \bibinfo{person}{Ana Sokolova}.} \bibinfo{year}{2006}\natexlab{}.
\newblock \showarticletitle{Generic Trace Theory}.
\newblock \bibinfo{journal}{\emph{Electronic Notes in Theoretical Computer Science}} \bibinfo{volume}{164}, \bibinfo{number}{1} (\bibinfo{year}{2006}), \bibinfo{pages}{47--65}.
\newblock


\bibitem[Hasuo et~al\mbox{.}(2007)]%
        {hjs:lmcs2007}
\bibfield{author}{\bibinfo{person}{Ichiro Hasuo}, \bibinfo{person}{Bart Jacobs}, {and} \bibinfo{person}{Ana Sokolova}.} \bibinfo{year}{2007}\natexlab{}.
\newblock \showarticletitle{Generic Trace Semantics via Coinduction}.
\newblock \bibinfo{journal}{\emph{Logical Methods in Computer Science}} \bibinfo{volume}{3}, \bibinfo{number}{4} (\bibinfo{year}{2007}).
\newblock


\bibitem[Hermida and Jacobs(1998)]%
        {hj:ic1998}
\bibfield{author}{\bibinfo{person}{Claudio Hermida} {and} \bibinfo{person}{Bart Jacobs}.} \bibinfo{year}{1998}\natexlab{}.
\newblock \showarticletitle{Structural Induction and Coinduction in a Fibrational Setting}.
\newblock \bibinfo{journal}{\emph{Information and Computation}} \bibinfo{volume}{145}, \bibinfo{number}{2} (\bibinfo{year}{1998}), \bibinfo{pages}{107--152}.
\newblock


\bibitem[Heunen and Jacobs(2011)]%
        {hj:entcs2011}
\bibfield{author}{\bibinfo{person}{Chris Heunen} {and} \bibinfo{person}{Bart Jacobs}.} \bibinfo{year}{2011}\natexlab{}.
\newblock \showarticletitle{Quantum Logic in Dagger Kernel Categories}.
\newblock \bibinfo{journal}{\emph{Electronic Notes in Theoretical Computer Science}} \bibinfo{volume}{270}, \bibinfo{number}{2} (\bibinfo{year}{2011}), \bibinfo{pages}{79--103}.
\newblock


\bibitem[Jacobs(1994)]%
        {jacobs:apal1994}
\bibfield{author}{\bibinfo{person}{Bart Jacobs}.} \bibinfo{year}{1994}\natexlab{}.
\newblock \showarticletitle{Semantics of Weakening and Contraction}.
\newblock \bibinfo{journal}{\emph{Annals of Pure and Applied Logics}} \bibinfo{volume}{69}, \bibinfo{number}{1} (\bibinfo{year}{1994}), \bibinfo{pages}{73--106}.
\newblock


\bibitem[Jacobs(2004)]%
        {jacobs:cmcs2004}
\bibfield{author}{\bibinfo{person}{Bart Jacobs}.} \bibinfo{year}{2004}\natexlab{}.
\newblock \showarticletitle{Trace Semantics for Coalgebras}.
\newblock \bibinfo{journal}{\emph{Electronic Notes in Theoretical Computer Science}}  \bibinfo{volume}{106} (\bibinfo{year}{2004}), \bibinfo{pages}{167--184}.
\newblock


\bibitem[Jacobs(2016)]%
        {jacobs:cmcs2016}
\bibfield{author}{\bibinfo{person}{Bart Jacobs}.} \bibinfo{year}{2016}\natexlab{}.
\newblock \showarticletitle{Affine Monads and Side-Effect-Freeness}. In \bibinfo{booktitle}{\emph{{CMCS}}} \emph{(\bibinfo{series}{Lecture Notes in Computer Science}, Vol.~\bibinfo{volume}{9608})}. \bibinfo{publisher}{Springer}, \bibinfo{pages}{53--72}.
\newblock


\bibitem[Johnstone(2002a)]%
        {johnstone:elephant1}
\bibfield{author}{\bibinfo{person}{Peter~T. Johnstone}.} \bibinfo{year}{2002}\natexlab{a}.
\newblock \bibinfo{booktitle}{\emph{Sketches of an Elephant (Volume 1)}}.
\newblock Number~43 in \bibinfo{series}{Oxford Logic Guides}. \bibinfo{publisher}{Oxford University Press}, \bibinfo{address}{New York}.
\newblock


\bibitem[Johnstone(2002b)]%
        {johnstone:elephant2}
\bibfield{author}{\bibinfo{person}{Peter~T. Johnstone}.} \bibinfo{year}{2002}\natexlab{b}.
\newblock \bibinfo{booktitle}{\emph{Sketches of an Elephant (Volume 2)}}.
\newblock Number~44 in \bibinfo{series}{Oxford Logic Guides}. \bibinfo{publisher}{Oxford University Press}, \bibinfo{address}{New York}.
\newblock


\bibitem[Jones(1995)]%
        {jones:afp1995}
\bibfield{author}{\bibinfo{person}{Mark~P. Jones}.} \bibinfo{year}{1995}\natexlab{}.
\newblock \showarticletitle{Functional Programming with Overloading and Higher-Order Polymorphism}. In \bibinfo{booktitle}{\emph{Advanced Functional Programming}} \emph{(\bibinfo{series}{Lecture Notes in Computer Science}, Vol.~\bibinfo{volume}{925})}. \bibinfo{publisher}{Springer}, \bibinfo{pages}{97--136}.
\newblock


\bibitem[Kelly(1982)]%
        {kelly:enrichedbook}
\bibfield{author}{\bibinfo{person}{Max Kelly}.} \bibinfo{year}{1982}\natexlab{}.
\newblock \bibinfo{booktitle}{\emph{Basic concepts of enriched category theory}}.
\newblock \bibinfo{publisher}{Cambridge University Press}.
\newblock


\bibitem[Kerstan and K{\"{o}}nig(2013)]%
        {kk:lmcs2013}
\bibfield{author}{\bibinfo{person}{Henning Kerstan} {and} \bibinfo{person}{Barbara K{\"{o}}nig}.} \bibinfo{year}{2013}\natexlab{}.
\newblock \showarticletitle{Coalgebraic Trace Semantics for Continuous Probabilistic Transition Systems}.
\newblock \bibinfo{journal}{\emph{Logical Methods in Computer Science}} \bibinfo{volume}{9}, \bibinfo{number}{4} (\bibinfo{year}{2013}).
\newblock


\bibitem[Kerstan et~al\mbox{.}(2014)]%
        {kkw:cmcs2014}
\bibfield{author}{\bibinfo{person}{Henning Kerstan}, \bibinfo{person}{Barbara K{\"{o}}nig}, {and} \bibinfo{person}{Bram Westerbaan}.} \bibinfo{year}{2014}\natexlab{}.
\newblock \showarticletitle{Lifting Adjunctions to Coalgebras to (Re)Discover Automata Constructions}. In \bibinfo{booktitle}{\emph{{CMCS}}} \emph{(\bibinfo{series}{Lecture Notes in Computer Science}, Vol.~\bibinfo{volume}{8446})}. \bibinfo{publisher}{Springer}, \bibinfo{pages}{168--188}.
\newblock


\bibitem[Klin(2011)]%
        {klin:tcs2011}
\bibfield{author}{\bibinfo{person}{Bartek Klin}.} \bibinfo{year}{2011}\natexlab{}.
\newblock \showarticletitle{Bialgebras for structural operational semantics: An introduction}.
\newblock \bibinfo{journal}{\emph{Theoretical Computer Science}} \bibinfo{volume}{412}, \bibinfo{number}{38} (\bibinfo{year}{2011}), \bibinfo{pages}{5043--5069}.
\newblock


\bibitem[Klin and Nachyla(2015)]%
        {kn:calco2015}
\bibfield{author}{\bibinfo{person}{Bartek Klin} {and} \bibinfo{person}{Beata Nachyla}.} \bibinfo{year}{2015}\natexlab{}.
\newblock \showarticletitle{Presenting Morphisms of Distributive Laws}. In \bibinfo{booktitle}{\emph{Proc.~{CALCO}}} \emph{(\bibinfo{series}{LIPIcs}, Vol.~\bibinfo{volume}{35})}, \bibfield{editor}{\bibinfo{person}{Lawrence~S. Moss} {and} \bibinfo{person}{Pawel Sobocinski}} (Eds.). \bibinfo{publisher}{Schloss Dagstuhl - Leibniz-Zentrum fuer Informatik}, \bibinfo{pages}{190--204}.
\newblock
\href{https://doi.org/10.4230/LIPIcs.CALCO.2015.190}{doi:\nolinkurl{10.4230/LIPIcs.CALCO.2015.190}}


\bibitem[Kock(1970)]%
        {kock:adm1970}
\bibfield{author}{\bibinfo{person}{Anders Kock}.} \bibinfo{year}{1970}\natexlab{}.
\newblock \showarticletitle{Monads on symmetric monoidal closed categories}.
\newblock \bibinfo{journal}{\emph{Archiv der Mathematik}} \bibinfo{volume}{21}, \bibinfo{number}{1} (\bibinfo{year}{1970}), \bibinfo{pages}{1--10}.
\newblock


\bibitem[Kock(1971)]%
        {kock:adm1971}
\bibfield{author}{\bibinfo{person}{Anders Kock}.} \bibinfo{year}{1971}\natexlab{}.
\newblock \showarticletitle{Bilinearity and Cartesian closed monads}.
\newblock \bibinfo{journal}{\emph{Math. Scand.}}  \bibinfo{volume}{29} (\bibinfo{year}{1971}), \bibinfo{pages}{161--174}.
\newblock


\bibitem[Kock(1972)]%
        {kock:adm1972}
\bibfield{author}{\bibinfo{person}{Anders Kock}.} \bibinfo{year}{1972}\natexlab{}.
\newblock \showarticletitle{Strong functors and monoidal monads}.
\newblock \bibinfo{journal}{\emph{Archiv der Mathematik}} \bibinfo{volume}{23}, \bibinfo{number}{1} (\bibinfo{year}{1972}), \bibinfo{pages}{113--120}.
\newblock


\bibitem[Lenisa et~al\mbox{.}(2000)]%
        {lpw:entcs2000}
\bibfield{author}{\bibinfo{person}{Marina Lenisa}, \bibinfo{person}{John Power}, {and} \bibinfo{person}{Hiroshi Watanabe}.} \bibinfo{year}{2000}\natexlab{}.
\newblock \showarticletitle{Distributivity for endofunctors, pointed and co-pointed endofunctors, monads and comonads}.
\newblock \bibinfo{journal}{\emph{Electronic Notes in Theoretical Computer Science}}  \bibinfo{volume}{33} (\bibinfo{year}{2000}), \bibinfo{pages}{230--260}.
\newblock


\bibitem[Lenisa et~al\mbox{.}(2004)]%
        {lpw:tcs2004}
\bibfield{author}{\bibinfo{person}{Marina Lenisa}, \bibinfo{person}{John Power}, {and} \bibinfo{person}{Hiroshi Watanabe}.} \bibinfo{year}{2004}\natexlab{}.
\newblock \showarticletitle{Category theory for operational semantics}.
\newblock \bibinfo{journal}{\emph{Theoretical Computer Science}} \bibinfo{volume}{327}, \bibinfo{number}{1-2} (\bibinfo{year}{2004}), \bibinfo{pages}{135--154}.
\newblock


\bibitem[Lindners(1979)]%
        {lindner:adm1979}
\bibfield{author}{\bibinfo{person}{Harald Lindners}.} \bibinfo{year}{1979}\natexlab{}.
\newblock \showarticletitle{Affine parts of monads}.
\newblock \bibinfo{journal}{\emph{Archiv der Mathematik}} \bibinfo{volume}{33}, \bibinfo{number}{1} (\bibinfo{year}{1979}), \bibinfo{pages}{437--443}.
\newblock


\bibitem[Litak(2014)]%
        {litak:ocl2014}
\bibfield{author}{\bibinfo{person}{Tadeusz Litak}.} \bibinfo{year}{2014}\natexlab{}.
\newblock \showarticletitle{Constructive modalities with provability smack}.
\newblock In \bibinfo{booktitle}{\emph{Leo Esakia on duality in modal and intuitionistic logics}}, \bibfield{editor}{\bibinfo{person}{Guram Bezhanishvili}} (Ed.). \bibinfo{series}{Outstanding Contributions to Logic}, Vol.~\bibinfo{volume}{4}. \bibinfo{publisher}{Springer}, \bibinfo{pages}{179--208}.
\newblock
\href{https://doi.org/10.1007/978-94-017-8860-1_8}{doi:\nolinkurl{10.1007/978-94-017-8860-1_8}}


\bibitem[Mac~Lane(1971)]%
        {maclane:cats}
\bibfield{author}{\bibinfo{person}{Saunders Mac~Lane}.} \bibinfo{year}{1971}\natexlab{}.
\newblock \bibinfo{booktitle}{\emph{Categories for the Working Mathematician}}.
\newblock \bibinfo{publisher}{Springer-Verlag}, \bibinfo{address}{Berlin}.
\newblock


\bibitem[Mac~Lane and Moerdijk(1994)]%
        {maclane:sheaves}
\bibfield{author}{\bibinfo{person}{Saunders Mac~Lane} {and} \bibinfo{person}{Ieke Moerdijk}.} \bibinfo{year}{1994}\natexlab{}.
\newblock \bibinfo{booktitle}{\emph{Sheaves in Geometry and Logic: a First Introduction to Topos Theory}}.
\newblock \bibinfo{publisher}{Springer-Verlag}.
\newblock
\showISBNx{0387977104}


\bibitem[M{\o}gelberg and Paviotti(2016)]%
        {mp:lics2015}
\bibfield{author}{\bibinfo{person}{Rasmus~Ejlers M{\o}gelberg} {and} \bibinfo{person}{Marco Paviotti}.} \bibinfo{year}{2016}\natexlab{}.
\newblock \showarticletitle{Denotational semantics of recursive types in synthetic guarded domain theory}. In \bibinfo{booktitle}{\emph{{LICS}}}. \bibinfo{publisher}{{ACM}}, \bibinfo{pages}{317--326}.
\newblock


\bibitem[Moggi(1989)]%
        {moggi:lics1989}
\bibfield{author}{\bibinfo{person}{Eugenio Moggi}.} \bibinfo{year}{1989}\natexlab{}.
\newblock \showarticletitle{Computational Lambda-Calculus and Monads}. In \bibinfo{booktitle}{\emph{{LICS}}}. \bibinfo{publisher}{IEEE Computer Society}, \bibinfo{pages}{14--23}.
\newblock


\bibitem[Moggi and Palumbo(1999)]%
        {mp:entcs1999}
\bibfield{author}{\bibinfo{person}{Eugenio Moggi} {and} \bibinfo{person}{F. Palumbo}.} \bibinfo{year}{1999}\natexlab{}.
\newblock \showarticletitle{Monadic Encapsulation of Effects: a Revised Approach}.
\newblock \bibinfo{journal}{\emph{Electronic Notes in Theoretical Computer Science}}  \bibinfo{volume}{26} (\bibinfo{year}{1999}), \bibinfo{pages}{121}.
\newblock


\bibitem[Moggi and Sabry(2001)]%
        {ms:jfc2001}
\bibfield{author}{\bibinfo{person}{Eugenio Moggi} {and} \bibinfo{person}{Amr Sabry}.} \bibinfo{year}{2001}\natexlab{}.
\newblock \showarticletitle{Monadic encapsulation of effects: a revised approach (extended version)}.
\newblock \bibinfo{journal}{\emph{Journal of Functional Programming}} \bibinfo{volume}{11}, \bibinfo{number}{6} (\bibinfo{year}{2001}), \bibinfo{pages}{591--627}.
\newblock


\bibitem[Mulry(1993)]%
        {mulry:mfps1993}
\bibfield{author}{\bibinfo{person}{Philip~S. Mulry}.} \bibinfo{year}{1993}\natexlab{}.
\newblock \showarticletitle{Lifting Theorems for Kleisli Categories}. In \bibinfo{booktitle}{\emph{{MFPS}}} \emph{(\bibinfo{series}{Lecture Notes in Computer Science}, Vol.~\bibinfo{volume}{802})}. \bibinfo{publisher}{Springer}, \bibinfo{pages}{304--319}.
\newblock


\bibitem[Panangaden(2009)]%
        {prakash:markovprocs}
\bibfield{author}{\bibinfo{person}{Prakash Panangaden}.} \bibinfo{year}{2009}\natexlab{}.
\newblock \bibinfo{booktitle}{\emph{Labelled Markov Processes}}.
\newblock \bibinfo{publisher}{Imperial College Press}, \bibinfo{address}{London, U.K.}
\newblock


\bibitem[Paviotti et~al\mbox{.}(2015)]%
        {pmb:entcs2015}
\bibfield{author}{\bibinfo{person}{Marco Paviotti}, \bibinfo{person}{Rasmus~Ejlers M{\o}gelberg}, {and} \bibinfo{person}{Lars Birkedal}.} \bibinfo{year}{2015}\natexlab{}.
\newblock \showarticletitle{A Model of {PCF} in Guarded Type Theory}.
\newblock \bibinfo{journal}{\emph{Electronic Notes in Theoretical Computer Science}}  \bibinfo{volume}{319} (\bibinfo{year}{2015}), \bibinfo{pages}{333--349}.
\newblock


\bibitem[Peressotti(2017)]%
        {peressotti:phdthesis}
\bibfield{author}{\bibinfo{person}{Marco Peressotti}.} \bibinfo{year}{2017}\natexlab{}.
\newblock \emph{\bibinfo{title}{Coalgebraic semantics of self-referential behaviours}}.
\newblock \bibinfo{thesistype}{Ph.\,D. Dissertation}. \bibinfo{school}{University of Udine, Italy}.
\newblock
\urldef\tempurl%
\url{https://opac.bncf.firenze.sbn.it/bncf-prod/resource?uri=TD17021083}
\showURL{%
\tempurl}


\bibitem[Power and Turi(1999)]%
        {pt:entcs1999}
\bibfield{author}{\bibinfo{person}{John Power} {and} \bibinfo{person}{Daniele Turi}.} \bibinfo{year}{1999}\natexlab{}.
\newblock \showarticletitle{A Coalgebraic Foundation for Linear Time Semantics}.
\newblock \bibinfo{journal}{\emph{Electronic Notes in Theoretical Computer Science}}  \bibinfo{volume}{29} (\bibinfo{year}{1999}), \bibinfo{pages}{259--274}.
\newblock


\bibitem[Power and Watanabe(2002)]%
        {pw:tcs2002}
\bibfield{author}{\bibinfo{person}{John Power} {and} \bibinfo{person}{Hiroshi Watanabe}.} \bibinfo{year}{2002}\natexlab{}.
\newblock \showarticletitle{Combining a monad and a comonad}.
\newblock \bibinfo{journal}{\emph{Theoretical Computer Science}} \bibinfo{volume}{280}, \bibinfo{number}{1-2} (\bibinfo{year}{2002}), \bibinfo{pages}{137--162}.
\newblock
\href{https://doi.org/10.1016/S0304-3975(01)00024-X}{doi:\nolinkurl{10.1016/S0304-3975(01)00024-X}}


\bibitem[Rutten(2000)]%
        {rutten:tcs2000}
\bibfield{author}{\bibinfo{person}{Jan~J.~M.~M. Rutten}.} \bibinfo{year}{2000}\natexlab{}.
\newblock \showarticletitle{Universal coalgebra: a theory of systems}.
\newblock \bibinfo{journal}{\emph{Theoretical Computer Science}} \bibinfo{volume}{249}, \bibinfo{number}{1} (\bibinfo{year}{2000}), \bibinfo{pages}{3--80}.
\newblock


\bibitem[Sangiorgi(2011)]%
        {sangiorgi:bisbook}
\bibfield{author}{\bibinfo{person}{Davide Sangiorgi}.} \bibinfo{year}{2011}\natexlab{}.
\newblock \bibinfo{booktitle}{\emph{Introduction to Bisimulation and Coinduction}}.
\newblock \bibinfo{publisher}{Cambridge University Press}.
\newblock


\bibitem[Scott(1972)]%
        {scott:lnm1972}
\bibfield{author}{\bibinfo{person}{Dana~S. Scott}.} \bibinfo{year}{1972}\natexlab{}.
\newblock \showarticletitle{Continuous lattices}.
\newblock \bibinfo{journal}{\emph{Lecture Notes in Mathematics}}  \bibinfo{volume}{274} (\bibinfo{year}{1972}), \bibinfo{pages}{97--136}.
\newblock


\bibitem[Smyth and Plotkin(1982)]%
        {ps:siam1982}
\bibfield{author}{\bibinfo{person}{Michael~B. Smyth} {and} \bibinfo{person}{Gordon~D. Plotkin}.} \bibinfo{year}{1982}\natexlab{}.
\newblock \showarticletitle{The Category-Theoretic Solution of Recursive Domain Equations}.
\newblock \bibinfo{journal}{\emph{SIAM J. Comput.}} \bibinfo{volume}{11}, \bibinfo{number}{4} (\bibinfo{year}{1982}), \bibinfo{pages}{761--783}.
\newblock


\bibitem[Sokolova(2011)]%
        {sokolova:tcs2011}
\bibfield{author}{\bibinfo{person}{Ana Sokolova}.} \bibinfo{year}{2011}\natexlab{}.
\newblock \showarticletitle{Probabilistic systems coalgebraically: A survey}.
\newblock \bibinfo{journal}{\emph{Theoretical Computer Science}} \bibinfo{volume}{412}, \bibinfo{number}{38} (\bibinfo{year}{2011}), \bibinfo{pages}{5095--5110}.
\newblock


\bibitem[{Stacks Project Authors}(2016)]%
        {stacks-project}
\bibfield{author}{\bibinfo{person}{The {Stacks Project Authors}}.} \bibinfo{year}{2016}\natexlab{}.
\newblock \bibinfo{title}{\itshape Stacks Project}.
\newblock \bibinfo{howpublished}{\url{http://stacks.math.columbia.edu}}.
\newblock


\bibitem[Svendsen et~al\mbox{.}(2016)]%
        {ssb:esop2016}
\bibfield{author}{\bibinfo{person}{Kasper Svendsen}, \bibinfo{person}{Filip Sieczkowski}, {and} \bibinfo{person}{Lars Birkedal}.} \bibinfo{year}{2016}\natexlab{}.
\newblock \showarticletitle{Transfinite Step-Indexing: Decoupling Concrete and Logical Steps}. In \bibinfo{booktitle}{\emph{{ESOP}}} \emph{(\bibinfo{series}{Lecture Notes in Computer Science}, Vol.~\bibinfo{volume}{9632})}. \bibinfo{publisher}{Springer}, \bibinfo{pages}{727--751}.
\newblock


\bibitem[Turi and Plotkin(1997)]%
        {tp:lics1997}
\bibfield{author}{\bibinfo{person}{Daniele Turi} {and} \bibinfo{person}{Gordon Plotkin}.} \bibinfo{year}{1997}\natexlab{}.
\newblock \showarticletitle{Towards a mathematical operational semantics}. In \bibinfo{booktitle}{\emph{Proc.~LICS}}. \bibinfo{publisher}{IEEE Computer Society}, \bibinfo{pages}{280--291}.
\newblock


\bibitem[Urabe and Hasuo(2015)]%
        {hu:calco2015}
\bibfield{author}{\bibinfo{person}{Natsuki Urabe} {and} \bibinfo{person}{Ichiro Hasuo}.} \bibinfo{year}{2015}\natexlab{}.
\newblock \showarticletitle{Coalgebraic Infinite Traces and Kleisli Simulations}. In \bibinfo{booktitle}{\emph{{CALCO}}} \emph{(\bibinfo{series}{LIPIcs}, Vol.~\bibinfo{volume}{35})}. \bibinfo{publisher}{Schloss Dagstuhl - Leibniz-Zentrum fuer Informatik}, \bibinfo{pages}{320--335}.
\newblock


\bibitem[Urabe et~al\mbox{.}(2016)]%
        {hsu:concur2016}
\bibfield{author}{\bibinfo{person}{Natsuki Urabe}, \bibinfo{person}{Shunsuke Shimizu}, {and} \bibinfo{person}{Ichiro Hasuo}.} \bibinfo{year}{2016}\natexlab{}.
\newblock \showarticletitle{Coalgebraic Trace Semantics for Buechi and Parity Automata}. In \bibinfo{booktitle}{\emph{{CONCUR}}} \emph{(\bibinfo{series}{LIPIcs}, Vol.~\bibinfo{volume}{59})}. \bibinfo{publisher}{Schloss Dagstuhl - Leibniz-Zentrum fuer Informatik}, \bibinfo{pages}{24:1--24:15}.
\newblock


\bibitem[van Breugel(2001)]%
        {breugel:tcs2001}
\bibfield{author}{\bibinfo{person}{Franck van Breugel}.} \bibinfo{year}{2001}\natexlab{}.
\newblock \showarticletitle{An introduction to metric semantics: operational and denotational models for programming and specification languages}.
\newblock \bibinfo{journal}{\emph{Theoretical Computer Science}} \bibinfo{volume}{258}, \bibinfo{number}{1-2} (\bibinfo{year}{2001}), \bibinfo{pages}{1--98}.
\newblock


\bibitem[van Breugel(2012)]%
        {breugel:ipl2012}
\bibfield{author}{\bibinfo{person}{Franck van Breugel}.} \bibinfo{year}{2012}\natexlab{}.
\newblock \showarticletitle{On behavioural pseudometrics and closure ordinals}.
\newblock \bibinfo{journal}{\emph{Inform. Process. Lett.}} \bibinfo{volume}{112}, \bibinfo{number}{19} (\bibinfo{year}{2012}), \bibinfo{pages}{715--718}.
\newblock


\bibitem[van Breugel and Worrell(2005)]%
        {bw:tcs2005}
\bibfield{author}{\bibinfo{person}{Franck van Breugel} {and} \bibinfo{person}{James Worrell}.} \bibinfo{year}{2005}\natexlab{}.
\newblock \showarticletitle{A behavioural pseudometric for probabilistic transition systems}.
\newblock \bibinfo{journal}{\emph{Theoretical Computer Science}} \bibinfo{volume}{331}, \bibinfo{number}{1} (\bibinfo{year}{2005}), \bibinfo{pages}{115--142}.
\newblock


\bibitem[van Glabbeek et~al\mbox{.}(1990)]%
        {vsst:lics1990}
\bibfield{author}{\bibinfo{person}{Rob~J. van Glabbeek}, \bibinfo{person}{Scott~A. Smolka}, \bibinfo{person}{Bernhard Steffen}, {and} \bibinfo{person}{Chris M.~N. Tofts}.} \bibinfo{year}{1990}\natexlab{}.
\newblock \showarticletitle{Reactive, Generative, and Stratified Models of Probabilistic Processes}. In \bibinfo{booktitle}{\emph{{LICS}}}. \bibinfo{publisher}{IEEE Computer Society}, \bibinfo{pages}{130--141}.
\newblock


\bibitem[Watanabe(2002)]%
        {watanabe:entcs2002}
\bibfield{author}{\bibinfo{person}{Hiroshi Watanabe}.} \bibinfo{year}{2002}\natexlab{}.
\newblock \showarticletitle{Well-behaved Translations between Structural Operational Semantics}.
\newblock \bibinfo{journal}{\emph{Electronic Notes in Theoretical Computer Science}} \bibinfo{volume}{65}, \bibinfo{number}{1} (\bibinfo{year}{2002}), \bibinfo{pages}{337--357}.
\newblock
\href{https://doi.org/10.1016/S1571-0661(04)80372-4}{doi:\nolinkurl{10.1016/S1571-0661(04)80372-4}}


\bibitem[Worrell(1999)]%
        {worrell:entcs1999}
\bibfield{author}{\bibinfo{person}{James Worrell}.} \bibinfo{year}{1999}\natexlab{}.
\newblock \showarticletitle{Terminal sequences for accessible endofunctors}.
\newblock \bibinfo{journal}{\emph{Electronic Notes in Theoretical Computer Science}}  \bibinfo{volume}{19} (\bibinfo{year}{1999}), \bibinfo{pages}{24--38}.
\newblock


\bibitem[Worrell(2000)]%
        {worrell:phdthesis}
\bibfield{author}{\bibinfo{person}{James Worrell}.} \bibinfo{year}{2000}\natexlab{}.
\newblock \emph{\bibinfo{title}{On Coalgebras and Final Semantics}}.
\newblock \bibinfo{thesistype}{Ph.\,D. Dissertation}. \bibinfo{school}{Computer Laborarory, Oxford University}.
\newblock


\bibitem[Worrell(2005)]%
        {worrell:tcs2005}
\bibfield{author}{\bibinfo{person}{James Worrell}.} \bibinfo{year}{2005}\natexlab{}.
\newblock \showarticletitle{On the final sequence of a finitary set functor}.
\newblock \bibinfo{journal}{\emph{Theoretical Computer Science}} \bibinfo{volume}{338}, \bibinfo{number}{1-3} (\bibinfo{year}{2005}), \bibinfo{pages}{184--199}.
\newblock


\end{thebibliography}

\appendix
\clearpage

\section{Category-valued sheaves over sites}
\label{sec:sheaves-sites}

In this section we recall basic concepts of sheaf theory; we refer the reader to
\cite{gray:top1965,stacks-project,maclane:sheaves,johnstone:elephant1,johnstone:elephant2,bgv:sheaves} for a thorough introduction to the topic.
All the material is not original and can be found in the referred works and in any textbook on sheaf theory. The only content we were not able to directly find in the literature is in \ref{sec:sheaf-enrichment} where categories enriched over categories of sheaves are described. Nonetheless, basic definitions contained in that section are instances of standard notions from enriched category theory \cite{kelly:enrichedbook}.

\subsection{Sites}
\label{sec:sites}

Sites are a categorical generalisation of topological spaces and locales. Roughly speaking, sites are categories equipped with additional data describing how their objects can be ``covered'' by families of objects: on one hand the firsts provide ``well-behaved quotients'' of the seconds and, on the other hand, the latter provide ``localizations'' of the former. 

\begin{definition}
	\label{def:coverage}
	A \emph{coverage} on a category $\CS$ consists of a rule $J$ assigning to each object $U$ in $\CS$ a collection of families of morphisms $\{p_i\colon U_i \to U\}_{i \in I}$ called \emph{covering families} with the following property:
	if $\{p_i\colon U_i \to U\}_{i \in I}$ is a covering family and $g\colon V \to U$ is a morphism in $\CS$, then there exists a covering family $\{q_k\colon V_k \to V\}_{k \in K}$ such that each composite $g \circ q_k$ factors through some $p_i$ as depicted in the diagram below.
	\begin{equation}
		\label{eq:covering-factoring-condition}
		\begin{tikzpicture}[
				auto, scale=1.6, diagram font,
				baseline=(current bounding box.center)
			]	
						
			\node (n0) at (0,1) {\(V_k\)};
			\node (n1) at (1,1) {\(U_i\)};
			\node (n2) at (0,0) {\(V\)};
			\node (n3) at (1,0) {\(U\)};
			
			\draw[->] (n0) to  (n1);
			\draw[->] (n0) to node[swap] {\(q_k\)} (n2); 
			\draw[->] (n1) to node {\(p_i\)} (n3); 
			\draw[->] (n2) to node[swap] {\(g\)} (n3);
		\end{tikzpicture}
	\end{equation}
\end{definition}

A good source of examples are topologies and topological bases. 
A topological base for a set $S$ is a collection $\bopens[S]$ of subsets of $S$ (called basic open sets) subject to the following two requirements: 
\begin{enumerate}
	\item $S = \bigcup \bopens[S]$;
	\item for any $U_1$, $U_2$ in $\bopens[S]$, if $s \in U_1 \cap U_2$ then there is $U_0 \in \bopens[S]$ such that $s \in U_0$ and $U_0 \subseteq U_1 \cap U_2$.
\end{enumerate}
Given a topological base $\bopens[S]$, let $\CS$ be the poset $(\bopens[S],\subseteq)$ regarded as a thin category\footnote{A category is called thin or posetal whenever all parallel morphisms are equal.} and define the coverage $J$ on it as the function mapping every basic open set to the set of its basic open covers:
\begin{equation*}
	J(U) = \left\{\{U_{i} \to U\}_{i \in I} \,\middle\vert\, U = {\textstyle\bigcup_{i \in I}} U_i\right\}
	\text{.}
\end{equation*}
In order to prove that this function is indeed a coverage, we construct for each covering family on $U$ and $V \subseteq U$, a suitable covering family on $V$. For $\{U_{i} \to U\}_{i \in I} \in J(U)$ and $V \to U$, consider a family $\{V_{i,x} \to V\}$ such that for each $i \in I$ and $x \in V \cup U_i$, $V_{i,x}$ is a basic open with the property that $x \in V_{i,x}$ and $V_{i,x} \subseteq V \cup U_i$ (which exists by definition of base). Because $V \cap U_i = \cup_{x \in V \cap U_i} V_{i,x}$ and $V = \cup_{i \in I} V \cap U_i$, the family $\{V_{i,x} \to V\}$ belongs to $J(V)$. Finally, note that each $V_{i,k} \to V$ satisfies \eqref{eq:covering-factoring-condition} by construction. Coverages for topological spaces are obtained via the same construction.

\
A special case of bases are those closed under finite intersections such as the set $\left\{\{s\}^\downarrow \,\middle\vert\, s \in S\right\}$ of \emph{cones} in a preorder $(S,\leq)$.
From the point of view of coverages, these topological bases are closed under finite intersections whenever their coverages have pullbacks.
\begin{definition}
	A coverage $J$ on a category $\CS$ with enough pullbacks is said to \emph{have pullbacks} whenever it has the following property:
	if $\{p_i\colon U_i \to U\}_{i \in I}$ is a covering family and $g\colon V \to U$ is a morphism, then the family of pullbacks of $g$ along each $p_i$ is a covering family of $V$.
\end{definition}

Coverages associated to topological spaces and complete Heyting algebras are instances of a stronger notion of coverage known as \emph{Grothendieck coverage} or \emph{Grothendieck topology}. These are more conveniently presented in terms of particular covering families called (covering) sieves.

\begin{definition}
	\label{def:sieve}
	For $U$ an object in a category $\CS$, a \emph{sieve} $S$ on $U$ is a covering family on $U$ that is closed by post-composition \ie, 
	for all $p$ and $q$ with suitable domain and codomain it holds that:
	\begin{equation*}
		p \in S \implies p \circ q \in S
		\textbf{.}
	\end{equation*}
	The set $\{p \mid \cod(p) = U\}$ is the \emph{maximal sieve} on $U$.
	For $S$ a sieve in $U$ and $g\colon V \to U$ a morphism in $\CS$, $g^*(S)$ is the sieve on $V$ consisting of all morphisms $h$ such that $g\circ h$ factors through some morphism in $S$.
\end{definition}

\begin{definition}
	\label{def:grothendieck-coverage}
	A \emph{Grothendieck coverage} (or \emph{Grothendieck topology}) on a category $\CS$ is a rule $J$ mapping each object $U$ of $\CS$ to a collection $J(U)$ of sieves called \emph{covering sieves} on $U$ with the following properties:
	\begin{itemize}
		\item the maximal sieve on $U$ belongs to the collection $J(U)$;
		\item if $S \in J(U)$, then $g^*(S) \in J(V)$ for any arrow $g\colon V \to U$ in $\CS$;
		\item if $S \in J(U)$ and $S'$ is any sieve on $U$ such that for all $g\colon V \to U$ in $S$ the sieve $g^*(S')$ belongs to $J(V)$, then $S' \in J(U)$.
	\end{itemize}
\end{definition}

Akin to how topological bases canonically induce topologies, coverages canonically induce Grothendieck topologies. For $\{p_i\colon U_i \to U \}_{i \in I}$ a family of morphisms of $\CS$, define the \emph{sieve on $U$ generated by $K$} as the smallest sieve $S$ on $U$ such that each $p_i\colon U_i \to U$ belongs to $S$.
For $J$ a coverage on a category $\CS$, define $\tilde{J}$ as the rule assigning each object $U$ in $\CS$ to the collection of sieves $\tilde{J}(U)$ on $U$ with the following property: there exists a covering family $K = \{p_i\colon U_i \to U\}_{i \in I}$ in $J(U)$ such that the sieve on $U$ generated by $K$ is contained in $\tilde{J}(U)$. 

\begin{proposition}
	For $J$ a coverage on $\CS$, $\tilde{J}$ is the smallest Grothendieck coverage on $\CS$ with the property that every covering family of $J$ generates a covering sieve.
\end{proposition}

In virtue of this result, for $J$ a coverage $\tilde{J}$ is called its \emph{associated Grothendieck coverage}.

\begin{definition}
	\label{def:site}
	A \emph{site} is pair $(\CS,J)$ where $\CS$ is a category (referred as the category \emph{underlying} the site) and $J$ a coverage for $\CS$.
	A site is said to \emph{have pullbacks} whenever its coverage has pullbacks.
	A site is called a \emph{Grothendieck site} whenever its coverage is a Grothendieck coverage.
	A site is called \emph{small} if its underlying category is small.
\end{definition}

Sites are equipped with a suitable notion of morphisms given by functors between their underlying categories that are well-behaved with respect to the associated coverages.

\begin{definition}
	For $(\CS,J)$ and $(\CS',J')$ sites, a \emph{morphism of sites} $f\colon (\CS,J) \to (\CS',J')$ is a functor, denoted by $f$ as well, between their underlying categories and subject to the following conditions:
	\begin{itemize}
		\item It is covering-flat \ie for every finite category $\cat{I}$, diagram $D\colon \cat{I} \to \CS$ and cone $C$ for $f \circ D$ in $\CS'$ with vertex $U$, the family of morphisms (actually a sieve) with target $U$
		\begin{equation*}\left\{
			p\colon V \to U \,\middle\vert\, p \text{ factors through the $f$-image of some cone for $D$}
			\right\}
		\end{equation*}
		belongs to the collection $J'(U)$.
		\item It preserves covering families \ie for $\{p_i\colon U_i \to U\}$ a covering family in $J$, $\{f(p_i)\colon f(U_i) \to f(U)\}$ is a covering family in $J'$.
	\end{itemize}
	The functor $f\colon \CS \to \CS'$ is called \emph{underlying}.
\end{definition}

For $\bopens$ a topological base on a set $S$ and $\opens$ its generated topology, let $(\CS_{\bopens},J_{\bopens})$ and $(\CS_{\opens},J_{\opens})$ their corresponding sites.
The coverage $J_{\bopens}$ extends to a coverage $J_{\bopens}'$ on  $\CS_{\opens}$ given on basic opens as $J_{\bopens}$:
\begin{equation*}
	J_{\bopens}'(U) = 
	\begin{cases}
		J_{\bopens}(U) & \text{if } U \in \bopens \\
		\emptyset & \text{otherwise}
	\end{cases}
\end{equation*}
and such that the Grothendieck coverage $\tilde{J'_{\bopens}}$ associated to $J_{\bopens}'$ is $J_{\opens}$. Then, there is a morphism of sites $i\colon (\CS_{\bopens},J_{\bopens}) \to (\CS_{\opens},J_{\opens})$ whose underlying functor is given by the inclusion $\bopens \subseteq \opens$.

\phantomLabel{def:alexandrov-topology}%
The \emph{Alexandrov topology} on a set $S$ equipped with a preorder relation $\leq$ is the topology $\alexT{S,\leq}$ whose open sets are all the subsets of $S$ that downward closed: $\alexT{S,\leq} = \left\{ {X^\downarrow} \,\middle\vert\, X \subseteq S\right\}$ where ${X^\downarrow} = \{s \mid \exists x \in X \text{ s.t. } s \leq x\}$ is the downward closure of the subset $S$ of $S$. The smallest topological base for $\alexT{S,\leq}$ is the set $\left\{\{s\}^\downarrow \,\middle\vert\, s \in S \land \forall X \subseteq S\setminus\{s\}(s \bigvee X) \right\}$ of primitive cones in $(S,\leq)$. %
In several examples we will consider ordinal numbers (seen as sets of ordinals) equipped with the standard ordering (whence omitted) and their Alexandrov topology. In particular, for $\alpha$ an ordinal, $\alexT{\alpha}$ is the successor ordinal $\alpha + 1$ and, since ordinals are already cones, the smallest base for $\alexT{\alpha}$ is the set $\{ \beta+1 \mid \beta + 1 < \alpha \}$ of all successor ordinals in $\alpha$. We refer to this topological base as the \emph{successors base} for $\alpha$.
For instance, consider the first infinite ordinal $\omega$, that is the set of natural numbers under natural ordering, $\alexT{\omega}$ is $\omega + 1$ the set of natural numbers together with $\omega$ itself and the successor base for $\omega$ is given by all natural numbers except $0$ (for it is a limit ordinal).

\subsection{Sheaves}
\label{sec:sheaves}

\phantomLabel{def:presheaf}%
\phantomLabel{def:funcat}%
Let $\CC$ and $\CS$ be categories. A \emph{($\CC$-valued) presheaf on $\CS$} is any contravariant functor from $\CS$ to $\CC$. Presheaves and natural transformations form the category $\PSh[\CC]{\CS}$ \ie the category of (covariant) functors $\Func{\CS\op}{\CC}$.
Objects of $\CS$ are called \emph{stages} and will be usually denoted by letter $U$, $V$ and variations thereof.
For $X$ a presheaf on $\CS$, the object $X(U)$ (written also as $X_U$) is called the \emph{object of sections} or \emph{value} of $X$ at stage $U$ and the morphism $X(r)\colon X(V) \to X(U)$ (written also as $X_r$) is called the \emph{restriction} morphism from $V$ to $U$. When values are taken in a concrete category, $x \in X(U)$ is called \emph{section} of $X$ at $U$.
\phantomLabel{def:morphism-in-thincat}%
\phantomLabel{def:section-restriction}%
If $\CS$ is thin, then we denote any $U \to V$ as $\res{U}{V}$ and, if in addition $\CC$ is concrete, we write $x|_U$, instead of $X_{\res{U}{V}}(x)$, for action of restriction maps on sections.

Sheaves on a site are presheaves that are well-behaved with respect to the coverage for their site where ``well-behaved'' means values at any stage $U$ are given by families of values on any covering on $U$. Reworded, they are presheaves that transport covering families to limits.
\begin{definition}
	\label{def:sheaf}
	A presheaf $X$ on a site $(\CS,J)$ with values on a category $\CC$ is called a \emph{sheaf} if for every covering family $\{p_i\colon U_i \to U\}_{i \in I}$ the family of restriction morphisms $\{X(p_i)\colon X(U) \to X(U_i)\}_{i \in I}$ induces an isomorphism: 
	\begin{equation*}
		X(U) \cong \varprojlim_{i \in I} X(U_i)
		\text{.}
	\end{equation*}
	Sheaves on a site $(\CS,J)$ form the full subcategory	$\Sh[\CC]{\CS,J} \embto \PSh[\CC]{\CS}$.
\end{definition}
In the sequel we adopt the standard convention of omitting the category of values for (pre)sheaves of sets and thus write $\PSh{\CS}$ and $\Sh{\CS,J}$ instead of $\PSh[\Set]{\CS}$ and $\Sh[\Set]{\CS,J}$, respectively.

\paragraph{Sheaves on topological spaces}
Sites defined from topological spaces are such that their covering families are colimiting cones. It follows that sheaves on sites of this kind are presheaves transporting colimits induced by open covers to limits.

\begin{proposition}
	Let $(\CS,J)$ be a site induced by some topological space.
	A presheaf $X$ of $\PSh[\CC]{\CS}$ is a sheaf of $\Sh[\CC]{\CS,J}$ if, and only if, for every complete full subcategory $I\colon \cat{U} \embto \CS$:
	\begin{equation*}
		X \varinjlim I  \cong \varprojlim X\circ I
		\text{.}
	\end{equation*}
\end{proposition}

Akin to topologies and topological bases, several properties of sheaves on topological spaces can be reduced to statements about sheaves on a base generating that topology. Formally, this situation corresponds to an equivalence of categories induced by the inclusion of a base into its generated topology, as stated by \cref{thm:shaves-topological-base-equivalence}.

\begin{proposition}
	\label{thm:shaves-topological-base-equivalence}
	For $\bopens$ a topological base on a set $S$, $\opens$ the topology generated by $\bopens$, and $i\colon (\CS_{\bopens},J_{\bopens}) \to (\CS_{\opens},J_{\opens})$ the inclusion morphisms for their associated sites, the following is an equivalence of categories:
	\begin{equation*}
		\Sh[\CC]{\CS_{\opens},J_{\opens}} \xrightarrow{(-\,\circ\,i)} \Sh[\CC]{\CS_{\bopens},J_{\bopens}}
		\text{.}
	\end{equation*}
\end{proposition}

Examples of this situation are sheaves on Alexandrov topologies for partial orders and sheaves on their bases of primitive cones. In particular, for $\alpha$ an ordinal, $\bopens(\alpha)$ the set $\{\beta+1 \mid \beta +1 < \alpha\}$ of all successor ordinals in $\alpha$, the inclusion $\bopens(\alpha)\subseteq \alexT{\alpha}$ yields an equivalence for the categories $\Sh[\CC]{\alexT{\alpha}}$ and $\Sh[\CC]{\bopens(\alpha)}$. For the sake of conciseness, we will write $\Sh[\CC]{\alpha}$ for the category of sheaves on $\alexT{\alpha}$ and on $\bopens(\alpha)$ when confusion is unlikely.

\paragraph{Sheaves on sites with pullbacks}
If the category of values has enough products, sheaves on sites that have pullbacks are precisely presheaves that satisfy the (familiar) \emph{descent condition}.
\begin{proposition}
	Let $(\CS,J)$ be a site that has pullbacks and $\CC$ be a category with products. A $\CC$-valued presheaf $X$ on $(\CS,J)$ is a sheaf if, and only if, for $\{p_i\colon U_i \to U\}_{i \in I}$ a covering family, the following diagram is an equaliser:
	\begin{equation*}
		X(U) \to \prod_{i \in I} X(U_i) \rightrightarrows \prod_{i,j \in I} X(U_i) \times_{U} X(U_j)
		\text{.}
	\end{equation*}
\end{proposition}

\paragraph{Sheaves of sets}
Sheaves taking sets as values are often introduced by the following component-wise definition based on \emph{matching families of sections} \ie families of sections that are pair-wise compatible with respect to restrictions sharing their target stage.

\begin{proposition}
	Let $(\CS,J)$ be a small site.
	A presheaf $X$ on $(\CS,J)$ and with values in $\Set$ is a sheaf if, and only if, for every covering family $\{p_i\colon U_i \to U\}$ and for every family of sections $\{x_i\}_{i \in I}$ with the property (called \emph{matching}) that $x_i \in X(U_i)$ and $X(g)(x_i) = X(h)(x_j)$ for all $g\colon V \to U_i$, $h\colon V \to U_j$, and $i,j \in I$, then	there is a unique element $x \in X(U)$ such that $X(p_i)(x) = x_i$.
\end{proposition}

\subsection{Associated and constant sheaves}
\label{sec:sheafification}

Let $(\CS,J)$ be a site and $\CC$ a category.
For a presheaf $X$ in $\PSh[\CC]{\CS}$ let $X^+$ denote, when defined, the presheaf given on each stage as: 
\begin{equation*}
	X^+(U) \cong \varinjlim_{K \in J(U)}\varprojlim_{V \to U \in K} X(V)
\end{equation*}
where the colimit is taken over covering families on $U$ and the limits over morphism in the covering. This construction is usually known as the \emph{plus construction} and, in general, it is not always possible for it relying on the existence of enough (co)limits in the category of values. A sufficient condition on $\CC$ is that it is a (co)complete category but weaker conditions can be derived for particular sites of interest. For instance, if $(\CS,J)$ corresponds to the Alexandrov topology for an ordinal $\alpha$, then it suffices to assume $\CC$ has limits of $\gamma$-sequences for $\gamma \leq \alpha$.

Iterating the plus construction always results in sheaves for the given site thus universally associating sheaves to presheaves.
For $X$ any presheaf, the sheaf $X^{++}$ is called the \emph{associated sheaf} or \emph{sheafification} of $X$.

\begin{proposition}
	\label{thm:sheafification}
	For $(\CS,J)$ a site and $\CC$ a category, the following are equivalent:
	\begin{itemize}
		\item every presheaf in $\PSh[\CC]{\CS}$ admits sheafification;
		\item the inclusion $\mathbf{i}\colon \Sh[\CC]{\CS,J} \to \PSh[\CC]{\CS}$ exhibits $\Sh[\CC]{\CS,J}$ as a reflective subcategory of $\PSh[\CC]{\CS}$.
	\end{itemize}
\end{proposition}

\phantomLabel{def:sheafification}%
Whenever it exists, the left adjoint to $\mathbf{i}$ is denoted as $\mathbf{a}$ and called \emph{associated sheaf functor} or \emph{sheafification functor}. 
For this functor to exists it suffices to assume $(\CS,J)$ small and all small (co)limits in $\CC$.

\phantomLabel{def:constant-sheaf}%
There is an inclusion of the category of values $\CC$ into $\PSh[\CC]{\CS}$ by means of $(-\circ!_{\CS})\colon \CC \to \PSh[\CC]{\CS}$ where $!_{\CS}\colon \CS \to \cat{1}$ is the final morphism in $\Cat$. This functor is denoted by $\Delta$ and called \emph{constant presheaf functor} for it assigns each object $Y$ in $\CC$ to the constant presheaf of $Y$:
\begin{equation*} \Delta(Y)(U) = Y \qquad \Delta(Y)(r) = \id_Y\end{equation*}
and each morphism $f\colon Y \to Y'$ to the natural transformation whose components are all $f$. Whenever $\CC$ has limits for diagrams of type $\CS\op$, the constant presheaf functor $\Delta$ has a right adjoint $\Gamma\colon \PSh[\CC]{\CS} \to \CC$ called \emph{global sections functor} for it maps each presheaf to the object of its global sections \ie it maps each presheaf $X$ the limit of a diagram $X\colon \CS\op \to \CC$.
By composition with the sheafification adjunction $(\mathbf{a} \dashv \mathbf{i})\colon \Sh[\CC]{\CS,J} \to \PSh[\CC]{\CS}$, the constant presheaf adjunction $(\Delta \dashv \Gamma)$ restricts to the subcategory of sheaves $\Sh[\CC]{\CS,J}$.
As common practice, we will abuse the notation and just write $\Delta$ and $\Gamma$ instead of $\mathbf{a} \circ \Delta$ and $\Gamma \circ \mathbf{i}$.
The terminology for $\Delta$ and $\Gamma$ is extended accordingly to sheaves:
for $X$ an object of $\CC$, the sheaf $\Delta (X)$ is called constant sheaf of $X$ and $\Delta$ constant sheaf functor, for $Y$ a $\CC$-valued sheaf on $(\CS,J)$, $\Gamma (Y)$ is the object of global sections of $Y$ and $\Gamma$ is called the global sections functor.

\subsection{Enrichment over sheaf categories}
\label{sec:sheaf-enrichment}

In this subsection we state some basic definitions about categories enriched over categories of sheaves which will be needed in order to introduce locally contractive functors and in the remaining of this work as well. Although we were not able to find published work that explicitly introduces this blend of enriched categories and functors, all definitions are obtained as instance of standard notions from enriched category theory; we refer the interested reader to \citeauthor{kelly:enrichedbook}'s body of work, especially \cite{kelly:enrichedbook}.

Let $(\CS,J)$ be a site. A category enriched over $\Sh{\CS,J}$ $\CC$ is characterised by the following data:
\begin{itemize}
	\item a collection $\obj(\CC)$ of objects,
	\item a sheaf $\CC(X,Y)$ over $(\CS,J)$ for each pair of objects $X,Y \in \obj(\CC)$,
	\item a point $\id_X\colon 1 \to \CC(X,X)$ in $\Sh{\CS,J}$ determining the identity for each $X \in \obj(\CC)$,
	\item a morphism $\circ_{X,Y,Z}\colon \CC(Y,Z) \times \CC(X,Y) \to \CC(X,Z)$ in $\Sh{\CS,J}$ determining composition for each $X,Y,Z \in \CC$ such that it is associative
	\begin{equation*}
		\begin{tikzpicture}[
				auto, xscale=3, yscale=1.6, diagram font,
				baseline=(current bounding box.center)
			]

			\node (n0) at (0,2) {\((\CC(Y,Z) \times \CC(X,Y)) \times \CC(W,X)\)};
			\node (n1) at (0,1) {\(\CC(X,Z) \times \CC(W,X)\)};
			\node (n2) at (1,0) {\(\CC(W,Z)\)};
			\node (n3) at (2,1) {\(\CC(Y,Z) \times \CC(W,Y)\)};
			\node (n4) at (2,2) {\((\CC(Y,Z) \times (\CC(X,Y) \times \CC(W,X))\)};
			
			\draw[]   (n0) to node {\(\cong\)} (n4);
			\draw[->] (n0) to node[swap] {\({\circ_{X,Y,Z}} \times \id_{\CC(W,X)}\)} (n1); 
			\draw[->] (n1) to node[swap] {\(\circ_{W,X,Z}\)} (n2); 
			\draw[<-] (n2) to node[swap] {\(\circ_{W,Y,Z}\)} (n3); 
			\draw[<-] (n3) to node[swap] {\(\id_{\CC(Y,Z)} \times {\circ_{W,X,Y}}\)} (n4); 
		\end{tikzpicture}
	\end{equation*}
	and has the points from above as identities
	\begin{equation*}
		\begin{tikzpicture}[
				auto, xscale=3, yscale=1.6, diagram font,
				baseline=(current bounding box.center)
			]	
			\node (n1) at (1,1) {\(\CC(Y,Y) \times \CC(X,Y)\)};
			\node (n2) at (1,0) {\(\CC(X,Y)\)};
			\node (n3) at (0,0) {\(1 \times \CC(X,Y)\)};
			
			\draw[->] (n1) to node {\(\circ_{X,Y,Y}\)} (n2); 
			\draw[->] (n3) to node {\(\id_{Y} \times \id_{\CC(X,Y)}\)} (n1); 
			\draw[->] (n3) to node[swap] {\(\cong\)} (n2); 
		\end{tikzpicture}
		\qquad
		\begin{tikzpicture}[
				auto, xscale=3, yscale=1.6, diagram font,
				baseline=(current bounding box.center)
			]	
			\node (n1) at (0,1) {\(\CC(X,Y) \times \CC(X,X)\)};
			\node (n2) at (0,0) {\(\CC(X,Y)\)};
			\node (n3) at (1,0) {\(\CC(X,Y) \times 1\)};
			
			\draw[->] (n1) to node[swap] {\(\circ_{X,X,Y}\)} (n2); 
			\draw[->] (n3) to node[swap] {\(\id_{\CC(X,Y)} \times \id_{X}\)} (n1); 
			\draw[->] (n3) to node {\(\cong\)} (n2); 
		\end{tikzpicture}
	\end{equation*}
\end{itemize}
Note that sheaf enriched categories do not have ``morphisms'' for their hom-objects do not have proper elements. It is convenient however convenient to be able to speak about ``morphisms'' especially in order to express ``diagrams'' in sheaf enriched categories in a convenient way. For an example of what a ``diagram'' looks like in this setting, see \eqref{eq:sheaf-enriched-naturality} which expresses naturality condition. besides these practical conveniences, the ability to express what a ``collection of morphisms'' is in sheaf enriched settings is of relevance to the definition of natural transformations. Elements of a sheaf $X$ are understood as its points \ie morphisms from the final objects $x\colon 1 \to X$. These are defined as the ``morphisms'' of a sheaf enriched category. It follows that there are several distinct enriched categories that present us with the same morphisms; this happens because restriction maps of hom-sheaves might not be subjections and hence hom-sheaves are not defined by their points. Nonetheless, there is a ``minimal'' category presenting a given collection of morphism: this category has the property that its hom-sheaves are subobjects of those of any other category with the same morphisms. If one is interested in morphisms per se, then restriction to these minimal categories does not introduce any loss of generality.

\phantomLabel{def:externalisation-sh-category}%
For $\CC$ a category enriched over $\Sh{\CS,J}$, we can associate to $\CC$ a category $\underlying{\CC}$, called its \emph{externalisation} or its \emph{underlying} category, defined by the following data:
\begin{itemize}
\item objects $\obj(\underlying{\CC}) \defeq \obj(\CC)$;
\item hom-set $\underlying{\CC}(X,Y) \defeq \Sh{\CS,J}(1,\CC(X,Y))$ for each pair of objects $X,Y$.
\end{itemize}
Note that hom-sets are defined as the sets of points into hom-sheaves \ie the notion of morphisms in sheaf enriched categories. From this perspective, the ``minimal'' category mentioned at the end of the previous paragraph is the initial object in the thin category whose objects of categories with the same externalisation and whose morphisms are inclusions. 
As common practice, we will often abuse notation and terminology by referring to an enriched category and its externalisation as if they were the same entity---provided the distinction is clear from the context. In practice, we will call a (genuine) category $\Sh{\CS,J}$-enriched if it is the externalisation of some $\Sh{\CS,J}$-enriched category and write $\CC$, instead of $\underlying{\CC}$, for the externalisation of $\CC$.

Examples of categories enriched over categories of $\Set$-valued sheaves are sheaf categories themselves. Formally, this means that for any $\Sh[\CC]{\CS,J}$ there is a $\Sh{\CS,J}$-enriched category  such that its externalisation is $\Sh[\CC]{\CS,J}$.

\begin{lemma}
	\label{thm:c-sheaves-sheaf-enriched}
	For $\CC$ a category and $(\CS,J)$ a site, the category $\Sh[\CC]{\CS,J}$ is enriched over $\Sh{\CS,J}$.
\end{lemma}

\begin{proofatend}
	For $X,Y \in \Sh[\CC]{\CS,J}$, the hom-sheaf $\Sh[\CC]{\CS,J}(X,Y)$ takes each object $U$ in $\CS$ to the value
	\begin{equation*}
		\Sh[\CC]{\alpha}(X,Y)(U) = \{(f,f_{U}) \mid f \in \Sh[\CC]{\CS,J}(X,Y)\}
	\end{equation*}
	and each morphism $p\colon U \to V$ in $\CS$ to the restriction
	\begin{equation*}
		\Sh[\CC]{\alpha}(X,Y)(p)(f,f_{V}) = (f,f_{U})
		\text{.}
	\end{equation*}
	We remark that the use of pairs morphism-component is crucial to the definition restriction maps: the fact that two morphisms share their component at a given stage, say $U$, does not imply that they do the same for components at each stage $V$ such that $V \to U \in \CS$ whence information about which transformation a component belongs to has to be included.
	The sheaf condition follows from definition of morphism of $\CC$-valued sheaves.
	For $X \in \Sh[\CC]{\CS,J}$, the point $\id_X\colon 1 \to \Sh[\CC]{\alpha}(X,X)$ determines at each stage $U$ to the identity on the object of sections at $U$:
	\begin{equation*}
		\id_{X,U}(*) = (\id_X, \id_{X(U)})
		\text{.}
	\end{equation*}
	For $X,Y,Z \in \Sh[\CC]{\CS,J}$, components of $(-\circ_{X,Y,Z}-)$ apply composition of arrows in $\CC$ as $(f \circ_{X,Y,Z} g)_U = (f \circ g, f_U \circ g_U)$.
	Diagrams for associativity and identities readily follow from the analogous properties of composition in $\CC$.
\end{proofatend}

A $\Sh{\CS,J}$-enriched functor $F\colon \CC \to \CD$ between $\Sh{\CS,J}$-enriched categories is a functorial mapping such that for every pair of objects $X, Y$ in $\CC$ the assignment $F_{X,Y}\colon \CC(X,Y) \to \CD(FX,FY)$ is a morphism of sheaves in $\Sh{\CS,J}$. Unless otherwise stated, we implicitly assume domain and codomain of an enriched functor to be similarly enriched. 
\phantomLabel{def:sh-cat}%
Categories and functors enriched over $\Sh{\CS,J}$ form the category $\VCat{\Sh{\CS,J}}$.
For $F,G\colon \CC \to \CD$ functors enriched over $\Sh{\CS,J}$, an enriched natural transformation $\rho\colon F \to G$ is a family of morphisms 
\begin{equation*}
	\{\rho_{X} \colon 1 \to \CD(FX,GX)\}_{X \in \CC}
\end{equation*}
indexed over $\obj(\CC)$ and such that for any pair of objects $X,Y \in \CC$, the following naturality diagram commutes:
\begin{equation}
  \label{eq:sheaf-enriched-naturality}
	\begin{tikzpicture}[
			auto, xscale=3.5, yscale=1.6, diagram font,
			baseline=(current bounding box.center)
		]	
		
		\foreach \x in {0,1,...,5}{
			\pgfmathsetmacro\a{-90-\x * 60}
			\coordinate (i\x) at (\a:1);
		}
		
		\node (n0) at (i0) {\(\CD(FX,GY)\)};
		\node (n1) at (i1) {\(\CD(FY,GY)\times\CD(FX,FY)\)};
		\node (n2) at (i2) {\(1 \times \CC(X,Y)\)};
		\node (n3) at (i3) {\(\CC(X,Y)\)};
		\node (n4) at (i4) {\(\CC(X,Y) \times 1\)};
		\node (n5) at (i5) {\(\CD(GX,GY)\times\CD(FX,FX)\)};
		
		\draw[->] (n1) to node[swap] {\(-\circ_{FX,FY,GY}-\)} (n0);
		\draw[->] (n5) to node {\(-\circ_{FX,GX,GY}-\)} (n0);
		\draw[->] (n2) to node[swap] {\(\rho_Y \times F_{X,Y}\)} (n1);
		\draw[->] (n4) to node {\(G_{X,Y} \circ \rho_X\)} (n5);
		\draw[->] (n3) to node[swap] {\(\cong\)} (n2);
		\draw[->] (n3) to node {\(\cong\)} (n4);
		
	\end{tikzpicture}
\end{equation}
\phantomLabel{def:sh-funcat}%
For $\CC$ and $\CD$ enriched over $\Sh{\CS,J}$, enriched functors and natural transformations form the $\Sh{\CS,J}$-enriched category $\VFunc{\Sh{\CS,J}}{\CC}{\CD}$ whose hom-sheaves are given on each pair of functors $F,G\colon \CC \to \CD$ as the sheaf taking value $\{(\rho,\{\rho_{X,U}\}_{X \in \CC}) \mid \rho\colon F \to G \}$ at stage $U \in \CS$. In order to keep the notation concise, we shall write $\Func{\CC}{\CD}$ instead of $\VFunc{\Sh{\CS,J}}{\CC}{\CD}$, provided enrichment is clear from the context.

\section{Algebraically compact functors and categories}
\label{sec:algebraically-compact-functors}
\label{sec:locally-contractive-functors}

In \cite{freyd:ct1991,barr:tcs1993} it is shown that for any given endofunctor there is a unique and canonical morphism from its initial to final invariants \ie (co)algebras---provided both exists. In \cite{freyd:ct1991}, \citeauthor{freyd:ct1991} termed \emph{algebraically compact} categories for which that morphism always exists and is an isomorphism. 
In \cite{barr:jpaa1992} \citeauthor{barr:jpaa1992} observed that a category rarely is algebraically compact in \citeauthor{freyd:ct1991}'s sense. Instead, he suggested to treat algebraic compactness as a property of individual functors, or classes of functors, since in several occasions it might be worth restricting the attention to classes of functors that are ``relevant'' to specific situations \eg, when working in enriched settings. In \cite{freyd:ct1991,freyd:lmms1992,barr:jpaa1992} the term \emph{algebraically complete} is introduced to indicate existence of initial invariants; clearly algebraic compactness implies algebraic completeness.

\begin{definition}
	An endofunctor $F$ is called:
	\begin{itemize}
		\item
			\emph{algebraically complete} if there is an initial $F$-algebra;
		\item
			\emph{coalgebraically cocomplete} if there is a final $F$-coalgebra;
		\item
			\emph{algebraically compact} if there is an initial $F$-algebra, a final $F$-coalgebra, and they are canonically isomorphic.
	\end{itemize}
\end{definition}

The terminology is extended to classes of functors and categories in the obvious way.

\begin{definition}
	For a class $\cat{E}$ of endofunctors over $\CC$, the category $\CC$ is said:
	\begin{itemize}
		\item
			\emph{algebraically complete with respect to \cat{E}} if every functor in $\cat{E}$ is algebraically complete;
		\item
			\emph{coalgebraically cocomplete with respect to \cat{E}} if every functor in $\cat{E}$ is coalgebraically cocomplete;
		\item
			\emph{algebraically compact with respect to \cat{E}} if every functor in $\cat{E}$ is algebraically compact.
	\end{itemize}
\end{definition}

The vast majority of works and results about algebraic compactness (or that rely on it) consider two main classes of functors: \emph{locally continuous functors} and \emph{locally contractive functors}.
Historically, the former is the first non-trivial class of algebraically compact functors identified (see \cite{barr:jpaa1992}) and was initially studied as part of a categorical generalisation of order-theoretic constructions used in domain theory, especially \citeauthor{scott:lnm1972}'s limit-colimit coincidence result \cite{scott:lnm1972}. Nowadays locally continuous functors are fruitfully applied to a broad class of problems besides domain theoretic ones, see \eg \cite{hjs:lmcs2007,hsu:concur2016,capretta:tcs2011,ak:fossacs2016,hs:mscs2010,hs:tcs2011,hj:entcs2011,bmp:jlamp2015,bp:lmcs2019,bp:concur2016}. The class of locally contractive functors was introduced more recently as the technical foundation of guarded recursion and guarded type theory \cite{bmss:lmcs2012,bbs:lmcs2013,am:icfp2013,litak:ocl2014,bbm:tlca2014,pmb:entcs2015,mp:lics2015,bbcgsv:csl2016,ssb:esop2016}.

\paragraph{Locally contractive functors}

Let $A$ be a complete Heyting algebra and write $\PSh[\CC]{A}$ and $\Sh[\CC]{A}$ for the categories of $\CC$-valued (pre)sheaves on the Grothendieck site associated to $A$. This topology is often called \emph{sup topology} for its sieves are such to cover their supremum: the coverage on $A$ (regarded as a thin category) is the function mapping an element $a \in A$ to the set $\left\{ (A')^{\downarrow} \,\middle\vert\, A' \subseteq A \land a = \bigvee A' \right\}$.
 
Following \cite{dgm:fossacs2004,bmss:lmcs2012} define the \emph{predecessor map} on $A$ $\mathbf{p}\colon A \to A$ as:
\begin{equation*}
	\mathbf{p}(a) \defeq \bigvee\left\{b \in B \mid b < a \right\}
	\text{.}
\end{equation*}
where $B$ is a base for $A$ \ie any subset of $A$ such that each element is a supremum for the set of elements from the base that are less or equal to it: 
\phantomLabel{def:predecessor-functor}
\begin{equation*}
	a \in A \implies a = \bigvee \left\{b \in B \mid b \leq a \right\}
	\text{.}
\end{equation*}
The predecessor map induces an endofunctor $\mathbf{p}^\ast$ over the presheaf category $\PSh[\CC]{A}$ and defined as its inverse image $\mathbf{p}^\ast(X) = X \circ \mathbf{p}$. Restriction morphisms induce a natural transformation $\nxt^{\mathbf{p}}\colon \Id \to \mathbf{p}^\ast$ whose components are given, on each presheaf $X$ and stage $a$, as $\nxt^{\mathbf{p}}_{X,a} = X_{\iota_{\mathbf{p}(a),a}}$ and such that $(\mathbf{p}^\ast,\nxt^{\mathbf{p}})$ is well-pointed\footnote{A well-pointed functor is any endofunctor $F$ equipped with a natural transformation $\eta\colon \Id \to F$ called point and such that $\eta \circ \id_F = \id_F \circ \eta$.}. 
Assume $(\mathbf{a} \dashv \mathbf{i})\colon \PSh[\CC]{A} \to \Sh[\CC]{A}$, then the predecessor endofunctor over $\PSh[\CC]{A}$ induces an endofunctor $\later$ over $\Sh[\CC]{A}$ as the restriction:
\phantomLabel{def:later-functor}
\begin{equation*}
	\later = \mathbf{a} \circ \mathbf{p}^\ast \circ \mathbf{i}\text{.}
\end{equation*}
This endofunctor is called \emph{later} in contexts where stages describe future words.
Similarly to the predecessor endofunctor $\mathbf{p}^\ast$, $\later$ is well-pointed when equipped with a point $\nxt\colon \Id \to \later$ defined as the composite $\eta \bullet \nxt^{\mathbf{p}}$ where $\bullet$ denotes vertical composition in the 2-category $\Cat$ and $\eta\colon \Id \to \mathbf{a} \circ \mathbf{i}$ is the unit of the associated sheaf adjunction. 
The functor $\later$ preserves 
all limits in $\Sh[\CC]{A}$.

For $\CC$ $\Sh{A}$-enriched define $\latercat{\CC}$ as the $\Sh{A}$-enriched category given by the following data:
\begin{itemize}
	\item
 		the objects of $\CC$, $\obj(\latercat{\CC}) = \obj(\CC)$,
	\item
		for any pair of objects $X, Y \in \obj(\CC)$, the sheaf $\latercat{\CC}(X,Y) = \later\CC(X,Y)$,
	\item
		for each object $X \in \obj(\CC)$, the point $\nxt_{\CC(X,Y)} \circ \id_X \colon 1 \to \latercat{\CC}(X,X)$ where $\id_X \colon 1 \to \CC(X,X)$ is the identity on $X$ in $\CC$,
	\item
		for each $X,Y,Z \in \CC$, the morphism
		\begin{equation*}
			\later\CC(Y,Z) \times \later\CC(X,Y) \xrightarrow{\cong} \later(\CC(Y,Z) \times \CC(X,Y)) \xrightarrow{\later(-\circ_{X,Y,Z}-)} \later\CC(X,Z)
		\end{equation*}
		where $\circ_{X,Y,Z}\colon \CC(Y,Z) \times \CC(X,Y) \to \CC(X,Z)$ is composition in $\CC$.
\end{itemize}
The natural transformation $\nxt\colon \Id_{\Sh{A}} \to \later$ induces a $\Sh{A}$-enriched functor $\nxt\colon \CC \to \latercat{\CC}$ acting as the identity on objects and as the components of $\nxt$ on hom-sheaves.

\begin{definition}
  \label{def:locally-contractive-functor}
	A \emph{locally contractive} functor is any $\Sh{A}$-enriched functor $F\colon \CC \to \CD$ that factors as a composition of functors enriched over $\Sh{\alpha}$:
	\begin{equation*}
		\CC \xrightarrow{\nxt\ } \latercat{\CC} \longrightarrow \CD \text{.}
	\end{equation*}
\end{definition}

The later endofunctor is always locally contractive and so is any composite $F \circ \later$ where $F$ is enriched over $\Sh{A}$. In general, local contractiveness is preserved by composition, dualisation, and products. 
\begin{lemma}
	\label{thm:locally-contractive-ops}
	For $A$ a complete Heyting algebra, the following statements are true.
	\begin{enumerate}
		\item The endofunctor $\later \colon \Sh[\CC]{A} \to \Sh[\CC]{A}$ is locally contractive.
		\item For $F$ and $G$ componible and $\Sh{A}$-enriched, if $F$ or $G$ is locally contractive then $F \circ G$ is locally contractive.
		\item For $F$ and $G$ locally contractive, $F \times G$ is locally contractive.		
		\item For $F$ locally contractive, $F\op$ is locally contractive.
	\end{enumerate}
\end{lemma}

For locally contractive functors, (co)algebraic (co)completeness and algebraic compactness coincide for these functors admit at most one invariant, up to isomorphism. In particular, for $F$ locally contractive and $X$ such that $X\cong F(X)$, the isomorphism identifies an initial $F$-algebra and a final $F$-coalgebra. Furthermore, $X \cong Y$ whenever $Y \cong F(Y)$.

\begin{lemma}[\cite{bmss:lmcs2012}]
	\label{thm:locally-contractive-unique}
	Let $F$ be a locally contractive endofunctor over a category $\CC$. If $F$ has an invariant object, then it is unique up to isomorphism.
\end{lemma}

\begin{definition}
	A sheaf enriched category $\CC$ is called \emph{contractively compact} if it is algebraically compact with respect to locally contractive functors.
\end{definition}

In \cite{bmss:lmcs2012} \Citeauthor{bmss:lmcs2012} identify conditions on the underlying category and on the Heyting algebra that are sufficient for algebraic compactness: completeness and well-foundedness.

\begin{proposition}[\cite{bmss:lmcs2012}]
	\label{thm:sh-contractively-complete}
	For $A$ a complete Heyting algebra with a well-founded base and $\CC$ a category enriched over $\Sh{A}$, if (the externalisation of) $\CC$ is complete then it is contractively complete.
\end{proposition}

Examples of contractively compact categories are categories of sheaves on Alexandrov topologies induced by ordinal numbers and that take values in a complete category such as $\Set$, $\Cpo$, $\Cpob$, $\cat{Top}$, and $\Meas$.

\begin{remark}
	Results and constructions described in this section were presented in \cite{bmss:lmcs2012,bbs:lmcs2013} in the more general setting of categories enriched over models of guarded terms. A concrete instance of these model categories are sheaves over well-founded complete Heyting algebras and we preferred to focus our exposition to these models because they can be presented without formally introducing models of guarded terms and because
	they provide a setting that is sufficiently general for the aims of this work.
\end{remark}

\omittedproofs[\section{Omitted proofs}\label{sec:omittedproofs}]

\twocolumn[\section*{Index of notation}]

\newlength\ionLengthA
\newlength\ionLengthB
\newcolumntype{P}[1]{>{\RaggedRight}p{#1}}
\tabcolsep=3pt
\xentrystretch{0.1}
\renewcommand{\arraystretch}{1.1}

\newcommand{\ionAdjustWidth}[1]{
	\settowidth\ionLengthA{#1}
	\setlength\ionLengthB{\dimexpr\columnwidth-0.5\columnsep-2\tabcolsep-\ionLengthA-1pt\relax}
}

\subsection*{Arrows}
\ionAdjustWidth{$(-)^{\downarrow}$}
\begin{xtabular}{ p{\ionLengthA} p{\ionLengthB} }
	$\monoto$ & monomorphism \\
	$\epito$  & epimorphism \\
	$\embto$  & inclusion \\
	$\xrightarrow{\cong}$   & isomorphism \\
	$\cong$   & isomorphism \\
	$\mapsto$ & effect of a map on an element \\
\end{xtabular}

\subsection*{Universals}
\ionAdjustWidth{$\langle f, g\rangle$}
\begin{xtabular}{ p{\ionLengthA} p{\ionLengthB} }
	$0$ & initial object\\
	$1$ & final object\\
	$?_X$ & initial morphism to $X$\\
	$!_X$ & final morphism from $X$\\
	$\langle f, g\rangle$ & morphism to product\\
	$[ f, g ]$ & morphism from coproduct
\end{xtabular}

\subsection*{Categories}
\ionAdjustWidth{$\VFunc{\CV}{\CC}{\CD}$}
\begin{xtabular}{ p{\ionLengthA} p{\ionLengthB} }
	$\cat{0}$ & initial category\\
	$\cat{1}$ & final category\\
	$\Set$ & small sets and functions\\
	$\Ord$ & ordinal and inclusions\\
	$\Pos$ & posets and monotonic functions\\
	$\Cpo$ & $\omega$-complete partially ordered sets and continuous functions\\
	$\Cpob$ & $\omega$-complete partially ordered sets with bottoms and strict continuous functions\\
	$\cat{Cppo}$ & $\omega$-complete pointed partially ordered sets and continuous functions, \pageref{def:cppocat}\\
	$\Meas$ & measurable spaces and measurable functions\\
	$(\CS,J)$ & site, \pageref{def:site}\\
	$\PSh[\CC]{\CS}$ & $\CC$-valued presheaves on $\CS$, \pageref{def:presheaf}\\
	$\PSh{\CS}$ & presheaves of sets on $\CS$, \pageref{def:presheaf}\\
	$\Sh[\CC]{\CS,J}$ & $\CC$-valued sheaves on $(\CS,J)$, \pageref{def:sheaf}\\
	$\Sh{\CS,J}$ & sheaves of sets on $(\CS,J)$, \pageref{def:sheaf}\\
	$\Sh[\CC]{\alpha}$ & $\CC$-valued sheaves on the Alexandrov topology of $\alpha$ an ordinal number\\	
	$\Alg{F}$ & $F$-algebras\\
	$\Coalg{F}$ & $F$-coalgebras\\
	$\Kl{T}$ & Kleisli category of $T$\\
	$\Cat$ & locally small categories and functors\\
	$\Func{\CC}{\CD}$ & functors from $\CC$ to $\CD$ and natural transformations, \pageref{def:funcat}\\
	$\Endo{\CC}$ & endofunctors over $\CC$ and natural transformations\\
	$\Mnd{\CC}$ & monads over $\CC$ and natural transformations\\
	$\MndEndo{\CC}$ & distributive laws of monads and endofunctors over $\CC$, \pageref{def:mndendocat}\\
	$\VCat{\CV}$ & categories and functors enriched over $\CV$, \pageref{def:sh-cat}\\ %
	$\VFunc{\CV}{\CC}{\CD}$ & $\CV$-enriched functors from $\CC$ to $\CD$, \pageref{def:sh-funcat}\\ %
	$\VEndo{\CV}{\CC}$ & $\CV$-enriched endofunctors over $\CC$\\
	$\underlying{\CC}$ & externalisation of $\CC$, \pageref{def:externalisation-sh-category}\\ %
\end{xtabular}

\subsection*{Functors}
\ionAdjustWidth{$F \mathrel{\lhd} G$}
\begin{xtabular}{ p{\ionLengthA} p{\ionLengthB} }
	$\later$ & later endofunctor, \pageref{def:later-functor}\\
	$\liftKl{F}$ & Kleisli lifting of $F$, \pageref{def:kleisli-liftings} \\
	$(\extP{-})$ & pointwise extension functor, \pageref{sec:extp-functor} \\
	$\underlying{F}$ & underlying functor, \pageref{def:externalisation-sh-category}\\ %
	$F \dashv G$ & adjunction\\
\end{xtabular}

\subsection*{Greek letters}
\ionAdjustWidth{$\alpha,\beta,\gamma$}
\begin{xtabular}{ p{\ionLengthA} p{\ionLengthB} }
	$\alpha,\beta,\gamma$ & ordinal numbers \\
	$\Gamma$ & global section functor, \pageref{def:constant-sheaf}\\
	$\Delta$ & constant (pre)sheaf functor, \pageref{def:constant-sheaf}\\
	$\varepsilon$ & counit of adjunction\\
	$\eta$ & unit of adjunction\\
	$\eta$ & unit of monad\\
	$\eta^T$ & unit of the monad $T$\\
	$\iota_{U,V}$ & $U \to V$ in a thin category, \pageref{def:morphism-in-thincat}\\
	$\lambda$ & distributive law, \pageref{def:kleisli-lifting-dist-law}\\
	$\lambda^{T,F}$ & distributive law of $T$ and $F$, \pageref{def:kleisli-lifting-dist-law}\\
	$\mu$ & multiplication of monad\\
	$\mu^T$ & multiplication of the monad $T$\\
	$\ifix F$ & initial $F$-algebra \\
	$\ffix F$ & final $F$-coalgebra \\
	$\pi$ & projection \\
	$\pi_d$ & projection at $d$ \\
	$\omega$ & first transfinite ordinal number
\end{xtabular}

\subsection*{Latin letters}
\ionAdjustWidth{$Ran_{F}(G)$}
\begin{xtabular}{ p{\ionLengthA} p{\ionLengthB} }
	$\mathbf{a}$ & associated sheaf functor, \pageref{def:sheafification}\\
	$\alexT{S}$ & Alexandrov topology of preorder $S$, \pageref{def:alexandrov-topology}\\
	$\cod(f)$ & codomain of $f$\\
	$\cstr$ & monad costrength, \pageref{def:monad-costrength}\\
	$\dom(f)$ & domain of $f$\\
	$\mathcal{D}$ & monad of probability distributions\\
	$\dstr$ & monad double strength, \pageref{def:monad-double-strength}\\
	$\eimg(F)$ & essential image of $F$\\
	$\fseq(F)$ & final sequence of $F$, \pageref{sec:final-sequence-sheaf} \\
	$\fix$ & fixed point \\
	$\mathcal{G}$ & Giry monad\\
	$\carr{h}$ & carrier of (co)algebra $h$\\
	$\mathbf{i}$ & right adjoint of $\mathbf{a}$, \pageref{def:sheafification}\\
	$\id$ & identity\\
	$\id_X$ & identity arrow over $X$\\
	$\Id_\CC$ & identity functor over $\CC$\\
  $\img(f)$ & image of $f$\\
	$J$ & coverage, \pageref{def:coverage}\\
	$\varinjlim$ & colimit\\
	$\varprojlim$ & limit\\
	$\nxt$ & point of $\later$, \pageref{def:later-functor}\\ 
	$\obj(\CC)$ & objects of $\CC$\\
	$\opens[X]$ & Open subsets of space $X$\\		
	$\mathbf{p}$ & predecessor functor, \pageref{def:predecessor-functor}\\
	$\mathcal{P}$ & powerset monad\\
	$\mathcal{P}^+$ & non-empty powerset\\
	$\mathcal{P}_f$ & finite powerset\\
	$Ran_{F}(G)$ & Right Kan extension of $G$ along $F$\\
	$\str$ & monad strength, \pageref{def:monad-strength}\\
	$(T,\mu,\eta)$ & monad \\
	$T^{\mathrm{a}}$ & affine part of monad $T$, \pageref{def:affine-part-monad} \\
	$x|_U$ & restriction at stage $U$ of section $x$, \pageref{def:section-restriction}\\
	$X^{+}$ & Grothendieck plus construction, \pageref{sec:sheafification}
\end{xtabular}

\onecolumn

\end{document}